\documentclass[12pt]{article}

\usepackage{tipa}
\usepackage{amsmath}
\usepackage{amssymb}
\usepackage{authblk}

\usepackage{natbib}
\usepackage{graphicx}

\newcommand{\vect}[1]{\boldsymbol{#1}}
\newcommand{\trace}{\operatorname{trace}}

\newcommand{\AS}{\operatorname{AS}}

\graphicspath{
	{./figure/}
}

\begin{document}
	
	\title{Robust Estimation for Multivariate Wrapped Models}
	
	\author[1]{Giovanni Saraceno}
	\author[1]{Claudio Agostinelli}
	\author[2]{Luca Greco}
	\affil[1]{Department of Mathematics, University of Trento, Trento, Italy \texttt{giovanni.saraceno@unitn.it, claudio.agostinelli@unitn.it}}
	\affil[2]{Department DEMM, University of Sannio, Benevento, Italy \texttt{luca.greco@unisannio.it}}
	
	\date{}
	
	\maketitle
	
	\begin{abstract}
		A weighted likelihood technique for robust estimation of a multivariate Wrapped Normal distribution for data points scattered on a $p-$dimensional torus is proposed.
		The occurrence of outliers in the sample at hand can badly compromise inference for standard techniques such as maximum likelihood method. Therefore, there is the need to handle such model inadequacies in the fitting process by a robust technique and an effective downweighting of observations not following the assumed model. Furthermore, the employ of a robust method could help in situations of hidden and unexpected substructures in the data. Here, it is suggested to build a set of data-dependent weights based on the Pearson residuals and solve the corresponding weighted likelihood estimating equations. In particular, robust estimation is carried out by using a Classification EM algorithm whose M-step is enhanced by the computation of weights based on current parameters' values.
		The finite sample behavior of the proposed method has been investigated by a Monte Carlo numerical studies and real data examples.   
	\end{abstract}
	
	\section{Introduction}
	\label{sec:introduction}
	Multivariate circular observations arise commonly in all those fields where a quantity of interest is measured as a direction or when instruments such as compasses, protractors, weather vanes, sextants or theodolites are used \citep{Mardia1972}.  
	Circular (or directional) data can be seen as points on the unit circle and represented by angles, provided that an initial direction and orientation of the circle have been chosen. 
	
	These data might be successfully modeled by using appropriate wrapped distributions such as the Wrapped Normal on the unit circle. The reader is pointed to \citet{MardiaJupp2000}, \citet{JammalamadakaSenGupta2001} and \citet{Batschelet1981} for modeling and inferential issues on circular data.

	When data come in a multivariate setting, we might extend the univariate wrapped distribution by using a component-wise wrapping of multivariate distributions. 
        The multivariate Wrapped Normal ($WN_p$) is obtained by component-wise wrapping of a $p$-variate Normal distribution ($N_p$) on a $p-$dimensional torus \citep{JohnsonWehrly1978, Baba1981}. Wrapping can be explained as the geometric translation of a distribution with support on $\mathbb{R}$ to a space defined on a \textit{circular} object, e.g., a unit circle \citep{MardiaJupp2000}.

        Let $\vect{X} \sim N_p(\vect{\mu}, \Sigma)$, where $\vect{\mu}$ is the vector mean and $\Sigma$ is the variance-covariance matrix. Then, the distribution of $\vect{Y} = \vect{X} \ \textrm{mod} \ 2\pi$ is denoted as $WN_p(\vect{\mu},\Sigma)$ with distribution function
	 	\begin{equation*}
	 F(\vect{y})= \sum_{\vect{j} \in \mathbb{Z}^p} [\Phi_p(\vect{y} + 2 \pi \vect{j}; \Omega)- \Phi_p(2 \pi \vect{j}; \Omega) ] \ ,
	 \end{equation*}
	 and density function
	  \begin{equation*}
	  f(\vect{y})= \sum_{\vect{j} \in \mathbb{Z}^p} \phi_p(\vect{y} + 2 \pi \vect{j}; \Omega) \ , 
	  \end{equation*}
	  with $\vect{y} \in(0,2\pi]^p$, $\vect{j}\in\mathbb{Z}^p$, $\Omega=(\vect{\mu},\Sigma)$,
	where $\Phi_p(\cdot)$ and $\phi_p(\cdot)$ are the distribution and density function of $\vect{X}$, respectively, and the modulus operator \textrm{mod} is applied component-wise.  An appealing property of the wrapped Normal distribution is its closure with respect to convolution \citep{Baba1981, JammalamadakaSenGupta2001}.

Likelihood based inference about the parameters of the $WN_p(\vect{\mu},\Sigma)$ distribution can be trapped in numerical and computational hindrances since the log-likelihood function 
	\begin{equation*} \label{equ:loglik}
	\ell(\Omega) = \sum_{i=1}^n \log \left[ \sum_{\vect{j} \in \mathbb{Z}^p} \phi_p(\vect{y}_i + 2 \pi \vect{j}; \Omega) \right] \,
	\end{equation*} 
	involves the evaluation of an infinite series. 
	 \citet{Agostinelli2007} proposed an Iterative Reweighted Maximum Likelihood Estimating Equations algorithm in the univariate setting, that is available in the \texttt{R} package \texttt{circular} \citep{AgostinelliLund2017}.
	  Algorithms bases on the Expectation-Maximization (EM) method have been used by \citet{Fisher1994} for parameter estimation for autoregressive models of Wrapped Normal distributions and by \citet{Coles1998}, \citet{Ravindran2011} and \citet{Ferrari2009} in a Bayesian framework according to a data augmentation approach to estimate the missing unobserved wrapping coefficients. 
	  An innovative estimation strategy based on EM and Classification EM algorithms has been discussed in \citet{nodehi2018estimation}. In order to perform maximum likelihood estimation, the wrapping coefficients are treated as latent variables. 
          
	 	Let $\vect{y}_1, \ldots, \vect{y}_n$ be a i.i.d. sample from a multivariate Wrapped Normal distribution $\vect{Y} \sim WN_p(\vect{\mu}, \Sigma)$ on the $p$-torus with mean vector $\vect{\mu}$ and variance-covariance matrix $\Sigma$. We can think of $\vect{y}_i = \vect{x}_i \mod 2\pi$ where $\vect{x}_i$ is a sample from $\vect{X}_i \sim N_p(\vect{\mu}, \Sigma)$.
	 The EM algorithm works with the complete log-likelihood function given by 
	 \begin{equation} \label{equ:completeloglik}
	 \ell_C(\Omega) = \sum_{i=1}^n \log\left[ \sum_{\vect{j} \in \mathbb{Z}^p} v_{i\vect{j}}  \phi(\vect{y}_i + 2 \pi \vect{j}; \Omega)\right] \ ,
	 \end{equation} 
	 that is characterized by the missing unobserved wrapping coefficients $\vect{j}$ and
	 $v_{i\vect{j}}$ is an indicator of the $i$th unit having the $\vect{j}$ vector as wrapping coefficients. The EM algorithm iterates between an Expectation (E) step and a Maximization (M) step. In the E-step, the conditional expectation of (\ref{equ:completeloglik}) is obtained by estimating the $v_{i\vect{j}}$ with the posterior probability that $\vect{y}_i$ has $\vect{j}$ as wrapping coefficients based on current parameters' values, i.e.
	 \begin{equation*}
	 v_{i\vect{j}} =  \frac{\phi(\vect{y}_i + 2 \pi \vect{j}; \Omega)}{\sum_{\vect{h} \in \mathbb{Z}^p} \phi(\vect{y}_i + 2 \pi \vect{h}; \Omega)} \ , \qquad \vect{j} \in \mathbb{Z}^p, \quad i=1,\ldots,n \ .
	 \end{equation*}  
	 In the M-step, the conditional expectation of (\ref{equ:completeloglik}) is maximized with respect to $\Omega$. The reader is pointed to \citet{nodehi2018estimation} for computational details about such maximization problem and updating formulas for $\Omega$. 
	 
	 An alternative estimation strategy is based on the CEM-type algorithm. The substantial difference is that the E-step is followed by a C-step (where C stands for classification) in which $v_{i\vect{j}}$  is estimated
	 as either 0 or 1 and so that each observation $\vect{y}_i$ is associated to the most likely wrapping coefficients $\vect{j}_i$ with $\vect{j}_i = \arg\max_{\vect{h} \in \mathbb{Z}^p} v_{i\vect{h}}$.
	 
	When the sample data is contaminated by
	the occurrence of outliers, it is well known that maximum likelihood estimation, also achieved through the implementation of the EM or CEM algorithm, is likely to lead to unreliable results \citep{farcomeni2016robust}. Then, there is the need for a suitable robust procedure providing protection against those unexpected anomalous values.
	An attractive solution would be to modify the likelihood equations in the M-step by introducing a set of weights aimed to bound the effect of outliers. Here, it is suggested to evaluate weights according to the weighted likelihood methodology \citep{markatou1998}. Weighted likelihood is an appealing robust technique for estimation and testing \citep{agostinelli2001test}. The methodology leads to a robust fit and gives the chance to detect possible substructures in the data. 
	Furthermore, the weighted likelihood methodology works in a very satisfactory fashion when combined with the EM and CEM algorithms, as in the case of mixture models \citep{greco2018weighted, greco2020weighted}.
	 
	The remainder of the paper is organized as follows. Section \ref{back} gives brief but necessary preliminaries on weighted likelihood. The weighted CEM algorithm for robust fitting of the multivariate Wrapped Normal model on data on a $p-$dimensional torus is discussed in 
	Section \ref{sec:estimation}.  
	Section \ref{sec:simulations} reports the results of some numerical studies, whereas a real data example is discussed in Section \ref{sec:examples}. Concluding remarks end the paper.

	\section{Preliminaries on weighted likelihood}
	\label{back}
	
	Let $\vect{y}_1, \cdots, \vect{y}_n$ be a random sample of size $n$ drawn from a r.v. $\vect{Y}$ with distribution function $F$ and probability (density) function $f$. Let $\mathcal{M} =  \{ M(\vect{y}; \vect{\theta}), \vect{\theta} \in \Theta \subseteq \mathbb{R}^d, d \geq 1, \vect{y} \in \mathcal{Y} \}$ be the assumed parametric model, with corresponding density $m(\vect{y};\vect{\theta})$, and 
	$\hat F_n$ the empirical distribution function. Assume that the support of $F$ is the same as that of $M$ and independent of $\vect{\theta}$. A measure of the agreement between the {\it true} and assumed model is provided by the Pearson residual function $\delta(\vect{y})$, with $\delta(\vect{y})\in [-1,+\infty)$, \citep{lindsay1994, markatou1998}, defined as 
	\begin{equation}
	\label{pearson}
	\delta(\vect{y}) = \delta(\vect{y}; \vect{\theta}, F) = \frac{f(\vect{y})}{m(\vect{y}; \vect{\theta})} - 1 \ .
	\end{equation}
The finite sample counterpart of (\ref{pearson}) can be obtained as
	 \begin{equation}
	 \label{residualfs}
	 \delta_n(\vect{y}) = \delta(\vect{y}; \vect{\theta}, \hat F_n) = \frac{\hat f_n(\vect{y})}{m(\vect{y}; \vect{\theta})} - 1 \ ,
	 \end{equation}
	where $\hat f_n(\vect{y})$ is a consistent estimate of the true density $f(\vect{y})$. 
	In discrete families of distributions, $\hat{f}_n(\vect{y})$ can be driven by the observed relative frequencies \citep{lindsay1994}, whereas in continuous models one could consider a non parametric density estimate based on the kernel function $k(\vect{y};\vect{t},h)$, that is
	 \begin{equation}
           \label{eqn:parametric-density}
	 \hat f_n(\vect{y})=\int_\mathcal{Y}k(\vect{y};\vect{t},h)d\hat F_n(\vect{t}) \ .
	 \end{equation}
	Moreover, in the continuous case, the model density in (\ref{residualfs}) can be replaced by a smoothed
	model density, obtained by using the same kernel involved in non-parametric density estimation \citep{basu1994minimum, markatou1998}, that is
	\begin{equation*}
	\hat m(\vect{y}; \vect{\theta})=\int_\mathcal{Y}k(\vect{y};\vect{t},h)m(\vect{t};\vect{\theta}) \ d\vect{t} \ 
	\end{equation*} 
	leading to
	\begin{equation}
	\label{residualfs2}
	\delta_n(\vect{y}) = \delta(\vect{y}; \vect{\theta}, \hat F_n) = \frac{\hat f_n(\vect{y})}{\hat m(\vect{y}; \vect{\theta})} - 1 \ .
	\end{equation}
	By smoothing the model, the Pearson residuals in (\ref{residualfs2}) converge to zero with probability one for every $\vect{y}$  under the assumed model and it is not required that the kernel bandwidth $h$ goes to zero as the sample size $n$ increases. 
	Large values of the Pearson residual function correspond to regions of the support $\mathcal{Y}$ where the model fits the data poorly, meaning that the observation is unlikely to occur under the assumed model.
	The reader is pointed to \cite{basu1994minimum}, \cite{markatou1998}, \cite{agostinelli2019weighted} and references therein for more details. 
	
	Observations leading to large Pearson residuals in (\ref{residualfs2}) are supposed to be down-weighted. Then, a weight in the interval $[0,1]$ is attached to each data point, that is  computed accordingly to the following
	weight function
	\begin{equation}
	\label{weight}
	w(\delta(\vect{y})) = \frac{\left[A(\delta(\vect{y})) + 1\right]^+}{\delta(\vect{y}) + 1} \ ,
	\end{equation}
	where $w(\delta)\in[0,1]$, $[\cdot]^+$ denotes the positive part and 
	$A(\delta)$ is the Residual Adjustment Function (RAF, \cite{basu1994minimum}).
	The weights $w(\delta_n(\vect{y}))$ are meant to be small for those data points that are in disagreement with the assumed model. Actually, 
	the RAF plays the role to bound the effect of large Pearson residuals on the fitting procedure.
	By using a RAF such that $|A(\delta)| \le |\delta|$ outliers are expected to be properly downweighted.
	The weight function (\ref{weight}) can be based on the families of RAF stemming from the 
	Symmetric Chi-squared divergence \citep{markatou1998}, the Generalized Kullback-Leibler divergence \citep{park+basu+2003}
        \begin{equation}
          \label{eq:raf-gkl}
	  A_{gkl}(\delta, \tau)=\frac{\log (\tau\delta+1)}{\tau}, \ 0\leq \tau \leq 1;
        \end{equation}
	or the
	Power Divergence Measure \citep{cressie+1984, cressie+1988}
	\begin{equation*}
	A_{pdm}(\delta, \tau) = \left\{
	\begin{array}{lc}
	\tau \left( (\delta + 1)^{1/\tau} - 1 \right) & \tau < \infty \\
	\log(\delta + 1) & \tau \rightarrow \infty \ .
	\end{array}
	\right .
	\end{equation*}
 The resulting weight function is unimodal and declines smoothly to zero as $\delta(\vect{y})\rightarrow -1$ or $\delta(\vect{y})\rightarrow\infty$.
	
Then, robust estimation can be based on
a Weighted Likelihood Estimating Equation (WLEE), defined as 
	\begin{equation} 
	\label{equ+wlee}
	\sum_{i=1}^n w(\delta_n(\vect{y}_i); \vect{\theta}, \hat{F}_n) s(\vect{y}_i;  \vect{\theta}) = 0 \ ,
	\end{equation}
	where $s(\vect{y}_i;\vect{\theta})$ is the individual contribution to the score function. Therefore, 
	weighted likelihood estimation can be thought as a root solving problem. Finding the solution of (\ref{equ+wlee}) requires an iterative weighting algorithm.
	
	The corresponding weighted likelihood estimator $\hat{\vect{\theta}}^w$ (WLE) is consistent, asymptotically normal and fully efficient at the assumed model, under some general regularity conditions pertaining the model, the kernel and the weight function \citep{markatou1998, agostinelli2001test, agostinelli2019weighted}. Its robustness properties have been established in \citet{lindsay1994} in connection with minimum disparity problems. It is worth to remark that under very standard conditions, one can build a simple WLEE matching a minimum disparity objective function, hence inheriting its robustness properties.
	
	In finite samples, the robustness/efficiency trade-off of weighted likelihood estimation can be tuned by varying the smoothing parameter $h$ in equation (\ref{eqn:parametric-density}). Large values of $h$ lead to Pearson residuals all close to zero and weights all close to one and, hence, large efficiency, since $\hat f_n(\vect{y})$ is stochastically close to the postulated model. On the other hand, small values of $h$ make $\hat f_n(\vect{y})$ more sensitive to the occurrence of outliers and the Pearson residuals become large for those data points that are in disagreement with the model. On the contrary, the shape of the kernel function $k(\vect{y};\vect{t},h)$ has a very limited effect.
	
	For what concerns the tasks of testing and setting confidence regions, a weighted likelihood counterparts of the classical likelihood ratio test, and its asymptotically equivalent Wald and Score versions, can be established. Note that, all share the standard asymptotic distribution at the model, according to the results stated in \citep{agostinelli2001test}, that is
	$$
	\Lambda(\vect{\theta})=2\sum_{i=1}^nw_i\left[\ell(\hat{\vect{\theta}}^w; \vect{y}_i)-\ell(\vect{\theta}; \vect{y}_i)\right]\stackrel{p}{\rightarrow}\chi^2_p \ ,
	$$
	with $w_i= w(\delta_n(\vect{y}_i); \hat{\vect{\theta}}^w, \hat{F}_n) $.
   Profile tests can be obtained as well.

	\section{A weighted CEM algorithm}
	\label{sec:estimation}
	
	As previously stated in the Introduction, \citet{nodehi2018estimation} provided effective iterative algorithms to fit a multivariate Wrapped normal distribution on the $p-$dimensional torus.
	Here, robust estimation is achieved by a suitable modification of their CEM algorithm, consisting in a weighting step 
	before performing the M-step, in which data-dependent weights are evaluated according to (\ref{weight}) yielding a WLEE (\ref{equ+wlee}) to be solved in the M-step. 
	
	The construction of Pearson residuals in (\ref{residualfs2}) involves a multivariate Wrapped Normal kernel with covariance matrix $h \Lambda$. Since the family of multivariate Wrapped Normal is closed under convolution, then the smoothed model density is still Wrapped Normal with covariance matrix $\Sigma+h\Lambda$. Here, we set $\Lambda = \Sigma$ so that $h$ can be a constant independent of the variance-covariance structure of the data.  
	
	The weighted CEM algorithm is structured as follows:
	\begin{itemize}
	\item[0] {\bf Initialization},
	Starting values can be obtained by maximum likelihood estimation evaluated over a randomly chosen subset. The subsample size is expected to be as small as possible in order to increase the probability to get an outliers' free initial subset but large enough to guarantee estimation of the unknown parameters. 
	A starting solution for $\mu$ can be obtained by
	the circular means, 
	whereas the diagonal entries of $\Sigma$ can be initialized as $-2\log(\hat{\rho}_r)$, where $\hat{\rho}_r$ is the sample mean resultant length and the off-diagonal elements by 
	$\rho_c(\vect{y}_r, \vect{y}_s) \sigma_{rr}^{(0)} \sigma_{ss}^{(0)}$ ($r \neq s$), where $\rho_c(\vect{y}_r, \vect{y}_s)$ is the circular correlation coefficient, $r=1,2,\ldots,p$ and $s=1,2,\ldots,p$, see \citet[][pag. 176, equation 8.2.2]{JammalamadakaSenGupta2001}.
	In order to avoid the algorithm to be dependent on initial values, a simple and common strategy is to run the algorithm from a number of starting values using the bootstrap root searching approach as in \citet{markatou1998}. A criterion to choose among different solutions will be illustrated in Section \ref{sec:examples}.
	
      \item[1] {\bf E-step}. Based on current parameters' values, first  evaluate posterior probabilities $$v_{i\vect{j}} =  \frac{\phi(\vect{y}_i + 2 \pi \vect{j}; \Omega)}{\sum_{\vect{h} \in \mathbb{Z}^p} \phi(\vect{y}_i + 2 \pi \vect{h}; \Omega)} \ , \qquad \vect{j} \in \mathbb{Z}^p, \quad i=1,\ldots,n \ ,$$

        \item[2] {\bf C-step} Set $\vect{j}_i = \arg\max_{\vect{h} \in \mathbb{Z}^p} v_{i\vect{h}}$, where $v_{i\vect{j}}=1$ for $\vect{j}=\vect{j}_i$, and $v_{i\vect{j}}=0$ otherwise.
	
	\item[3] {\bf W-step} (Weighting step) Based on current parameters' values, compute Pearson residuals according to (\ref{residualfs2}) based on a multivariate Wrapped Normal kernel with bandwidth $h$ and evaluate the weights as
          \begin{equation*}
            w_i=w(\delta_n(\vect{y}_i), \Omega, \hat F_n).
          \end{equation*}
	
	\item[4] {\bf M-step} Update parameters'  values by computing a weighted mean and variance-covariance matrix with weights $w_i$, used to compute estimates, by maximizing the classification log-likelihood conditionally on $\hat{\vect{j}}_i$ $(i = 1,\ldots,n)$, given by
          \begin{align*}
            \hat{\vect{\mu}}_i &= \frac{\sum_{i=1}^n w_i \hat{\vect{x}}_i}{\sum_{i=1}^n w_i}, \\ 
	    \Sigma_{ij} &= \frac{\sum_{i=1}^n (\hat{\vect{x}}_i - \hat{\vect{\mu}}_i)^\top (\hat{\vect{x}}_j - \hat{\vect{\mu}}_j)w_i}{\sum_{i=1}^n w_i}.
          \end{align*}
          Note that, at each iteration the classification algorithm provides also an estimate of the original unobserved sample obtained as $\hat{\vect{x}}_i = \vect{y}_i + 2 \pi \hat{\vect{j}}_i$, $i = 1, \ldots, n$.
	\end{itemize}

	\section{Numerical studies}
	\label{sec:simulations}
	
	The finite sample behavior of the proposed weighted CEM has been investigated by some numerical studies based on 500 Monte Carlo trials each. Data have been drawn from a $WN_p(\vect{\mu},\Sigma)$. We set $\vect{\mu}=0$, whereas in order to account for the lack of affine equivariance of the Wrapped Normal model \citep{nodehi2018estimation}, we consider different covariance structures $\Sigma$ as in \citet{AgostinelliEtAll2015}.
        In particular, for fixed condition number $CN = 20$, we construct a random correlation matrix $R$. Then, we convert the correlation matrix $R$ into the covariance matrix $\Sigma = D^{1/2} R D^{1/2}$, with $D=\textrm{diag}(\sigma\vect{1}_p)$, where $\sigma$ is a chosen constant and $\vect{1}_p$ is a $p$-dimensional vector of ones. 
	Outliers have been obtained by shifting a proportion $\epsilon$ of randomly chosen data points by an amount $k_\epsilon$ in the direction of the smallest eigenvalue of $\Sigma$.
	We consider sample sizes $n=50,100,500$, dimensions $p=2,5$, contamination level $\epsilon=0, 5\%, 10\%, 20\%$,
	contamination size $k_\epsilon=\pi/4, \pi/2, \pi$ and $\sigma=\pi/8, \pi/4, \pi/2$.

        For each combination of the simulation parameters, we are going to compare the performance of CEM and weighted CEM algorithms. The weights used in the W-step are computed using the Generalized Kullback-Leibler RAF in equation (\ref{eq:raf-gkl}) with $\tau = 0.1$. According to the strategy described in \cite{agostinelli2001test}, the bandwidth $h$ has been selected by setting $\Lambda = \Sigma$, so that $h$ is a constant independent of the scale of the model. 
	Here, $h$ is obtained so that any outlying observation located at least three standard deviations away from the mean in a componentwise fashion, is attached a weight not larger than 0.12 when the rate of contamination in the data has been fixed equal to $20\%$. The algorithm has been initialized according to the root search approach described in \cite{markatou1998} based on $15$ subsamples of size $10$.  

  The weighted CEM is assumed to have reached convergence when at the $(k+1)$th iteration 
	$$
	\max \left(\sqrt{2(1-\cos(\hat{\vect{\mu}}^{(k)}-\hat{\vect{\mu}}^{(k+1)}))}, \max|\hat{\Sigma}^{(k)}-\hat{\Sigma}^{(k+1)} | \right) <10^{-6}.
	$$
	The algorithm has been implemented so that $\mathbb{Z}^p$ is replaced by the Cartesian product $\times_{s=1}^p \vect{\mathcal{J}}$ where $\vect{\mathcal{J}} = (-J, -J+1, \ldots, 0, \ldots, J-1, J)$ for some $J$ providing a good approximation. Here we set $J=3$.
	The algorithm runs on \texttt{R} code \citep{cran} available from the authors upon request.
	
\begin{figure}
	 \begin{center}
	 \includegraphics[height=0.3\textheight, width=0.3\textwidth]{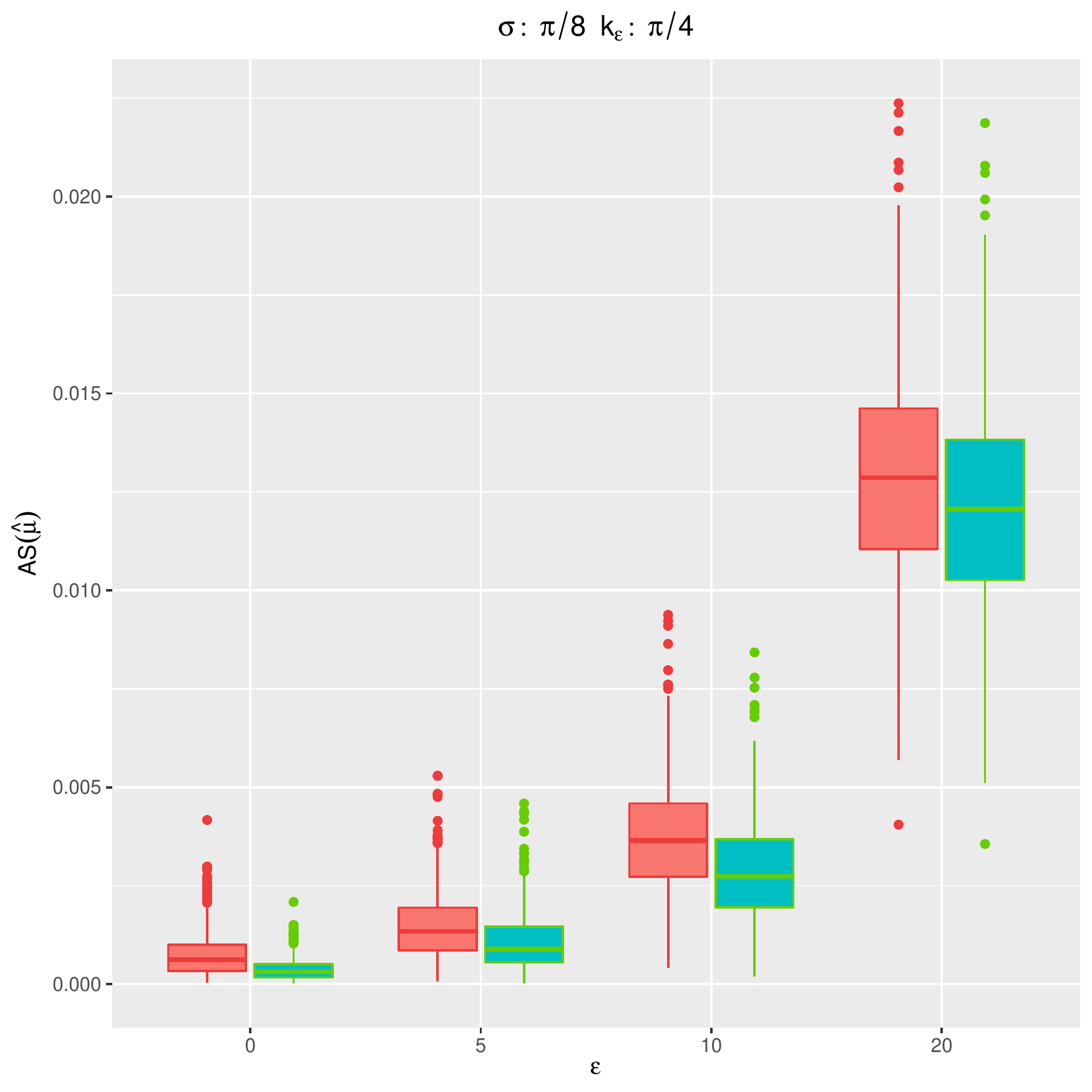} 
	 \includegraphics[height=0.3\textheight, width=0.3\textwidth]{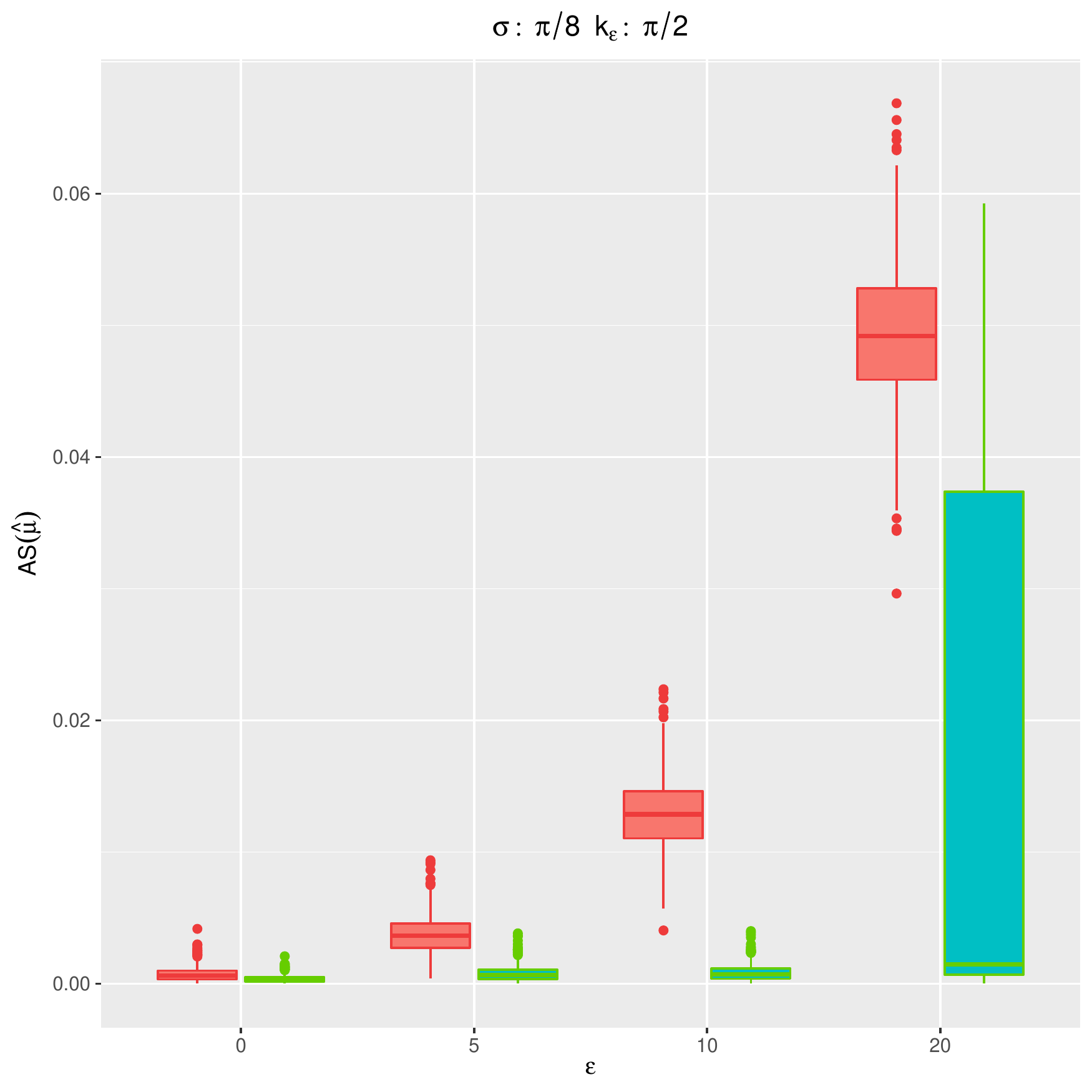} 
	 \includegraphics[height=0.3\textheight, width=0.3\textwidth]{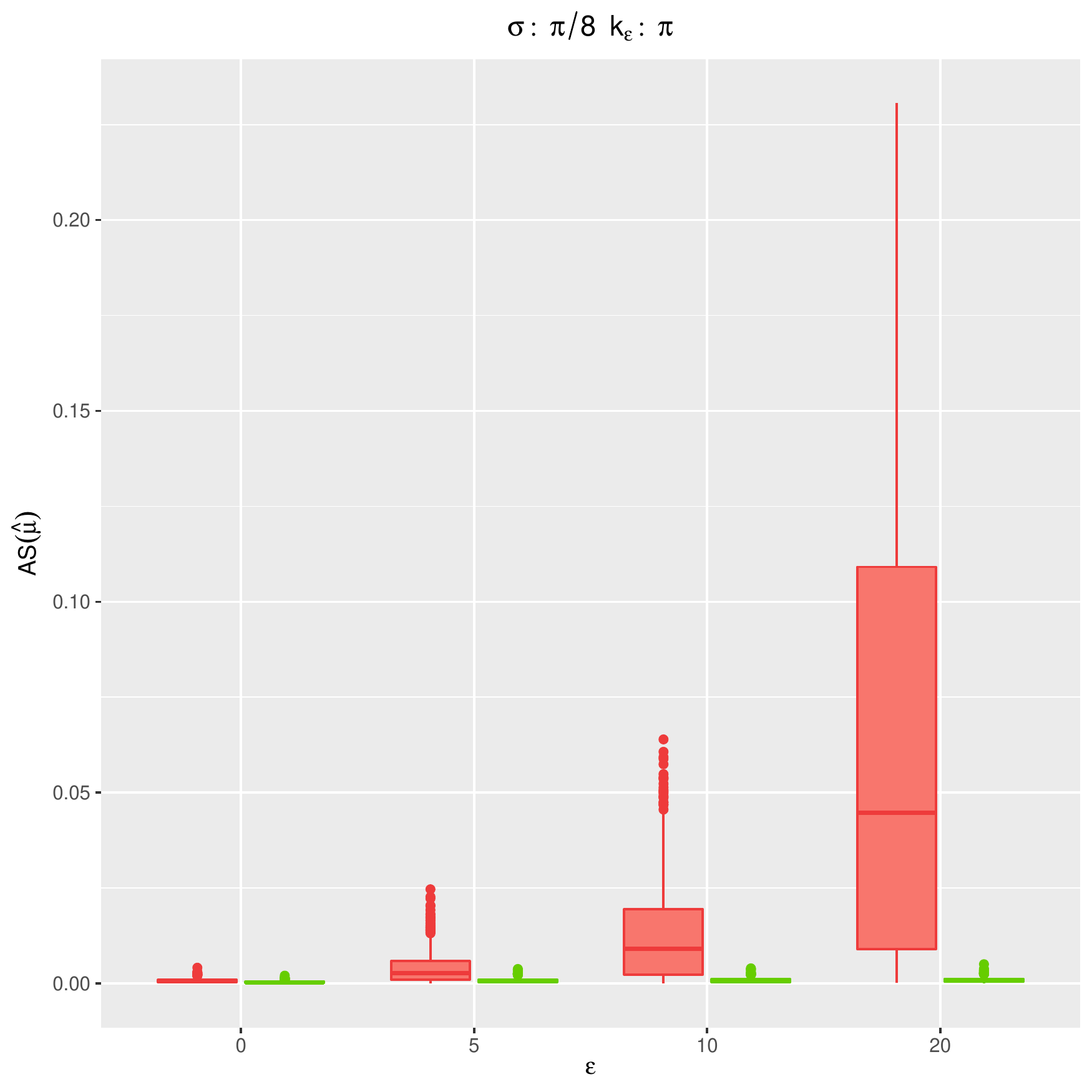} \\
	\includegraphics[height=0.3\textheight, width=0.3\textwidth]{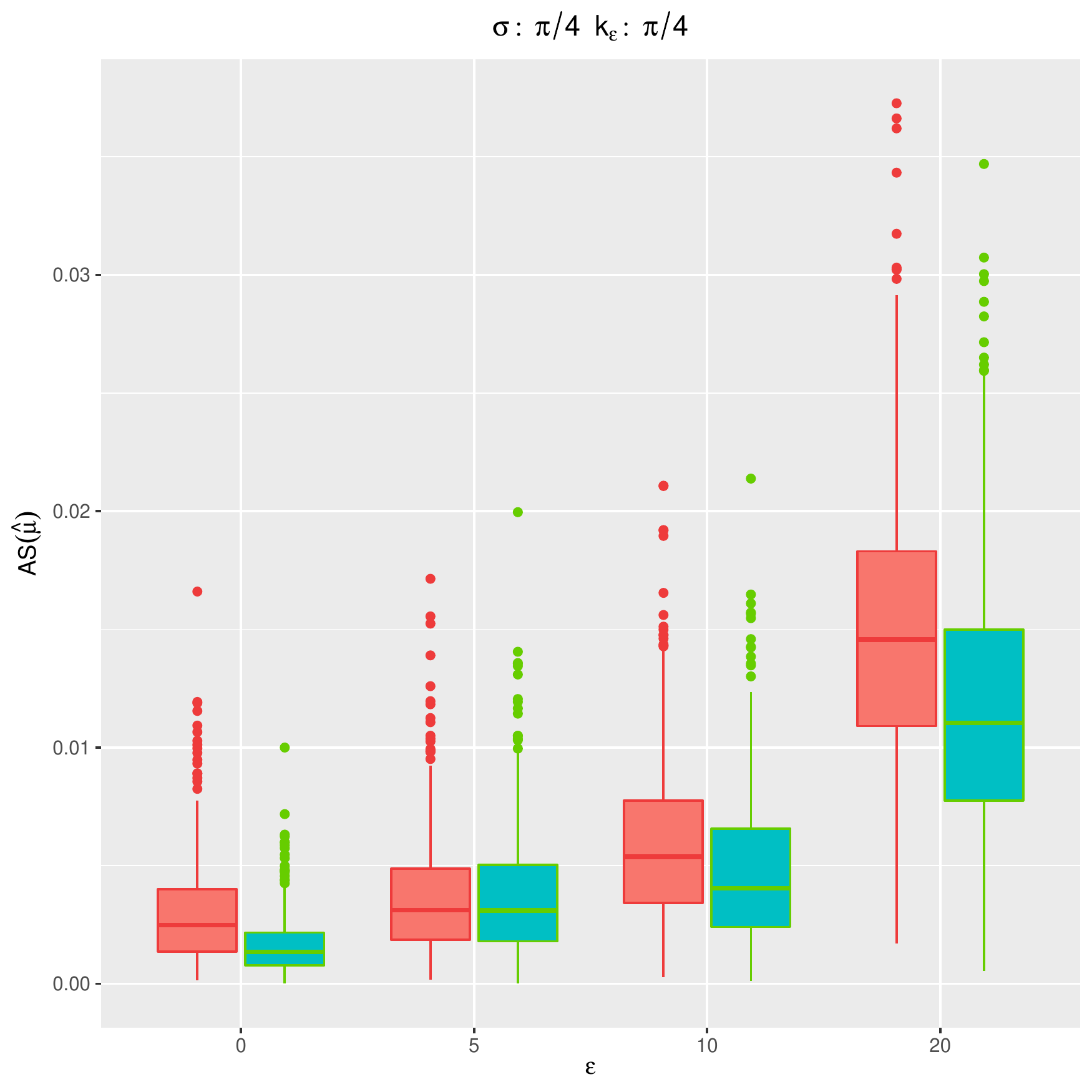} 
	 \includegraphics[height=0.3\textheight, width=0.3\textwidth]{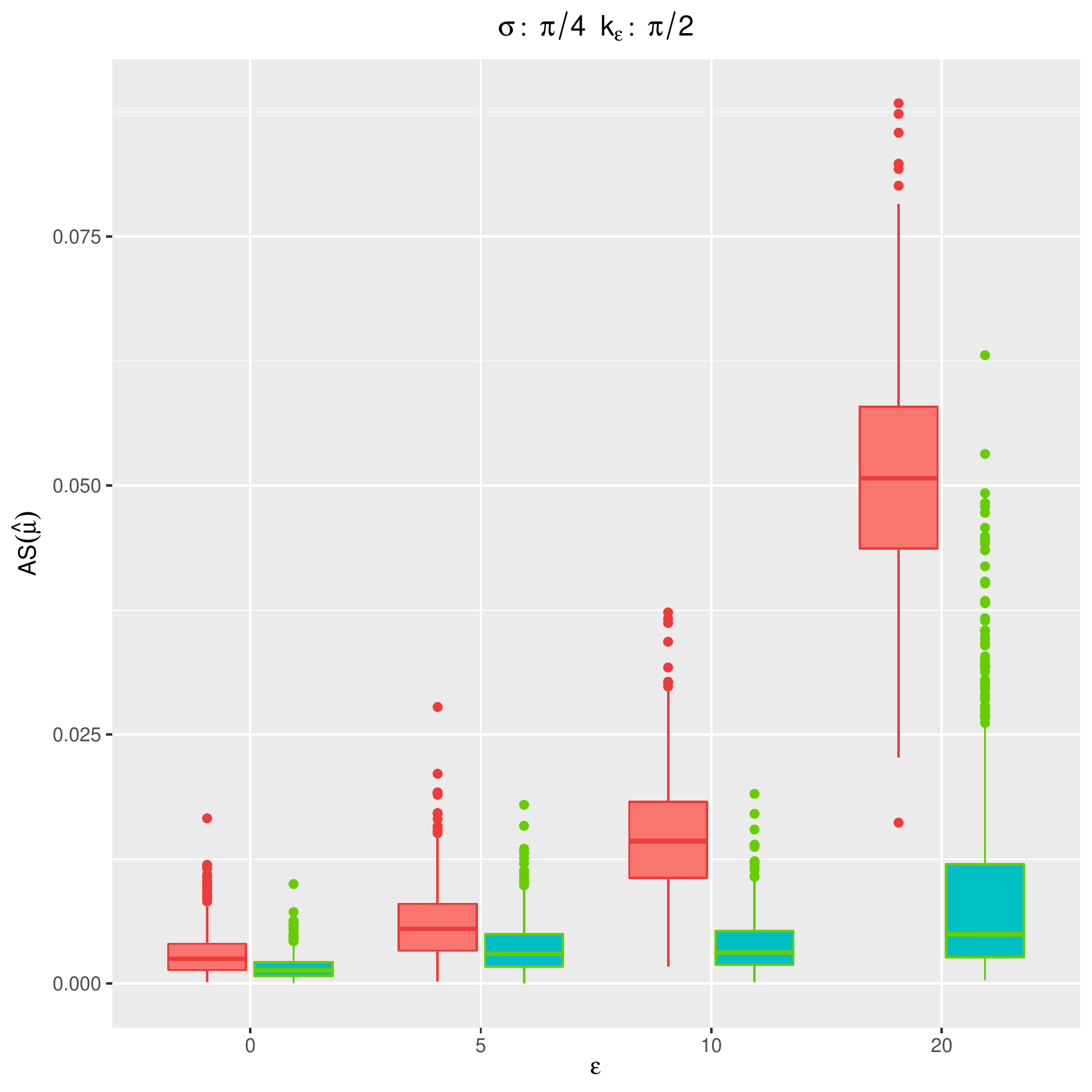} 
	 \includegraphics[height=0.3\textheight, width=0.3\textwidth]{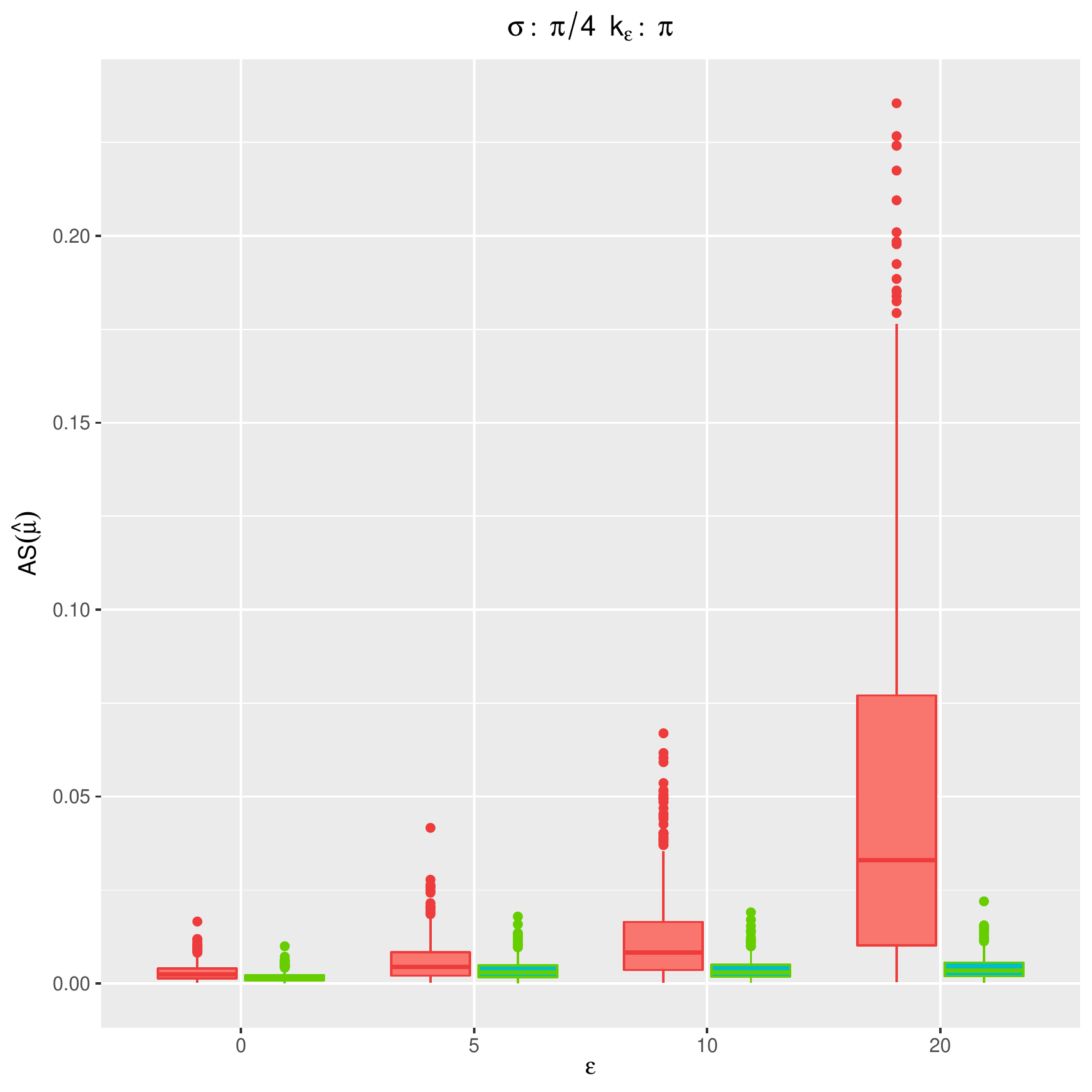} \\
	 \includegraphics[height=0.3\textheight, width=0.3\textwidth]{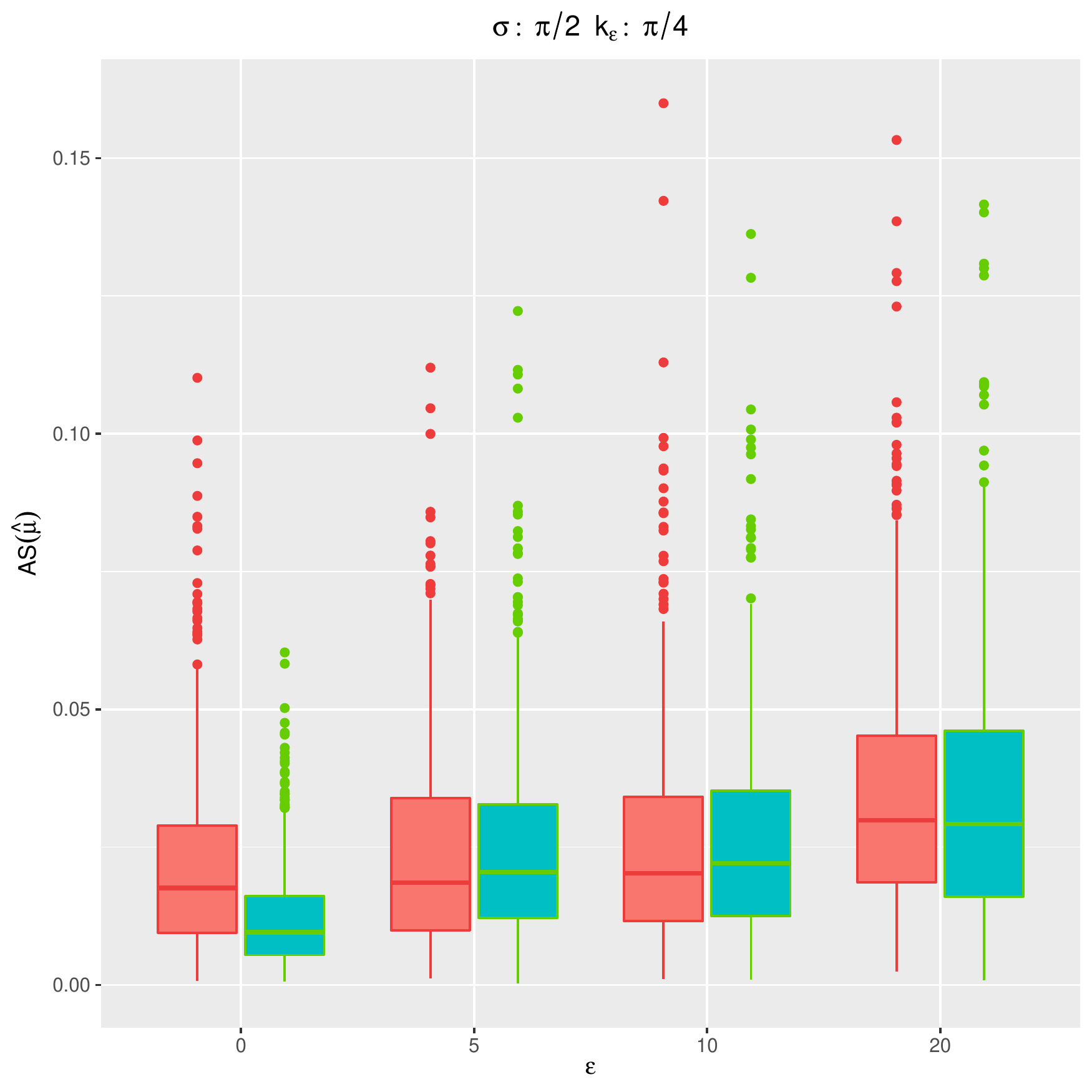} 
	 \includegraphics[height=0.3\textheight, width=0.3\textwidth]{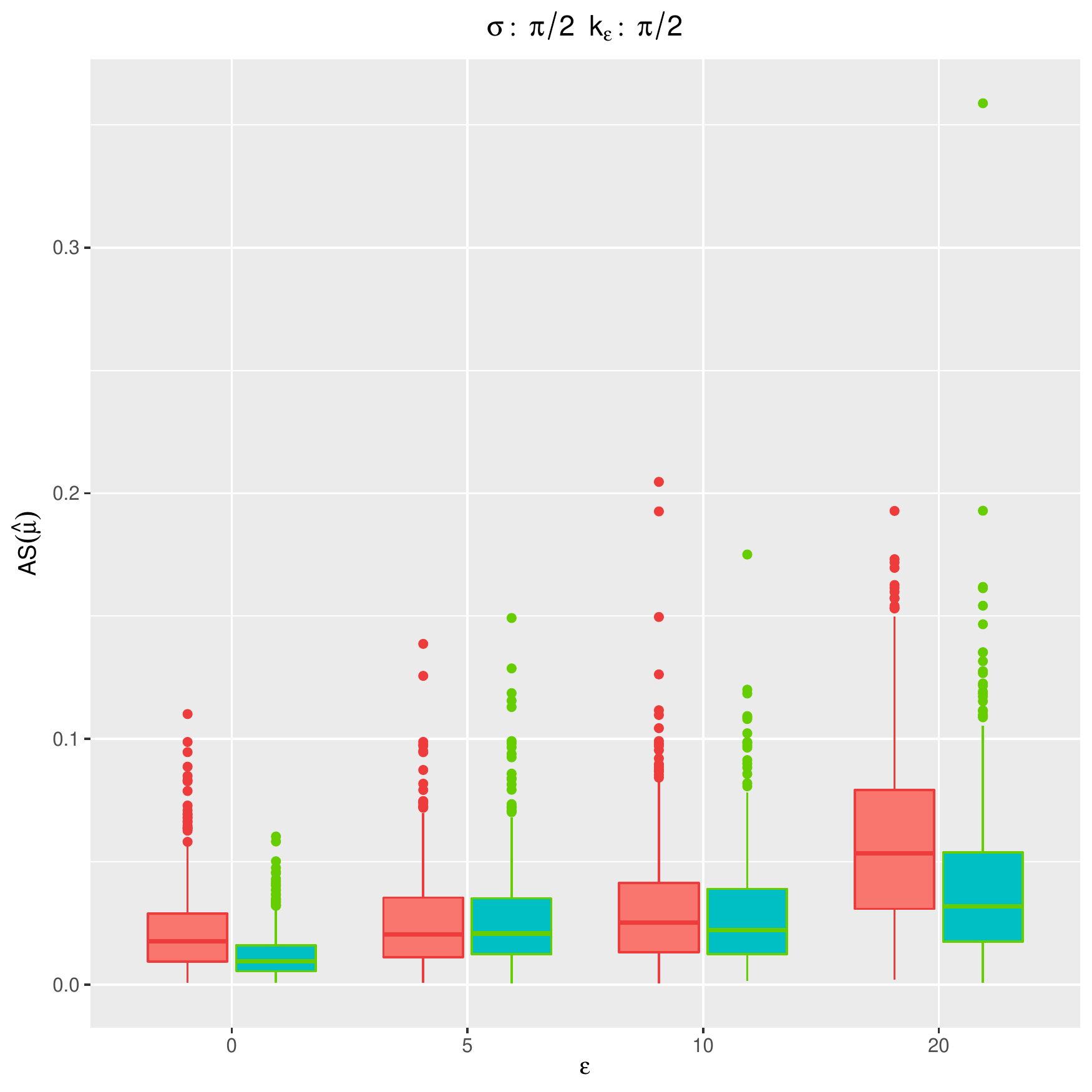} 
	 \includegraphics[height=0.3\textheight, width=0.3\textwidth]{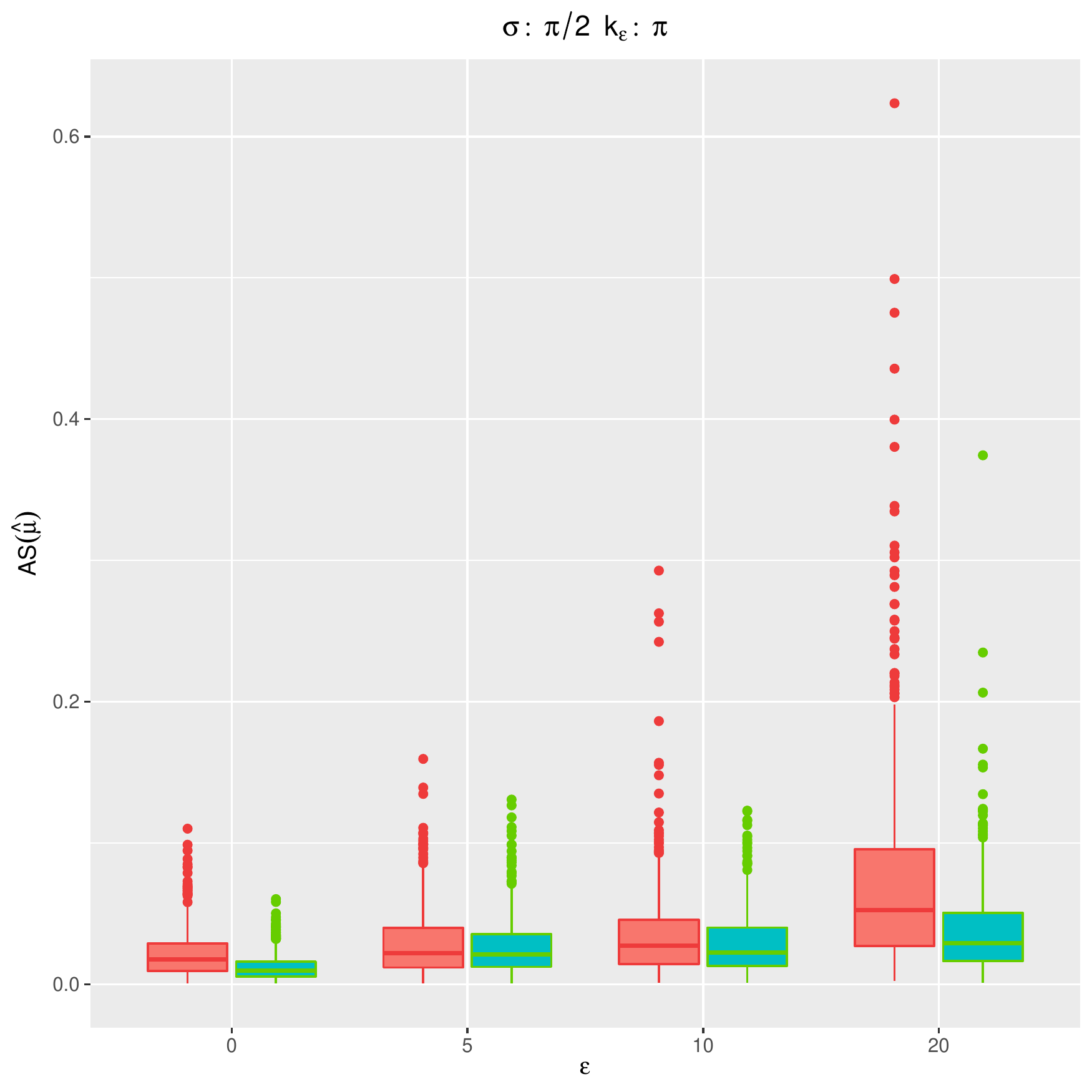} \\
	 \end{center}
	 \caption{Distribution of average angle separation for $n=100$ and $p=5$ using weighted CEM (in green) and the CEM (in red). The contamination rate $\epsilon$ is given on the horizontal axis. Increasing contamination size $k_\epsilon$ from left to right, increasing $\sigma$ from top to bottom.}
	 \label{fig:one}
	\end{figure}

\begin{figure}
 	\begin{center}
 		\includegraphics[height=0.3\textheight, width=0.3\textwidth]{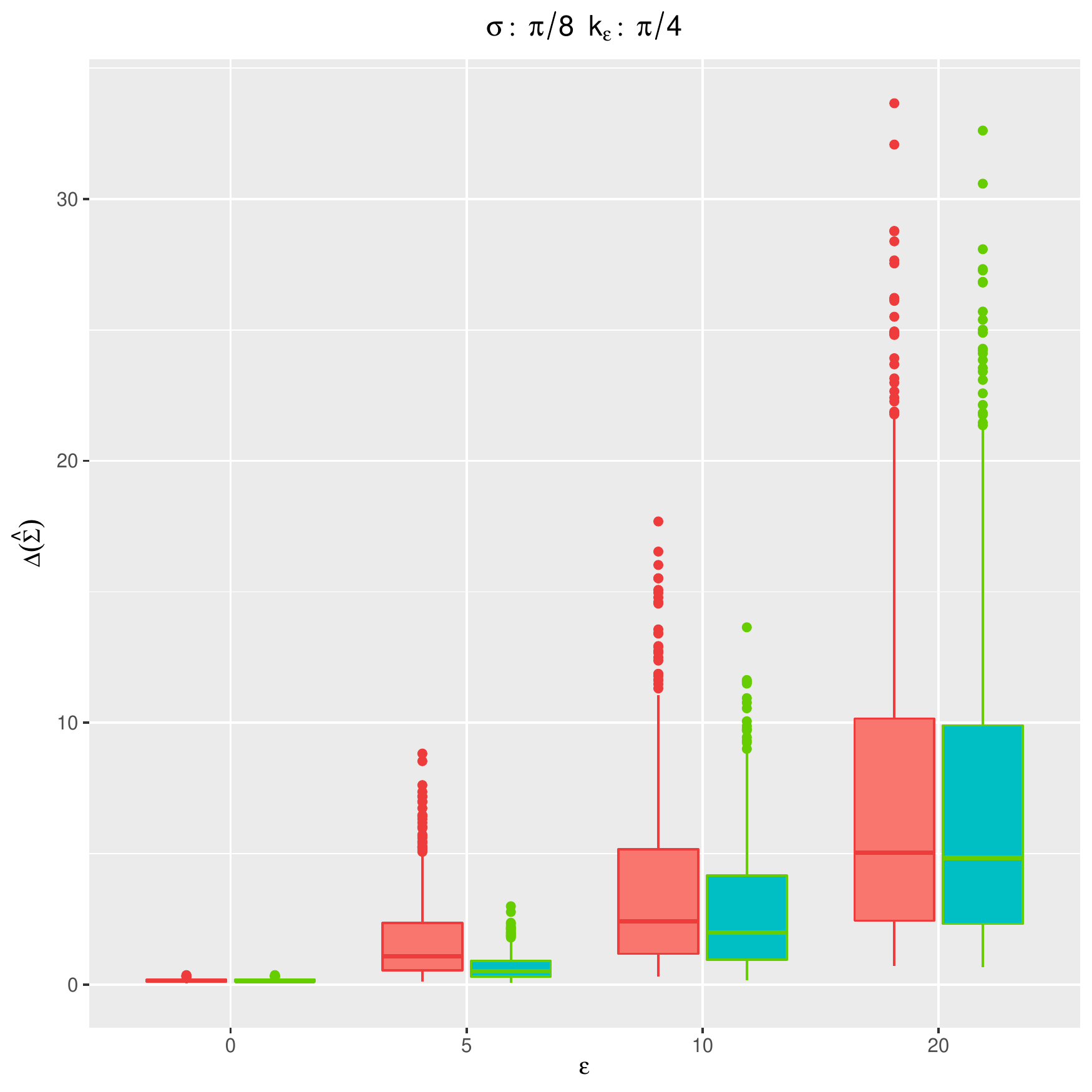} 
 		\includegraphics[height=0.3\textheight, width=0.3\textwidth]{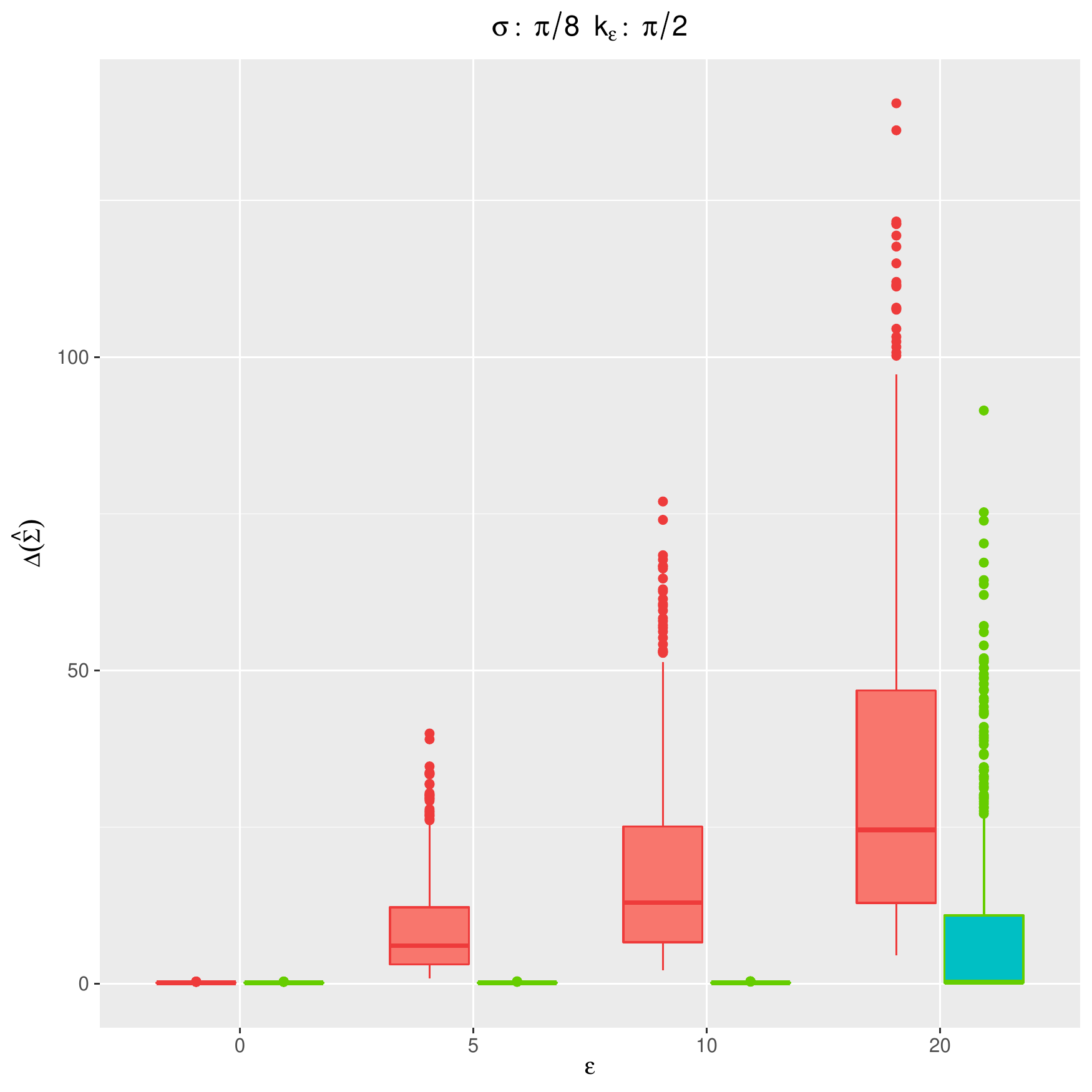} 
 		\includegraphics[height=0.3\textheight, width=0.3\textwidth]{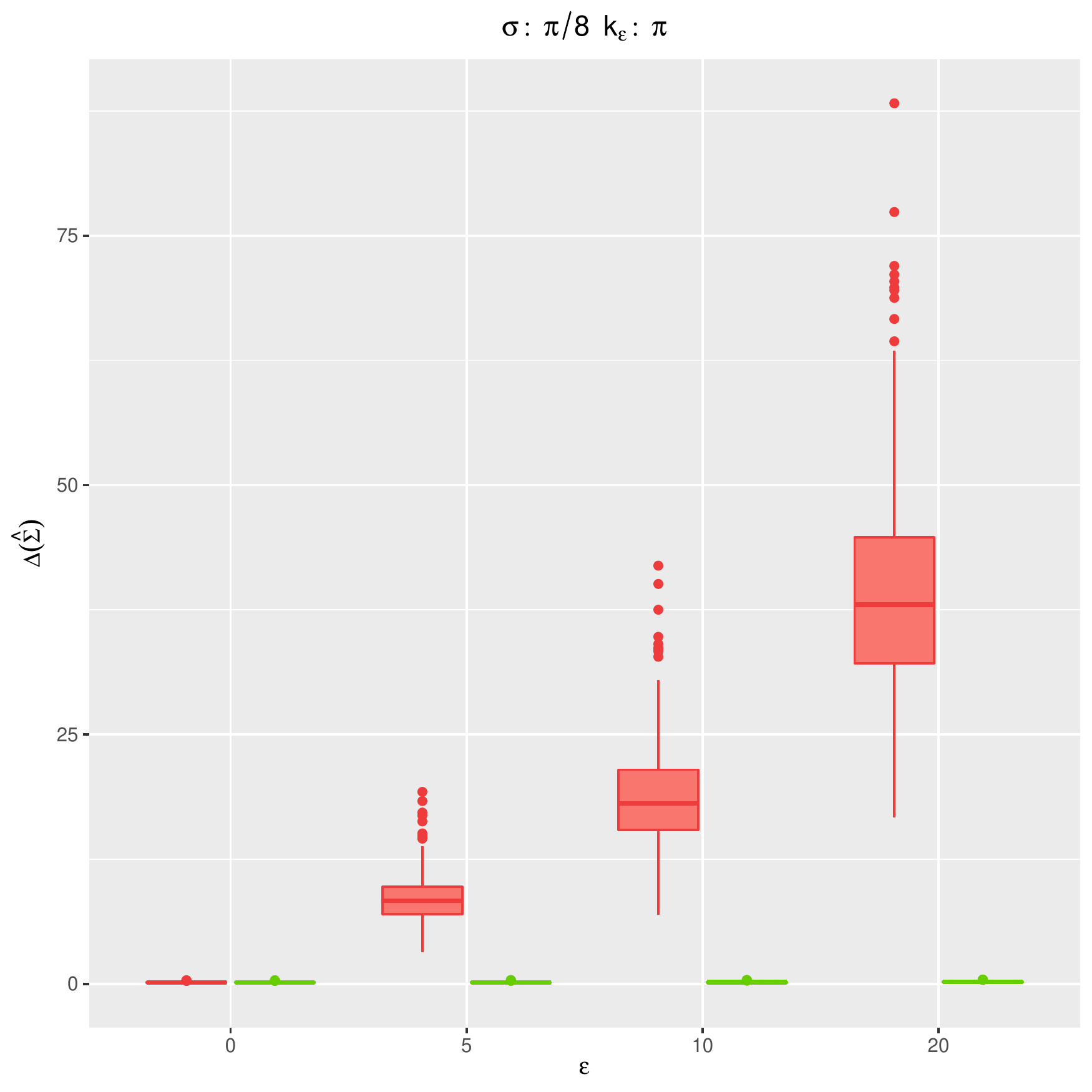} \\
 		\includegraphics[height=0.3\textheight, width=0.3\textwidth]{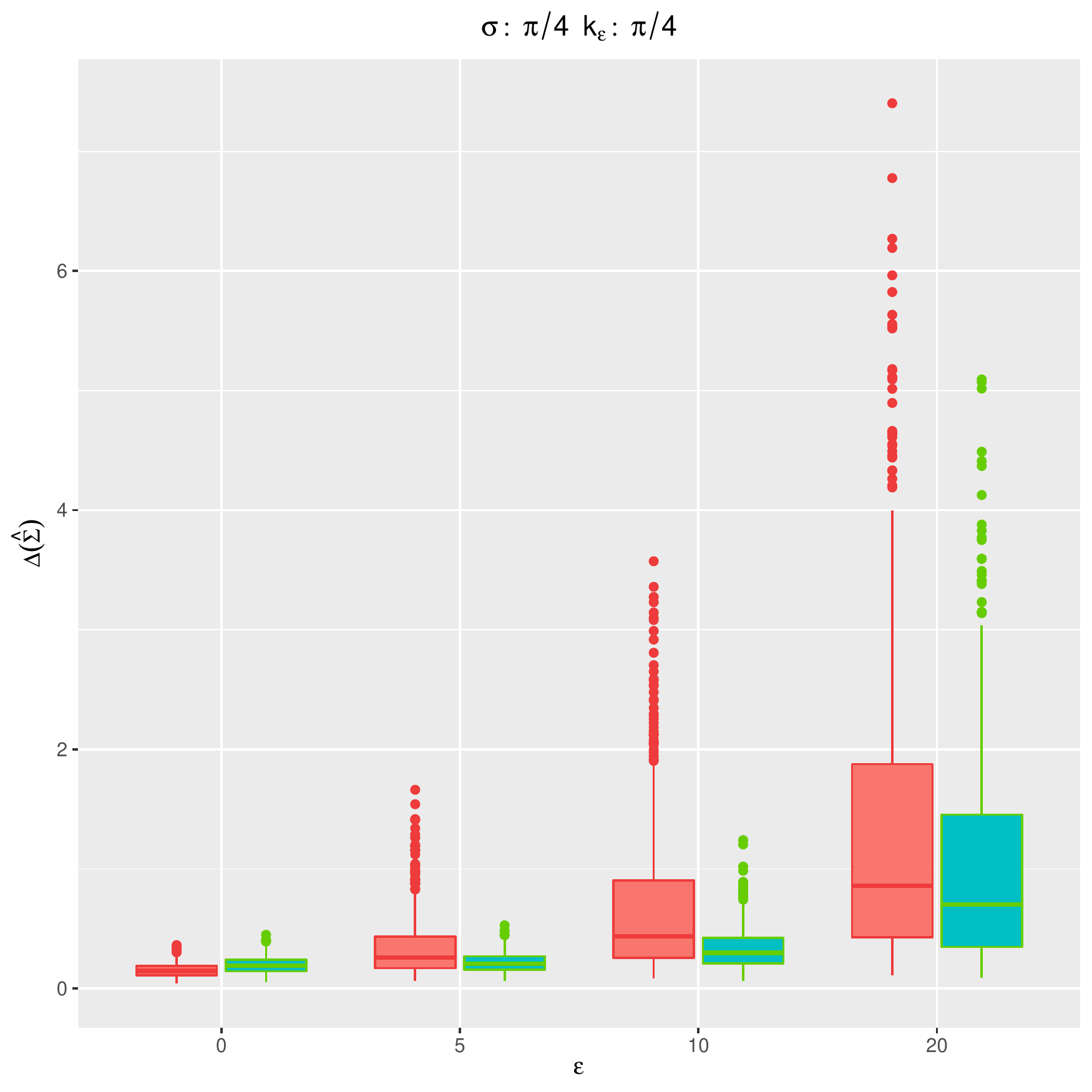} 
 		\includegraphics[height=0.3\textheight, width=0.3\textwidth]{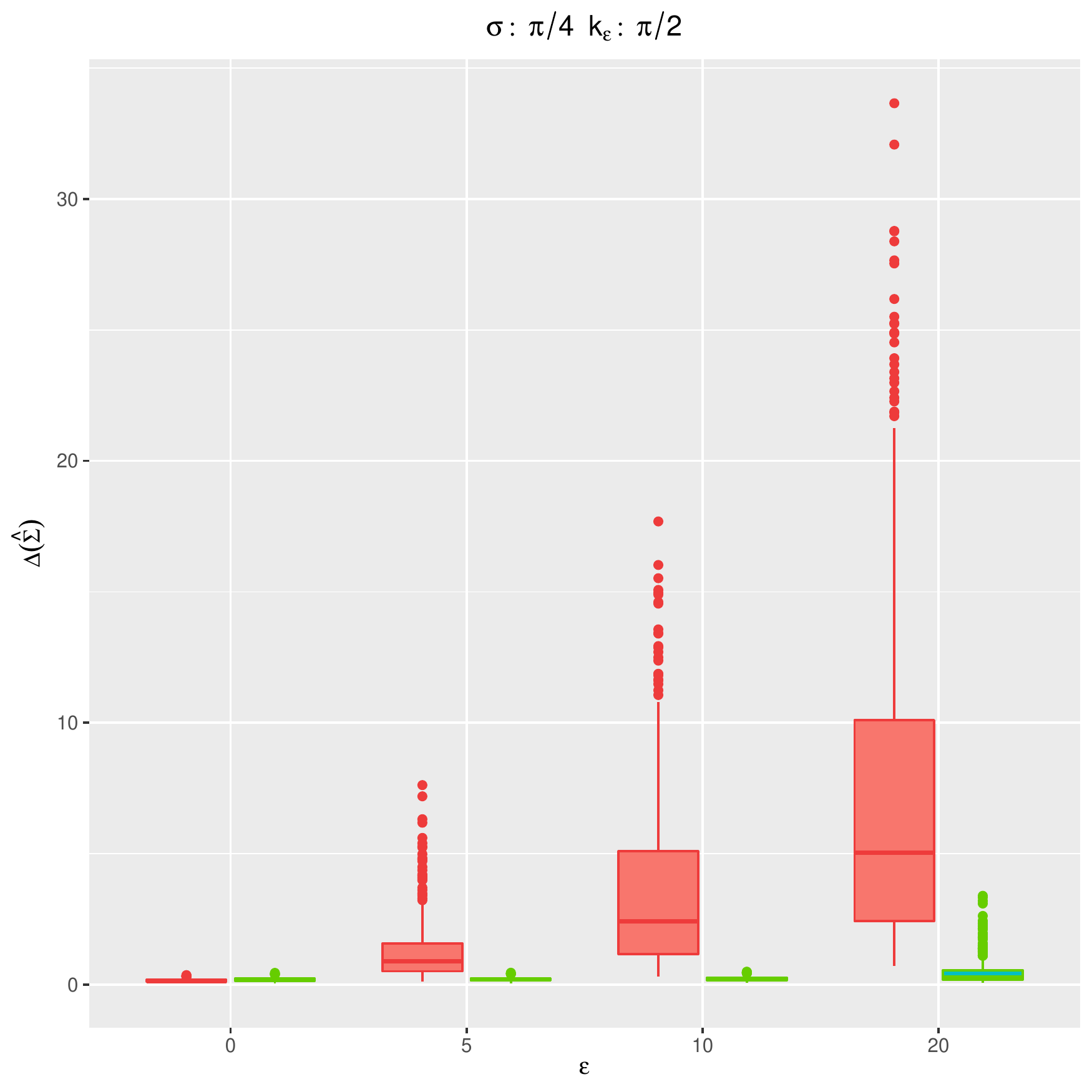} 
 		\includegraphics[height=0.3\textheight, width=0.3\textwidth]{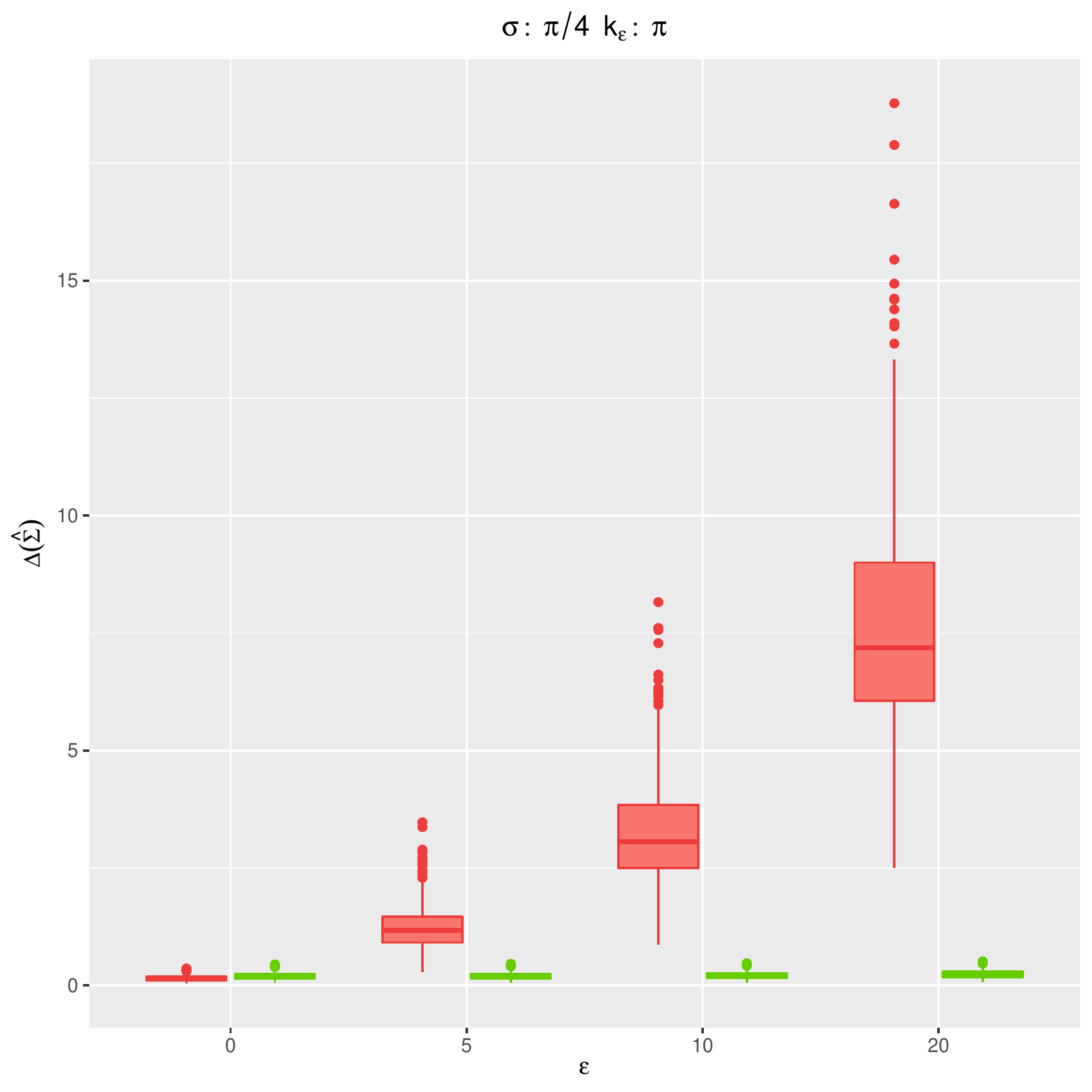} \\
 		\includegraphics[height=0.3\textheight, width=0.3\textwidth]{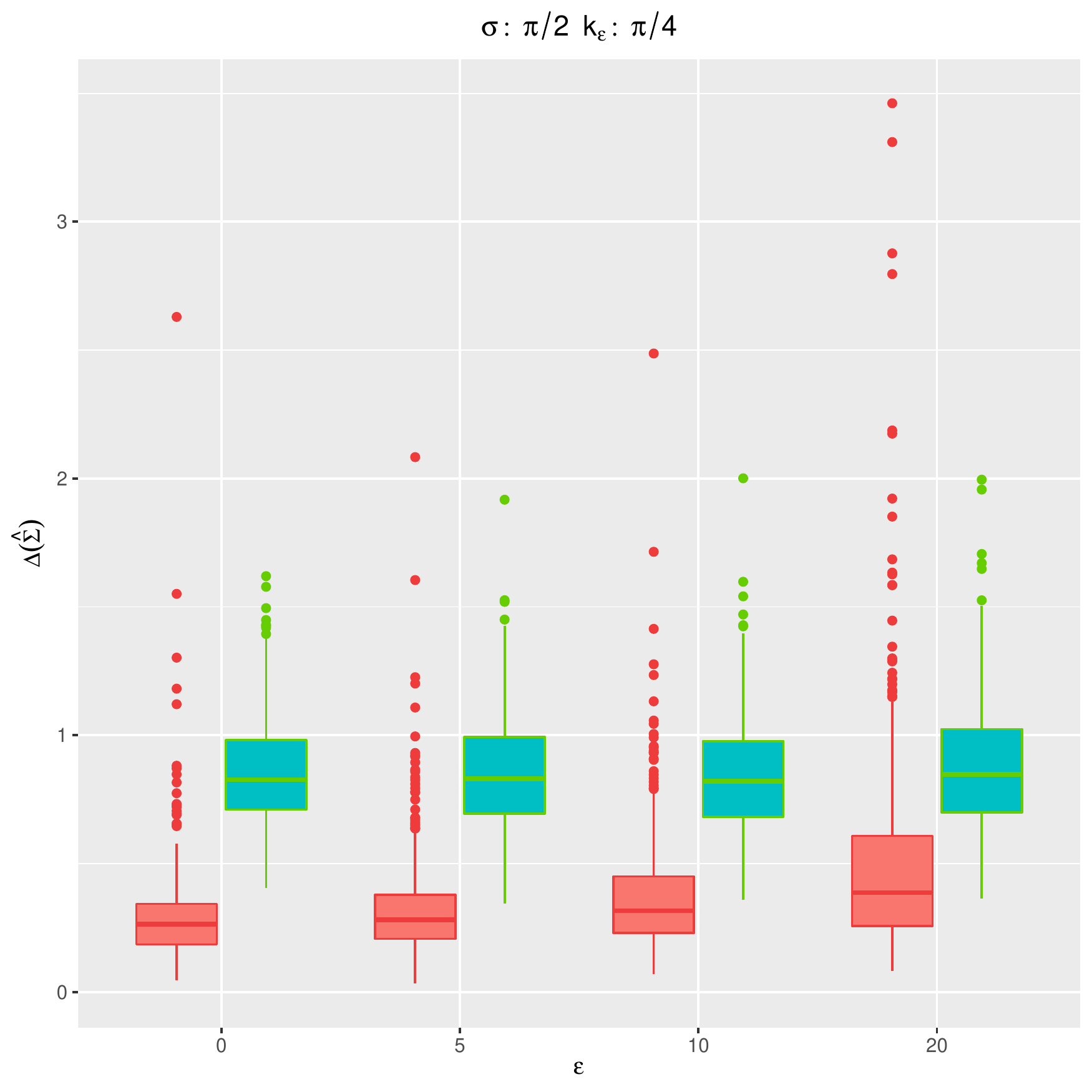} 
 		\includegraphics[height=0.3\textheight, width=0.3\textwidth]{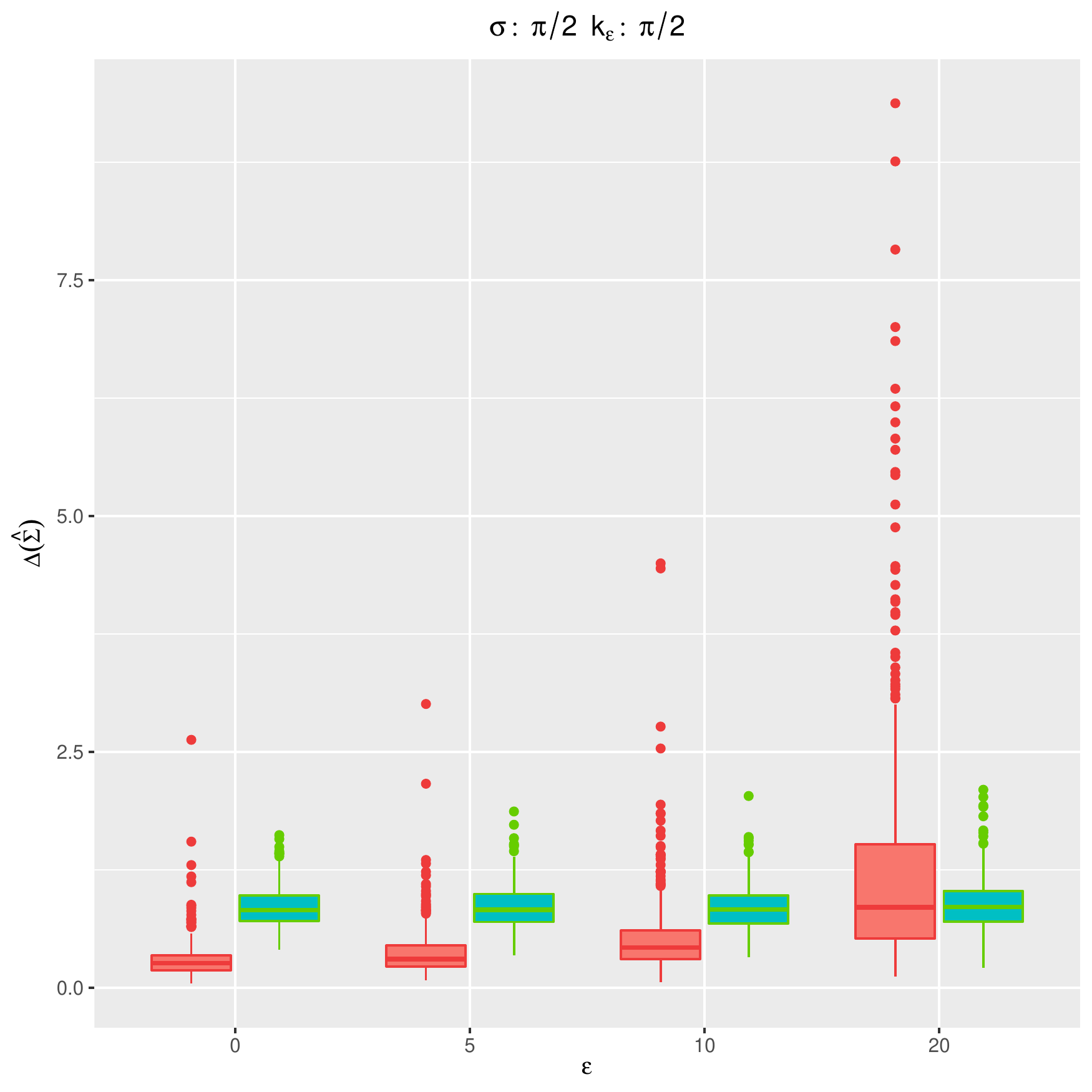} 
 		\includegraphics[height=0.3\textheight, width=0.3\textwidth]{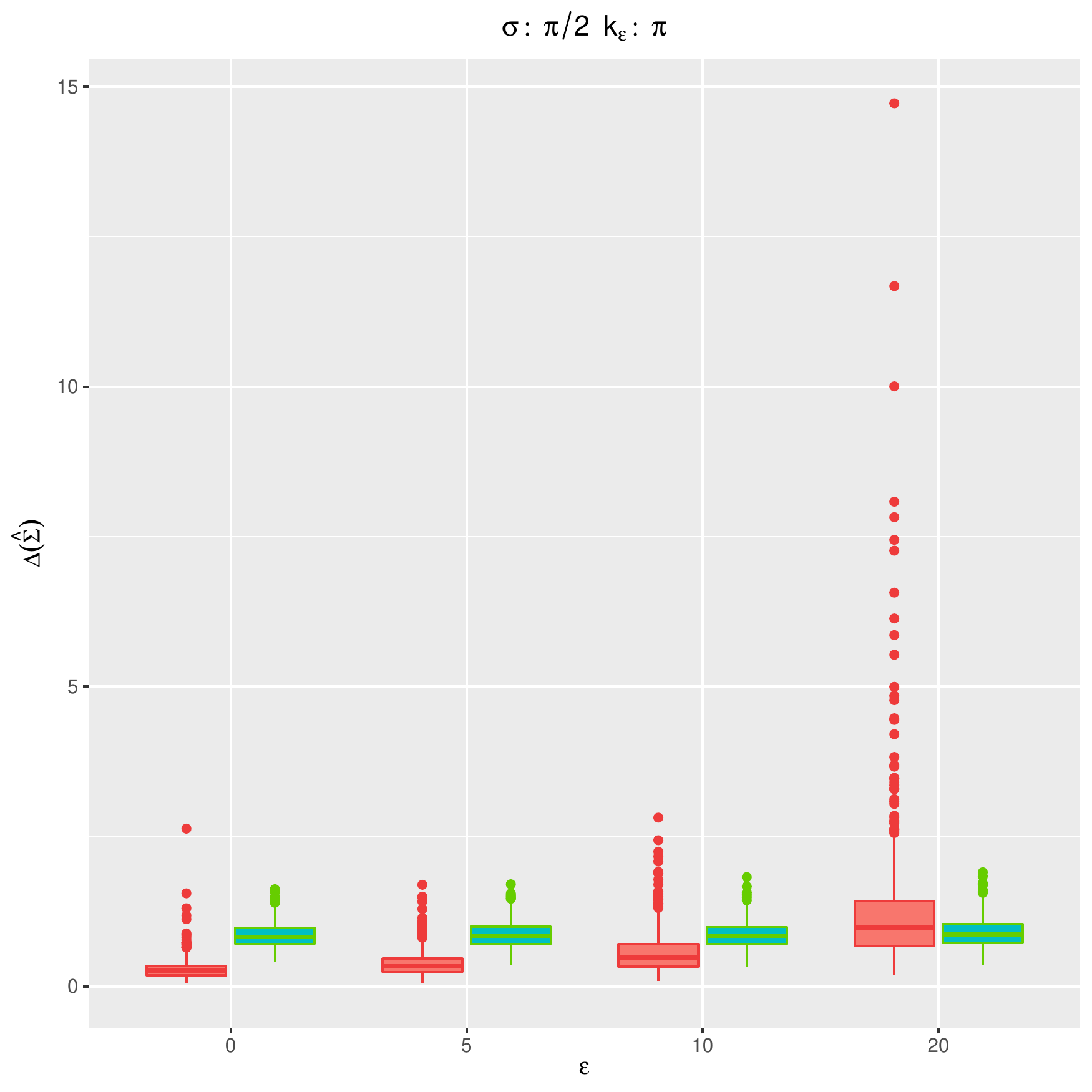} \\
 	\end{center}
 	\caption{Distribution of the divergence measure for $n=100$ and $p=5$ using the weighted CEM (in green) and the CEM (in red). The contamination rate $\epsilon$ is given on the horizontal axis. Increasing contamination size $k_\epsilon$ from left to right, increasing $\sigma$ from top to bottom.}
 	\label{fig:two}
 \end{figure}

     Fitting accuracy has been evaluated according to
	\begin{itemize}
	\item[(i)] the average angle separation
	\begin{equation*}
		\AS(\hat{\vect{\mu}}) = \frac{1}{p} \sum_{i=1}^p (1 - \cos(\hat{\vect{\mu}}_i - \vect{\mu}_{i})) \ , 
	\end{equation*}
	which ranges in $[0, 2]$, for the mean vector;
	\item[(ii)] the divergence 
	\begin{equation*}
		\Delta(\hat{\Sigma}) = \trace(\hat{\Sigma} \Sigma^{-1}) - \log(| \hat{\Sigma} \Sigma^{-1} |) - p \ ,
	\end{equation*}
	\end{itemize}
for the variance-covariance matrix.
	Here, we only report the results stemming from the challenging situation with $n=100$ and $p=5$.
	Figure \ref{fig:one} displays the average angle separation whereas Figure \ref{fig:two} gives the divergence to measure the accuracy in estimating the variance-covariance matrix for the weighted CEM (in green) and CEM (in red). The weighted CEM exhibits a fairly satisfactory fitting accuracy both under the assumed model (i.e. when the sample at hand is not corrupted by the occurrence of outliers) and under contamination. The robust method outperforms the CEM method, especially in the estimation of the variance-covariance components. The algorithm results in biased estimates for both the mean vector and the variance-covariance matrix only for the large contamination rate $\epsilon=20\%$, with small contamination size and a large $\sigma$. Actually, in this data constellation outliers are not well separated from the group of genuine observations. A similar behavior has been observed for the other sample sizes.  
	Complete results are made available in the Supplementary Material.
		
	\section{Real data example: Protein data}
	\label{sec:examples}
	
	The data under consideration \citep{Najibi2017} contain bivariate information about $63$ protein domains that were randomly selected from three remote Protein classes in the Structural Classification of Proteins (SCOP).
	In the following, 
	we consider the data set corresponding to the 39-th protein domain.
	A  bivariate Wrapped Normal has been fitted to the data at hand by using the weighted CEM algorithm, based on a Generalized Kullback-Leibler RAF with $\tau=0.25$ and $J=6$.
	 The tasks of bandwidth selection and initialization have been resolved according to the same strategy described above in Section~\ref{sec:simulations}. 
	
\begin{figure}[!ht]
		\begin{center}
			\includegraphics[width=0.9\textwidth,height=0.5\textheight]{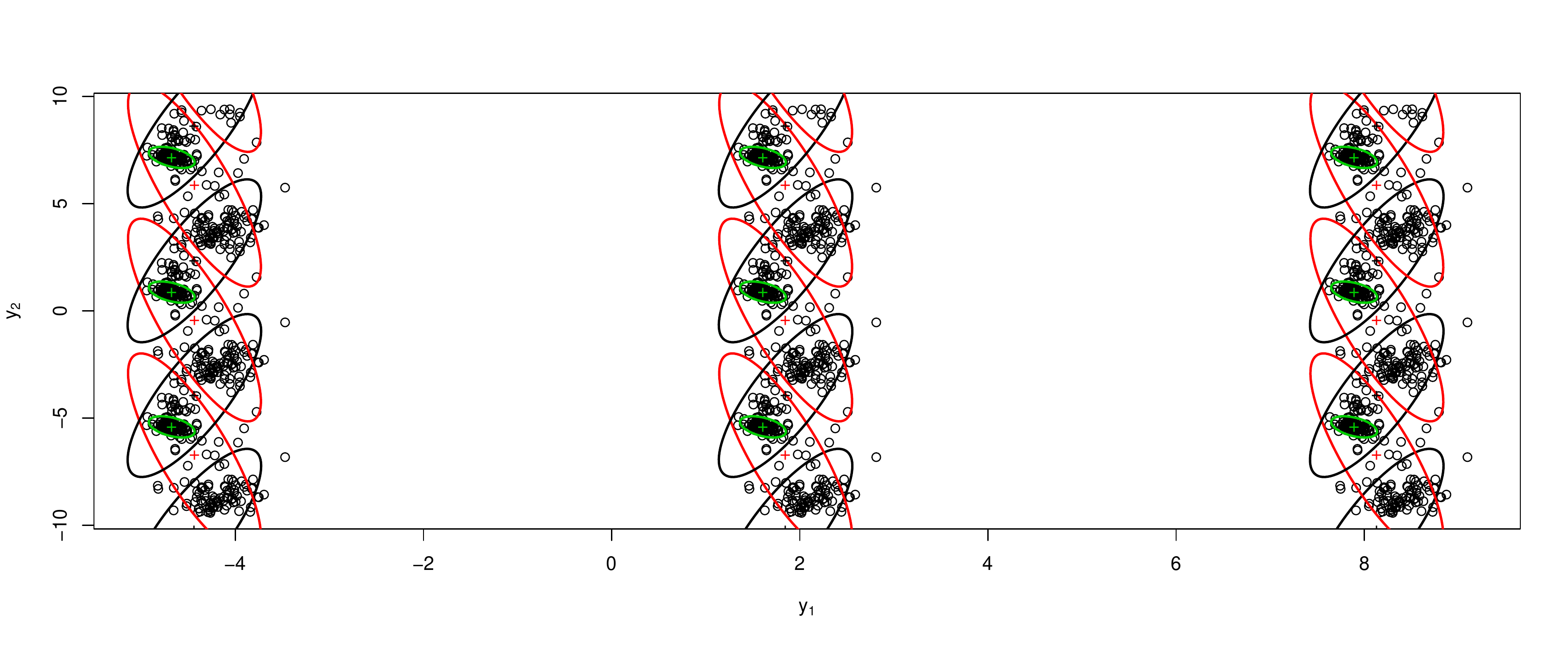} 
				\end{center}
		\caption{Protein data. Fitted means ($+$) and $95\%$ confidence regions corresponding to three different roots from weighted CEM ($J=6$).}		
		\label{fig:three}
	\end{figure}

\begin{figure}[!ht]
			\begin{center}
				\includegraphics[width=0.9\textwidth, height=0.5\textheight]{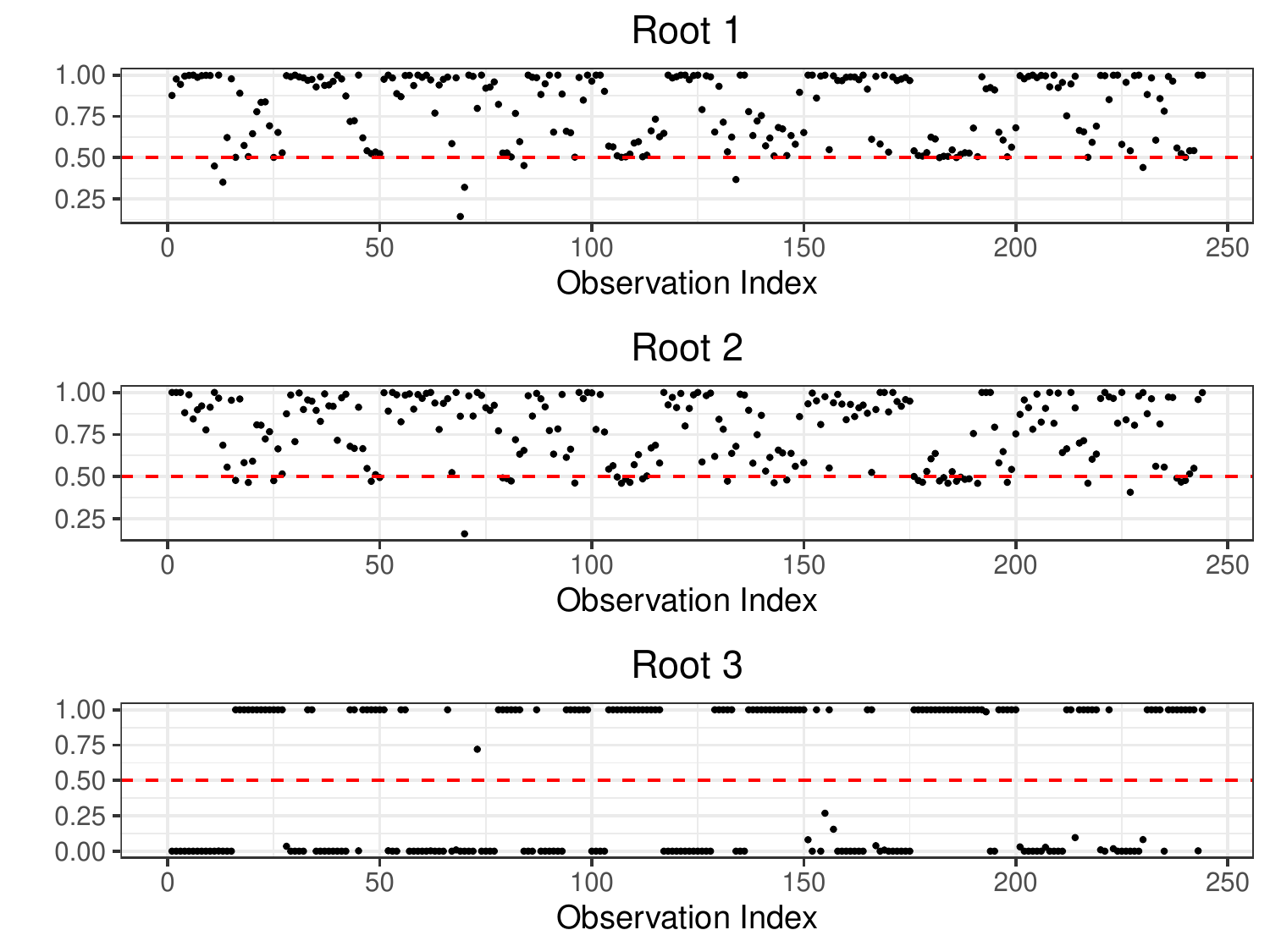} 
			\end{center}
				\caption{Protein data. Weights corresponding to three different roots from weighted CEM.}		
			\label{fig:four}
		\end{figure}

	The inspection of the data suggests the presence of at least a couple of clusters that make the data non homogeneous. 
	Figure \ref{fig:three} displays the data on a flat torus together with fitted means and $95\%$ confidence regions corresponding to three different roots of the WLEE (that are illustrated by different colors): one root gives location estimate $\vect{\mu}_1=(1.85, 2.34)$ and a positive correlation $\rho_1=0.79$; the second root gives location estimate $\vect{\mu}_2=(1.85, 5.86)$ and a negative correlation $\rho_2=-0.80$; the third root gives location estimate $\vect{\mu}_3=(1.61, 0.88)$ and correlation $\rho_3=-0.46$.
	The first and second roots are very close to maximum likelihood estimates obtained in different directions when unwrapping the data: this is evident from the shift in the second coordinate of the mean vector and the change in the sign of the correlation. In both cases the data exhibit weights larger than 0.5, except in few cases, corresponding to the most extreme observations, as displayed in the first two panels of Figure \ref{fig:four}.  In none of the two cases the bulk of the data corresponds to an homogeneous sub-group. On the contrary, the third root is able to detect an homogeneous substructure in the sample, corresponding to the most dense region in the data configuration.
	Almost half of the data points is attached a weight close to zero, as shown in the third panel of Figure \ref{fig:four}.
	This findings still confirm the ability of the weighted likelihood methodology to tackle such uneven patterns as a diagnostic of hidden substructures in the data.  
	In order to select one of the three roots we have found, we consider the strategy discussed in \cite{agostinelli2006}, that is, we select the root leading to the lowest fitted probability
	\begin{equation*}
	\mathrm{Prob}_{\hat\Omega}\left(\delta_n(\vect{y}; \hat{\Omega}, \hat F_n)<-0.95\right) \ .
	\end{equation*}
	This probability has been obtained by drawing 5000 samples from the fitted bivariate Wrapped Normal distribution for each of the three roots. The criterion correctly leads to choose the third root, for which an almost null probability is obtained, wheres the fitted probabilities for the first and second root are 0.204 and 0.280, respectively.
	
	\section{Conclusions}
	\label{sec:conclusions}
	In this paper an effective strategy for robust estimation of a multivariate Wrapped normal on a $p-$dimensional torus has been presented. The method inherits the good computational properties of a CEM algorithm developed  in \cite{nodehi2018estimation} jointly with the robustness properties stemming from the employ of Pearson residuals and the weighted likelihood methodology. In this respect, it is particularly appealing the opportunity to work with a family of distribution that is close under convolution and allows to parallel the procedure one would have developed on the real line by using the multivariate normal distribution. 
	The proposed weighted CEM works satisfactory at least in small to moderate dimensions, both on synthetic and real data. It is worth to stress that
	the method can be easily extended to other multivariate wrapped models. 
	
	\bibliography{robtorus}
	
\end{document}


\title{Supplemental Material: Robust Estimation for Multivariate Wrapped Models}
	
	\author[1]{Giovanni Saraceno}
	\author[1]{Claudio Agostinelli}
	\author[2]{Luca Greco}
	\affil[1]{Department of Mathematics, University of Trento, Trento, Italy \texttt{giovanni.saraceno@unitn.it, claudio.agostinelli@unitn.it}}
	\affil[2]{Department DEMM, University of Sannio, Benevento, Italy \texttt{luca.greco@unisannio.it}}
	
	\date{\today}
	
	\maketitle

        The Supplemental Material contains complete results of the numerical study described in Section 4 of the manuscript.

        \section*{Additional Numerical results}
        \label{sm:sec:numericalstudy}

\begin{figure}
	 \begin{center}
	 \includegraphics[height=0.3\textheight, width=0.3\textwidth]{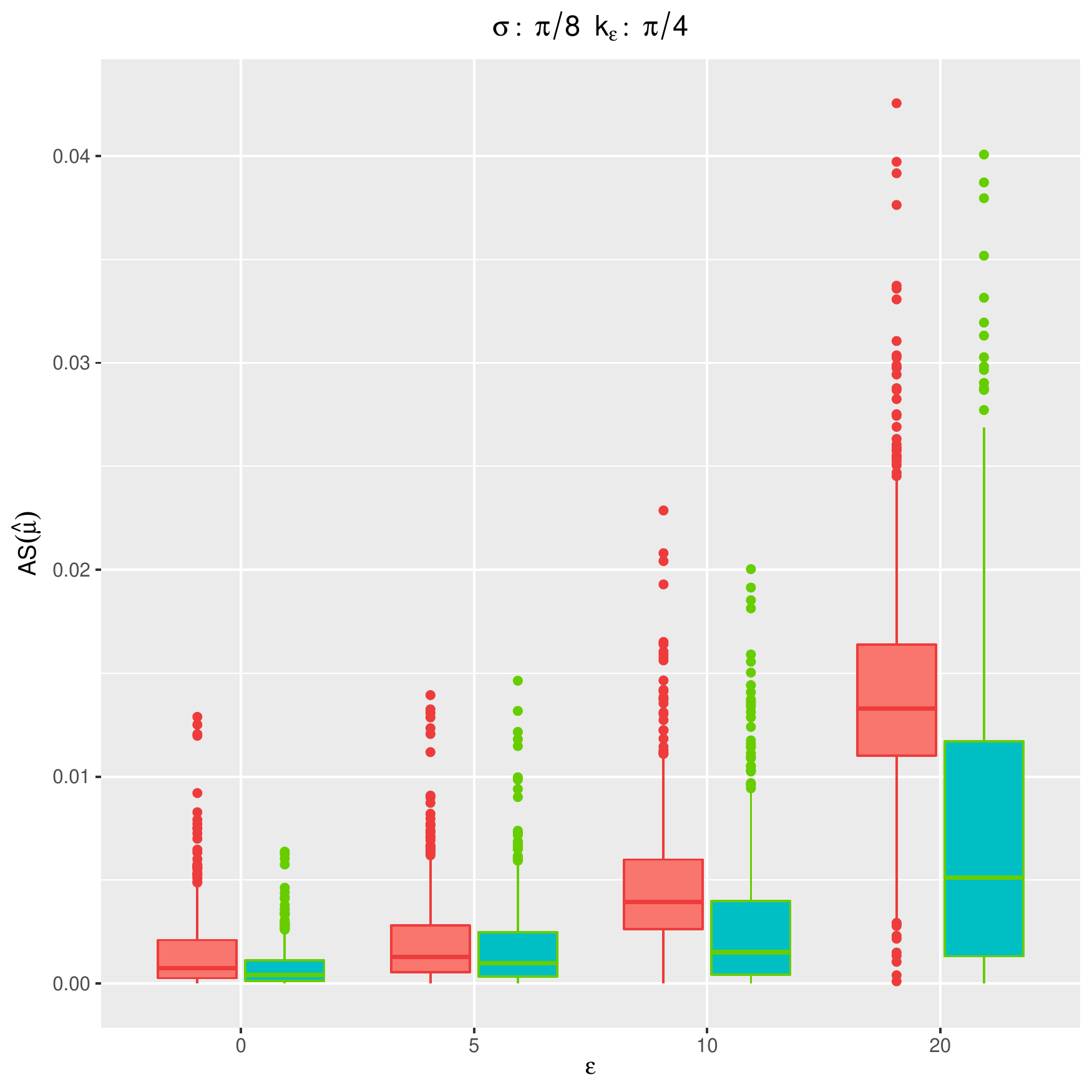} 
	 \includegraphics[height=0.3\textheight, width=0.3\textwidth]{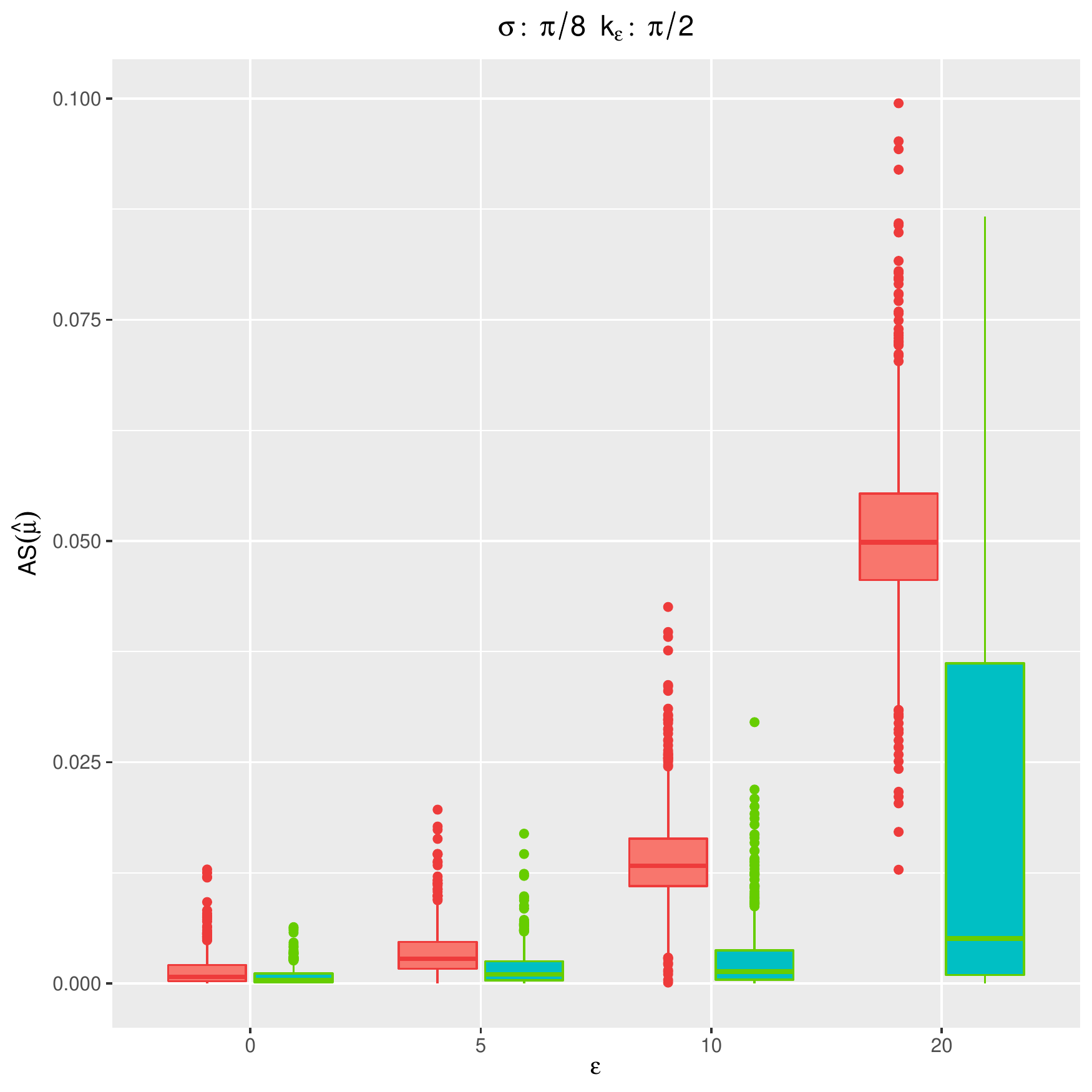} 
	 \includegraphics[height=0.3\textheight, width=0.3\textwidth]{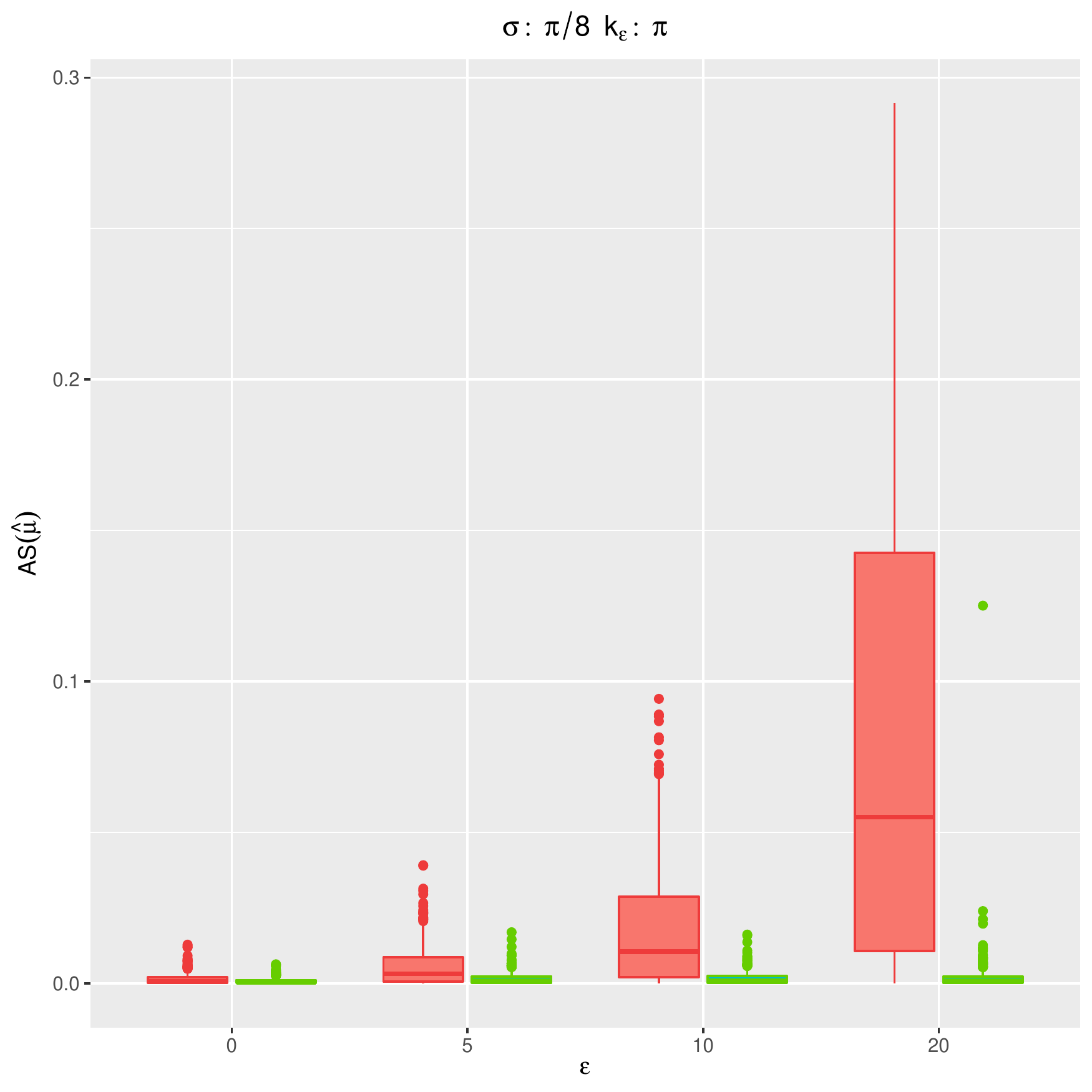} \\
	\includegraphics[height=0.3\textheight, width=0.3\textwidth]{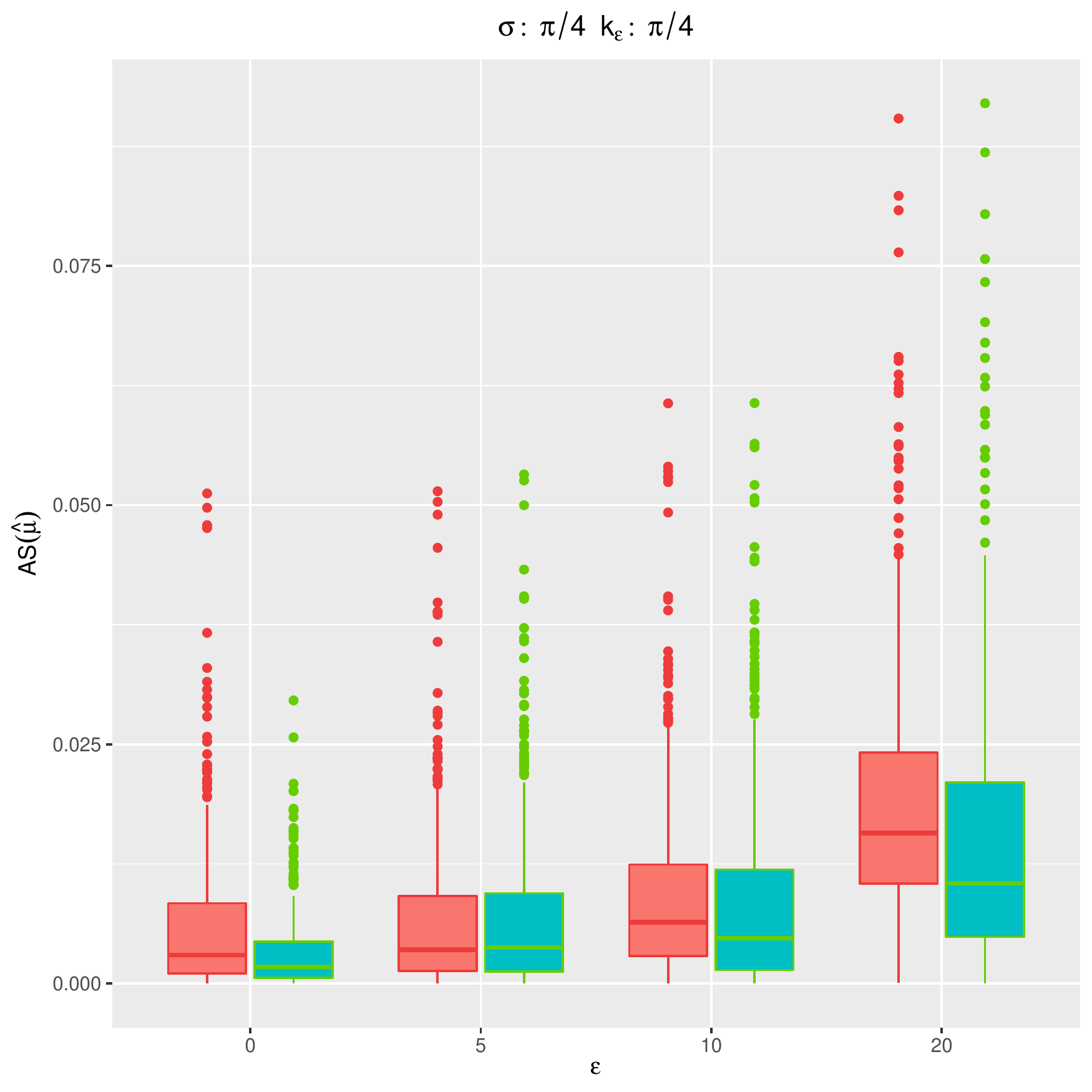} 
	 \includegraphics[height=0.3\textheight, width=0.3\textwidth]{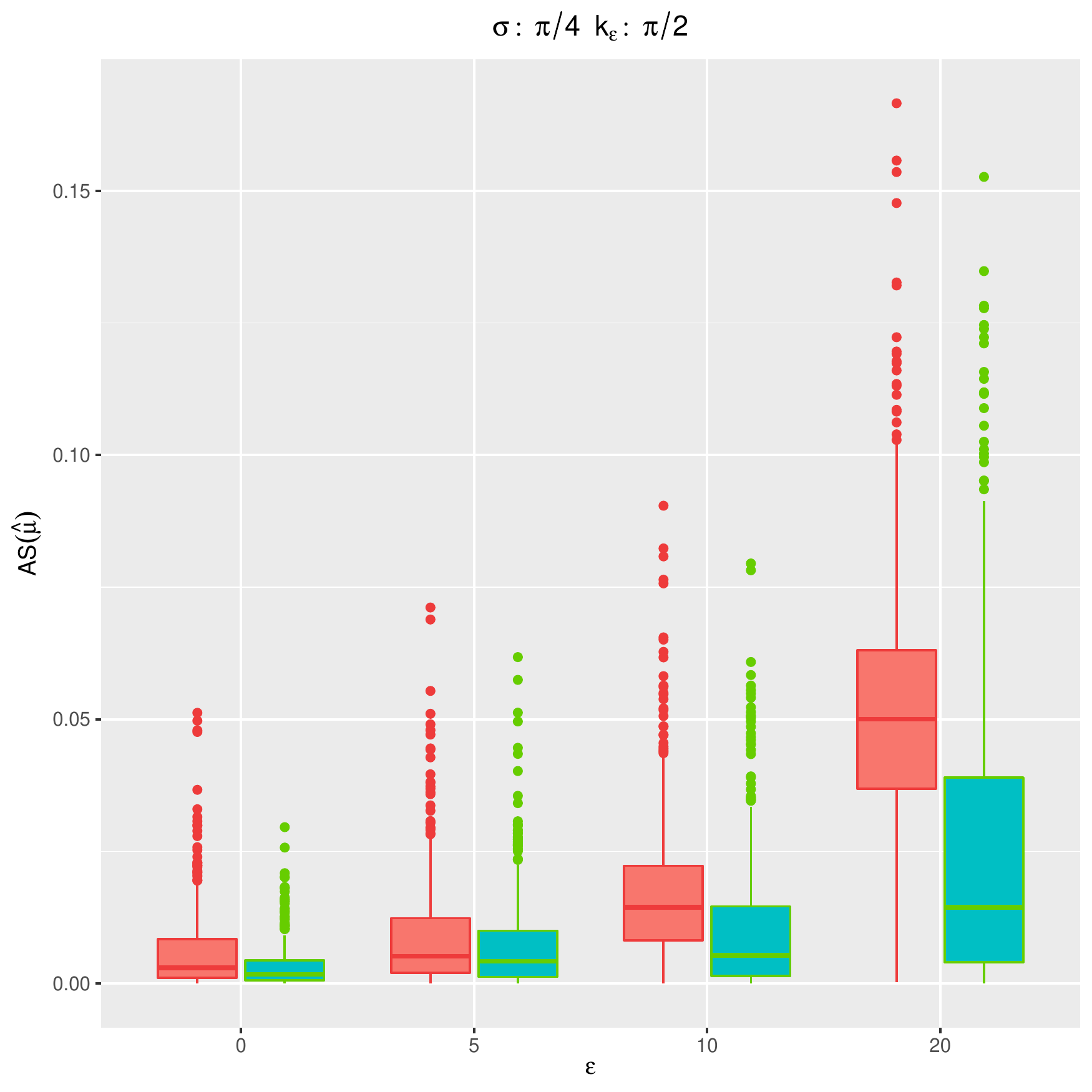} 
	 \includegraphics[height=0.3\textheight, width=0.3\textwidth]{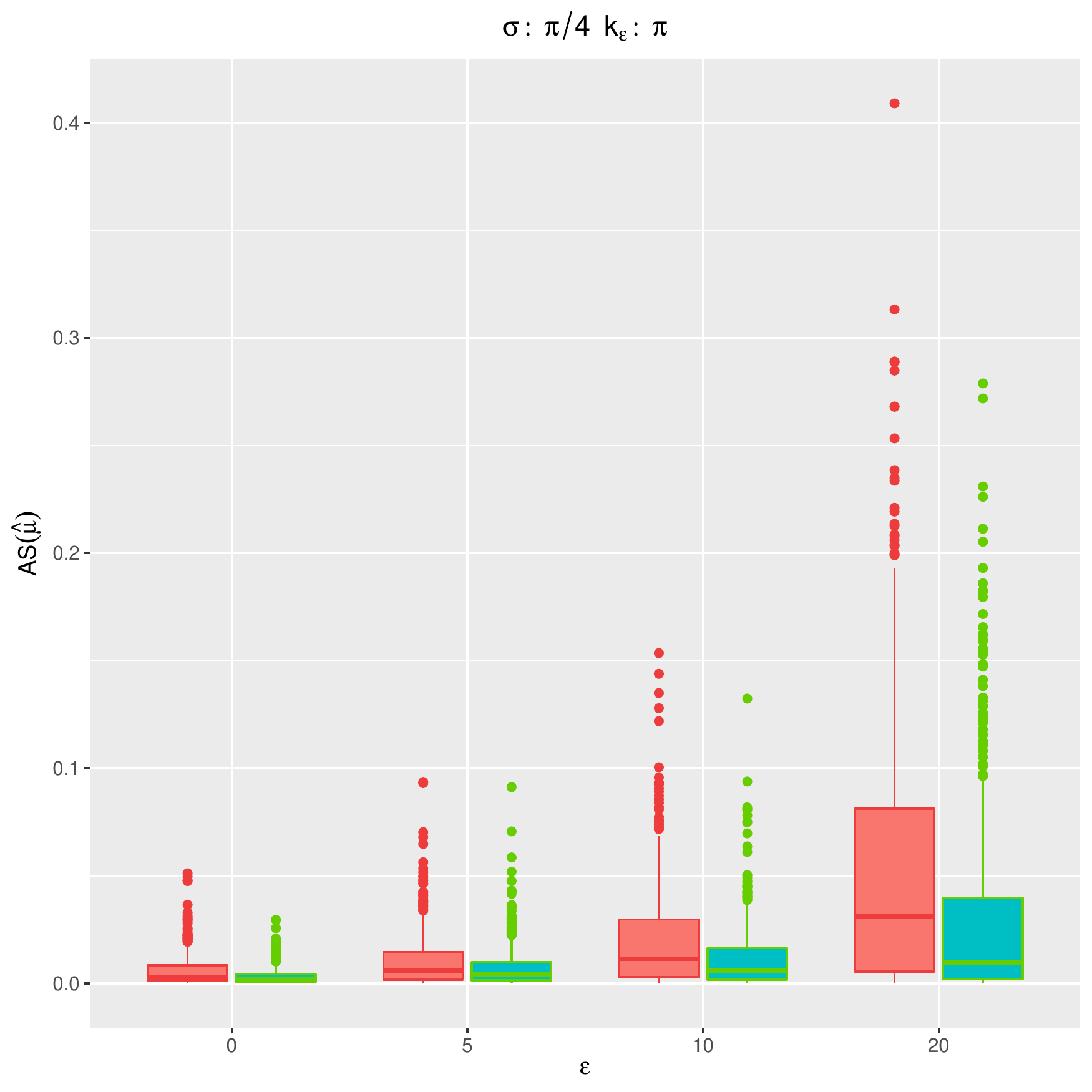} \\
	 \includegraphics[height=0.3\textheight, width=0.3\textwidth]{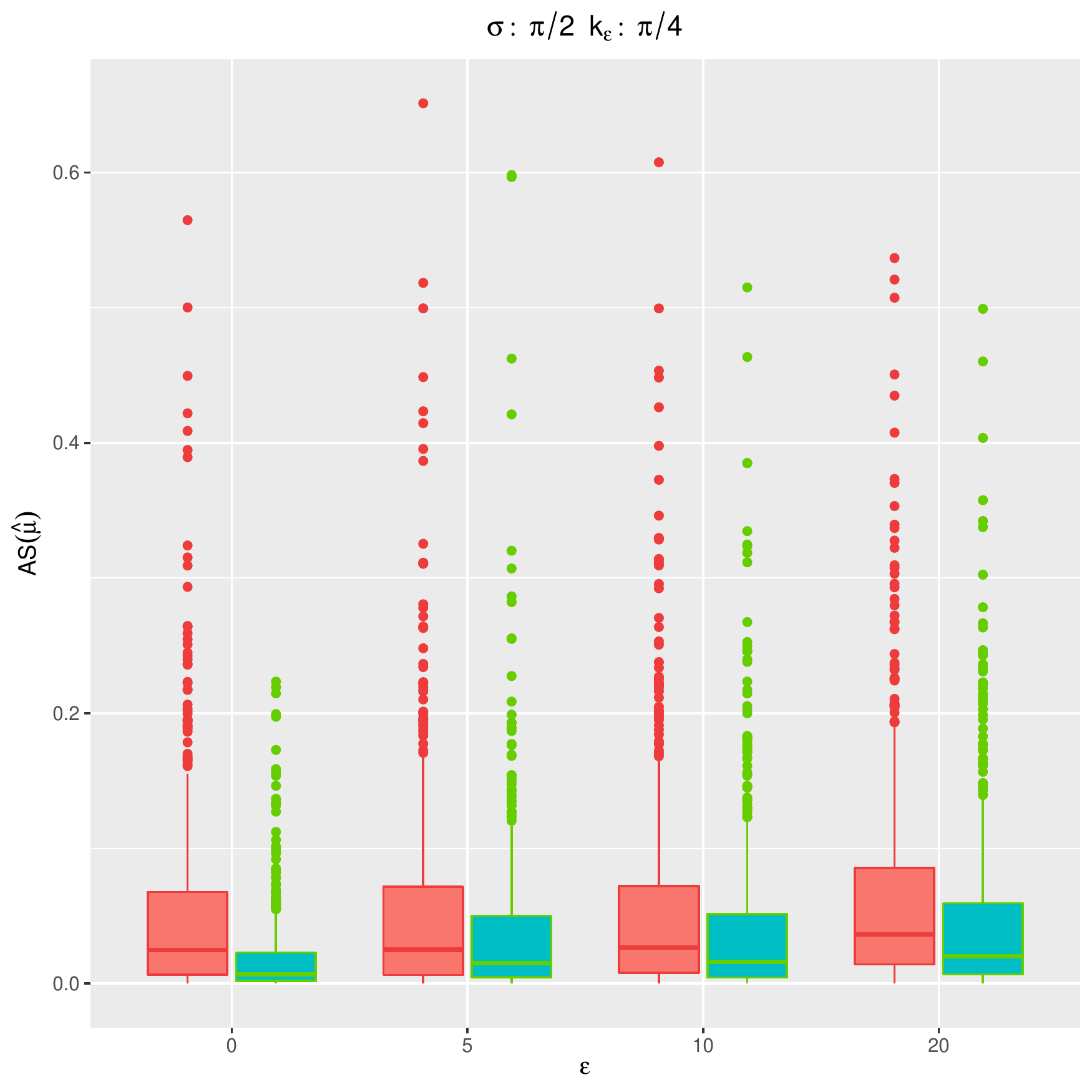} 
	 \includegraphics[height=0.3\textheight, width=0.3\textwidth]{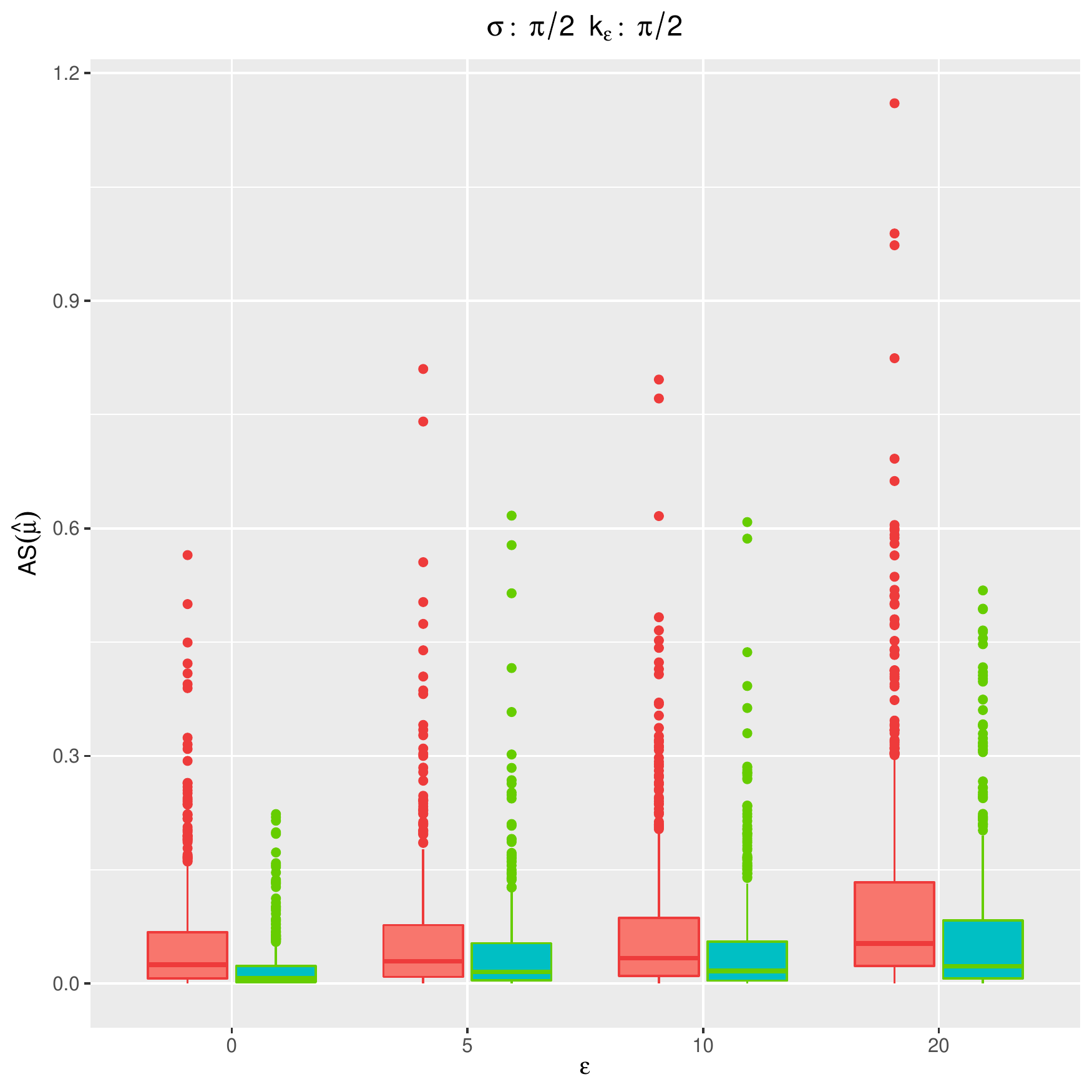} 
	 \includegraphics[height=0.3\textheight, width=0.3\textwidth]{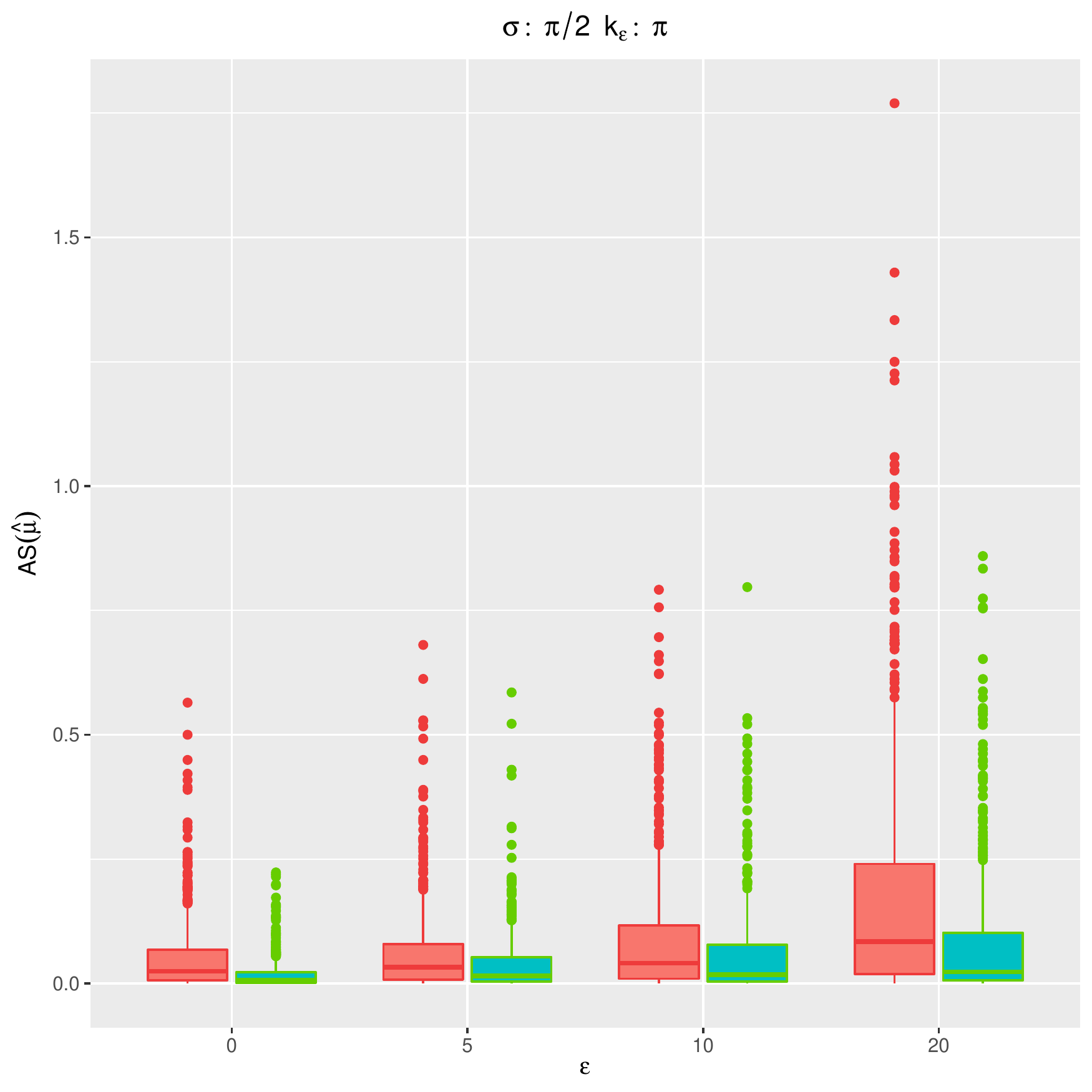} \\
	 \end{center}
	 \caption{Distribution of angle separation for $n=50$ and $p=2$ using weighted CEM (in green) and the CEM (in red). The contamination rate $\epsilon$ is given on the horizontal axis. Increasing contamination size $k_\epsilon$ from left to right, increasing $\sigma$ from top to bottom.}
	 \label{fig:sm:1}
	\end{figure}

\begin{figure}
 	\begin{center}
 		\includegraphics[height=0.3\textheight, width=0.3\textwidth]{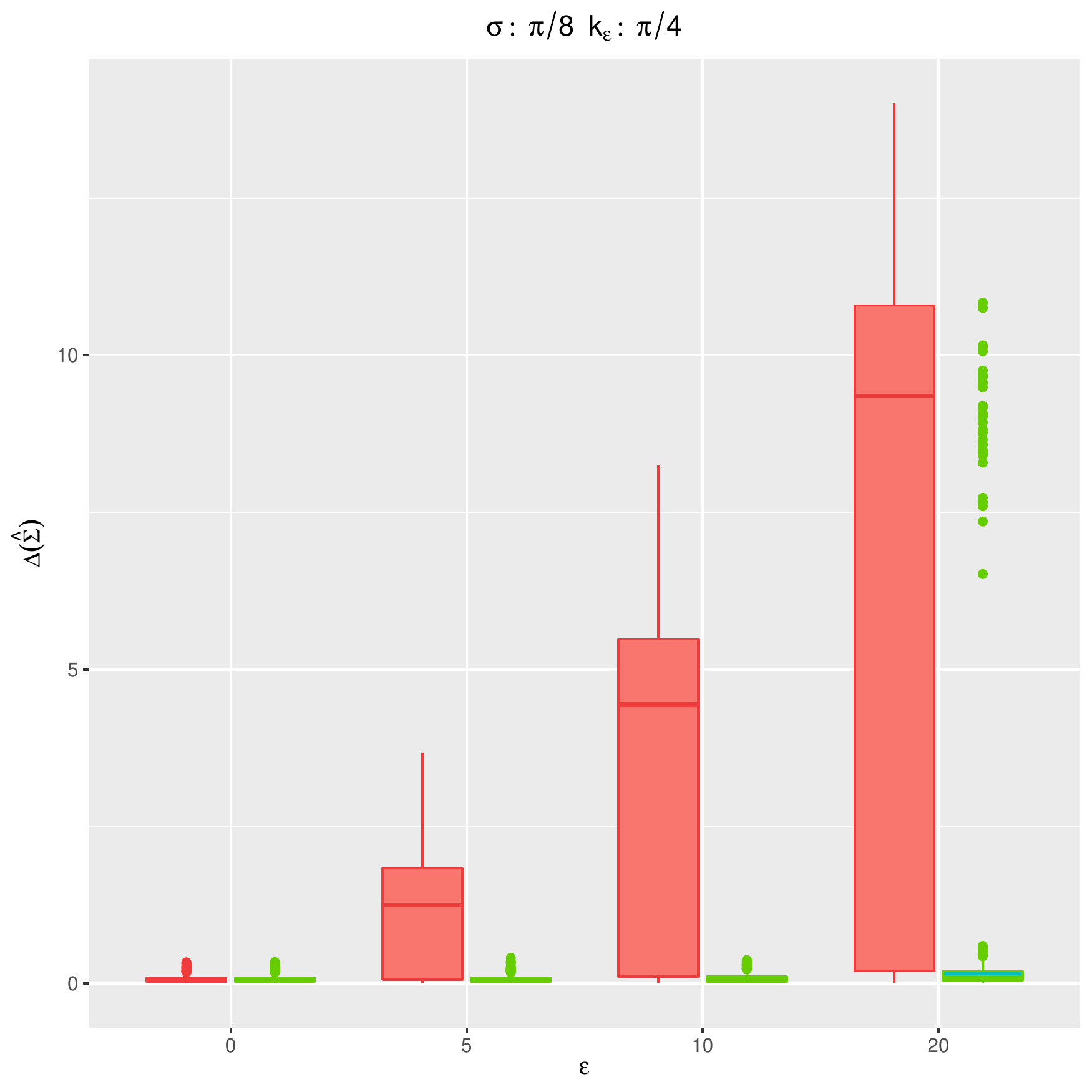} 
 		\includegraphics[height=0.3\textheight, width=0.3\textwidth]{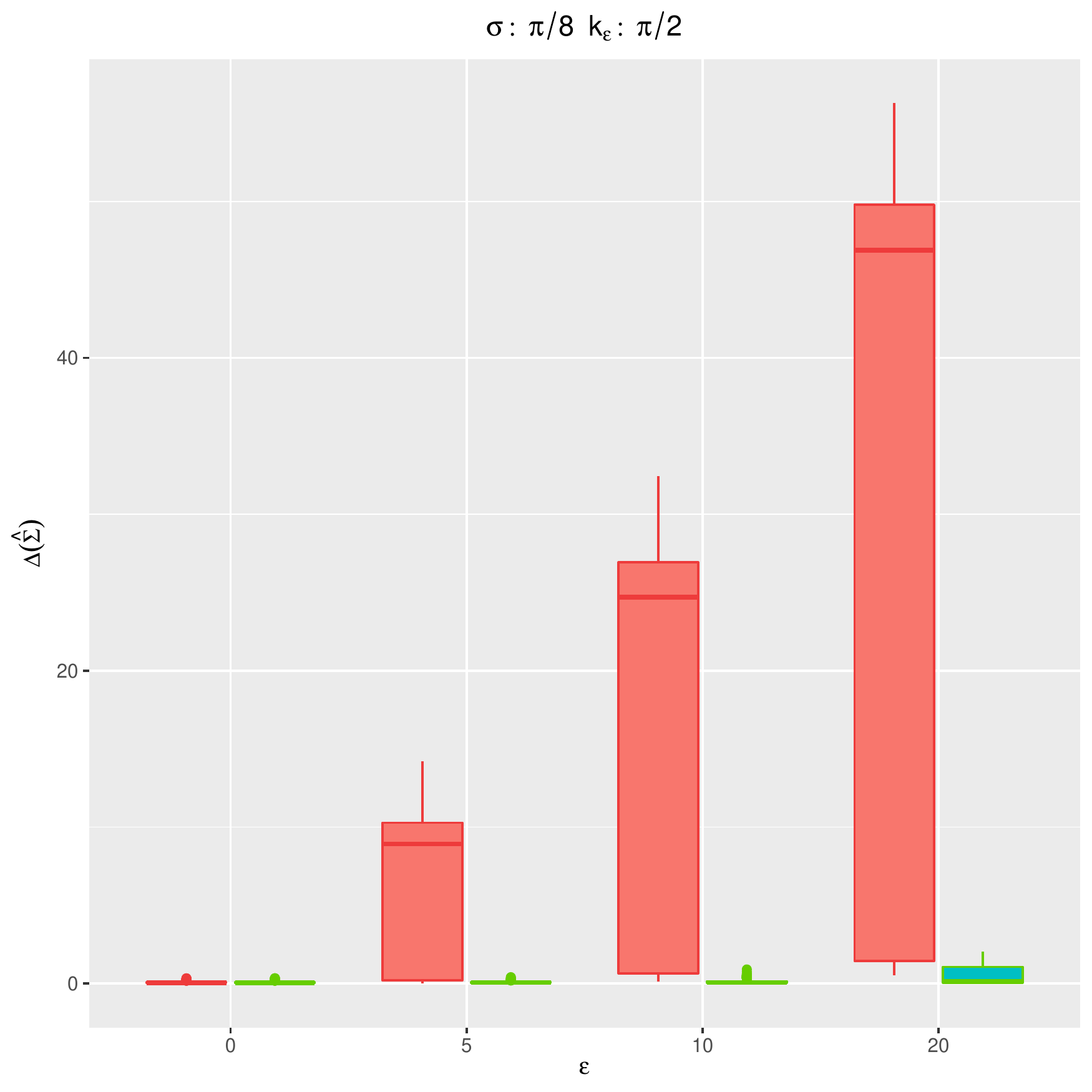} 
 		\includegraphics[height=0.3\textheight, width=0.3\textwidth]{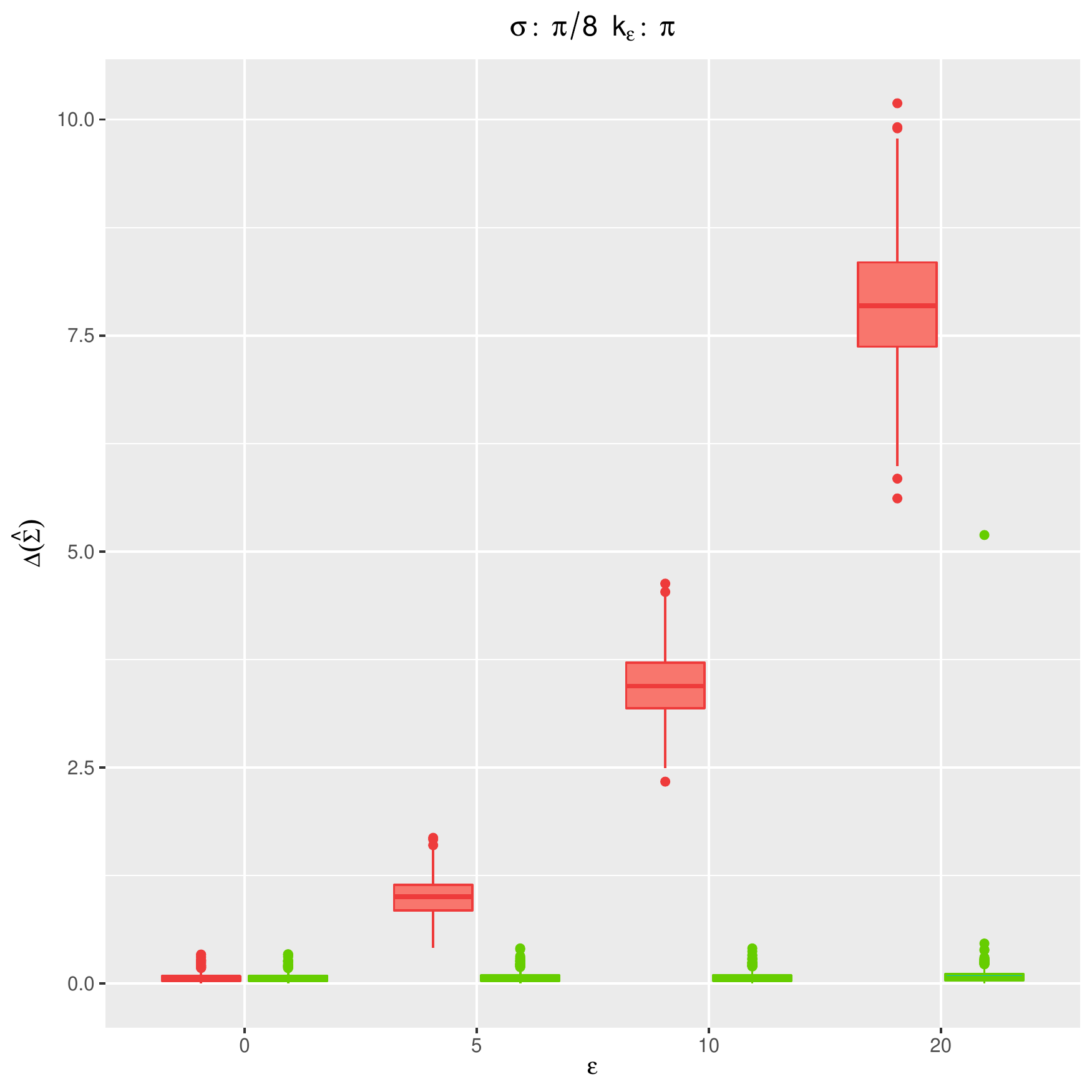} \\
 		\includegraphics[height=0.3\textheight, width=0.3\textwidth]{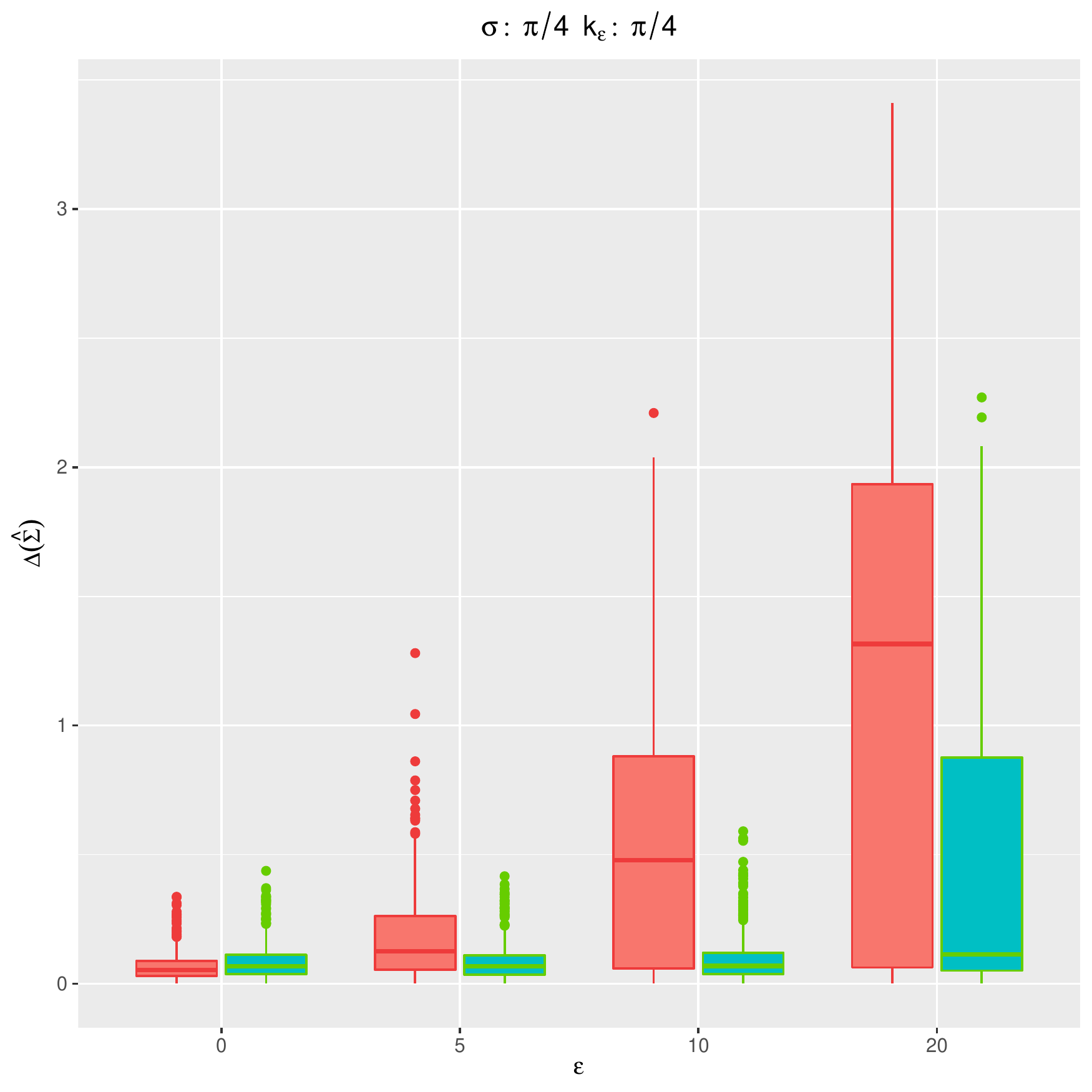} 
 		\includegraphics[height=0.3\textheight, width=0.3\textwidth]{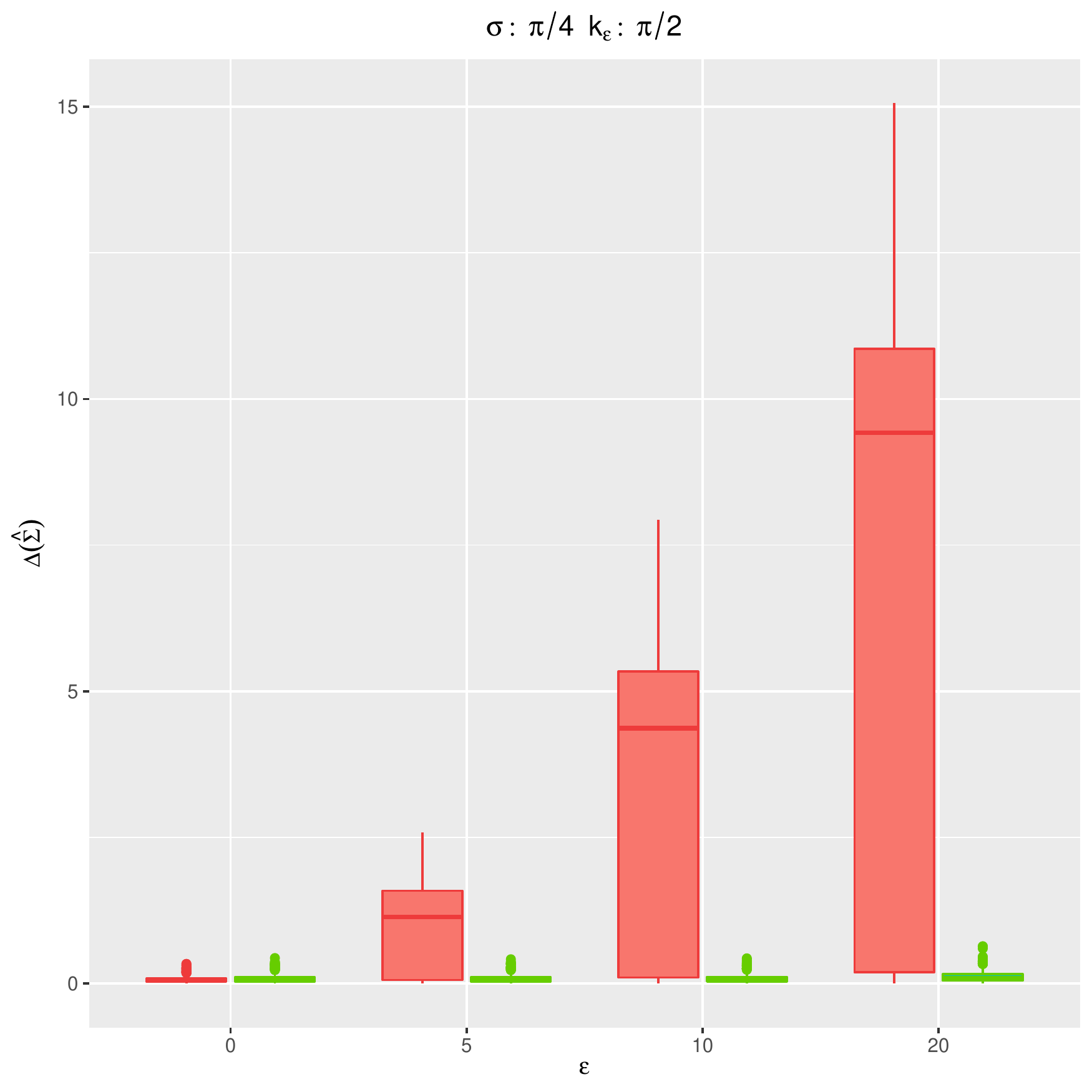} 
 		\includegraphics[height=0.3\textheight, width=0.3\textwidth]{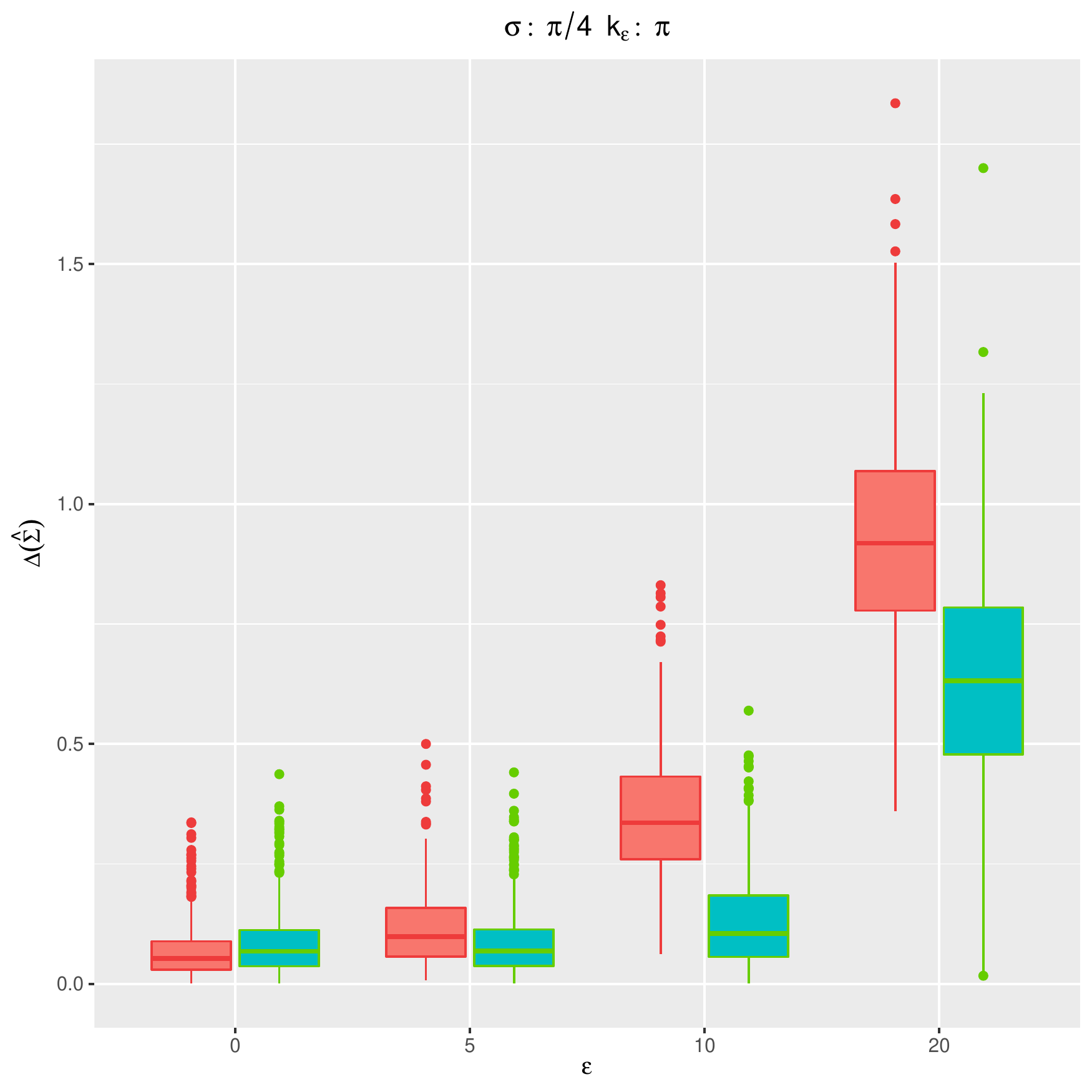} \\
 		\includegraphics[height=0.3\textheight, width=0.3\textwidth]{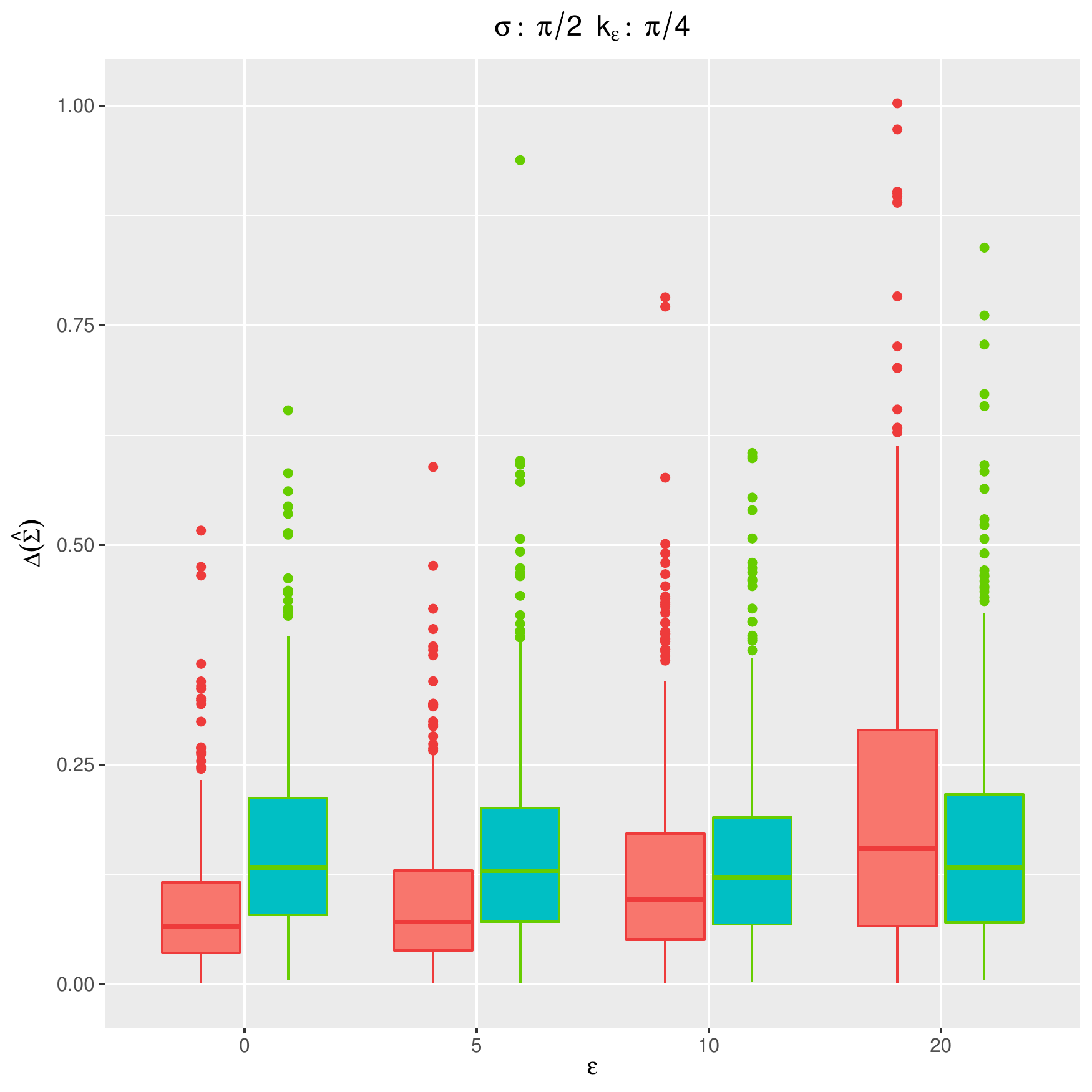} 
 		\includegraphics[height=0.3\textheight, width=0.3\textwidth]{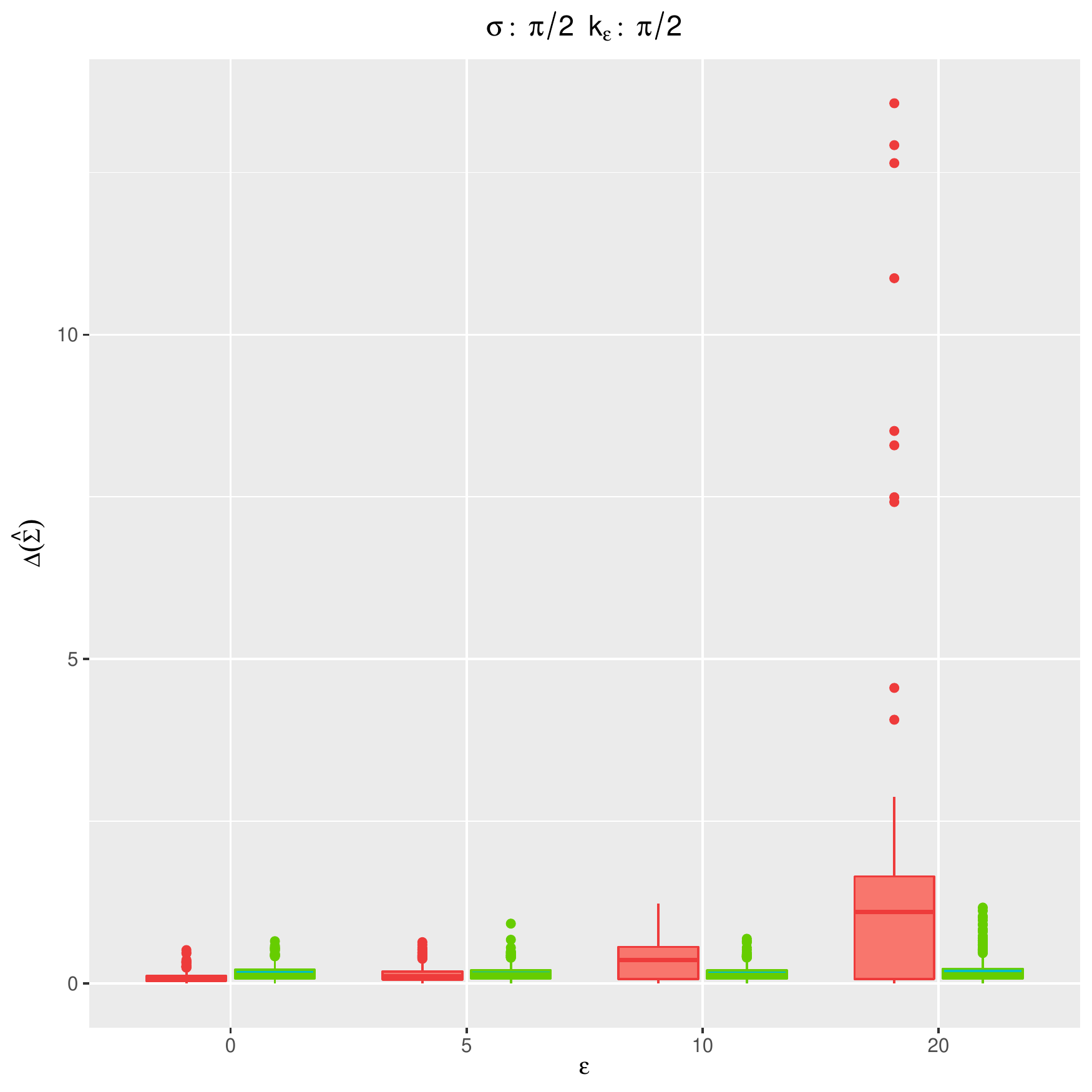} 
 		\includegraphics[height=0.3\textheight, width=0.3\textwidth]{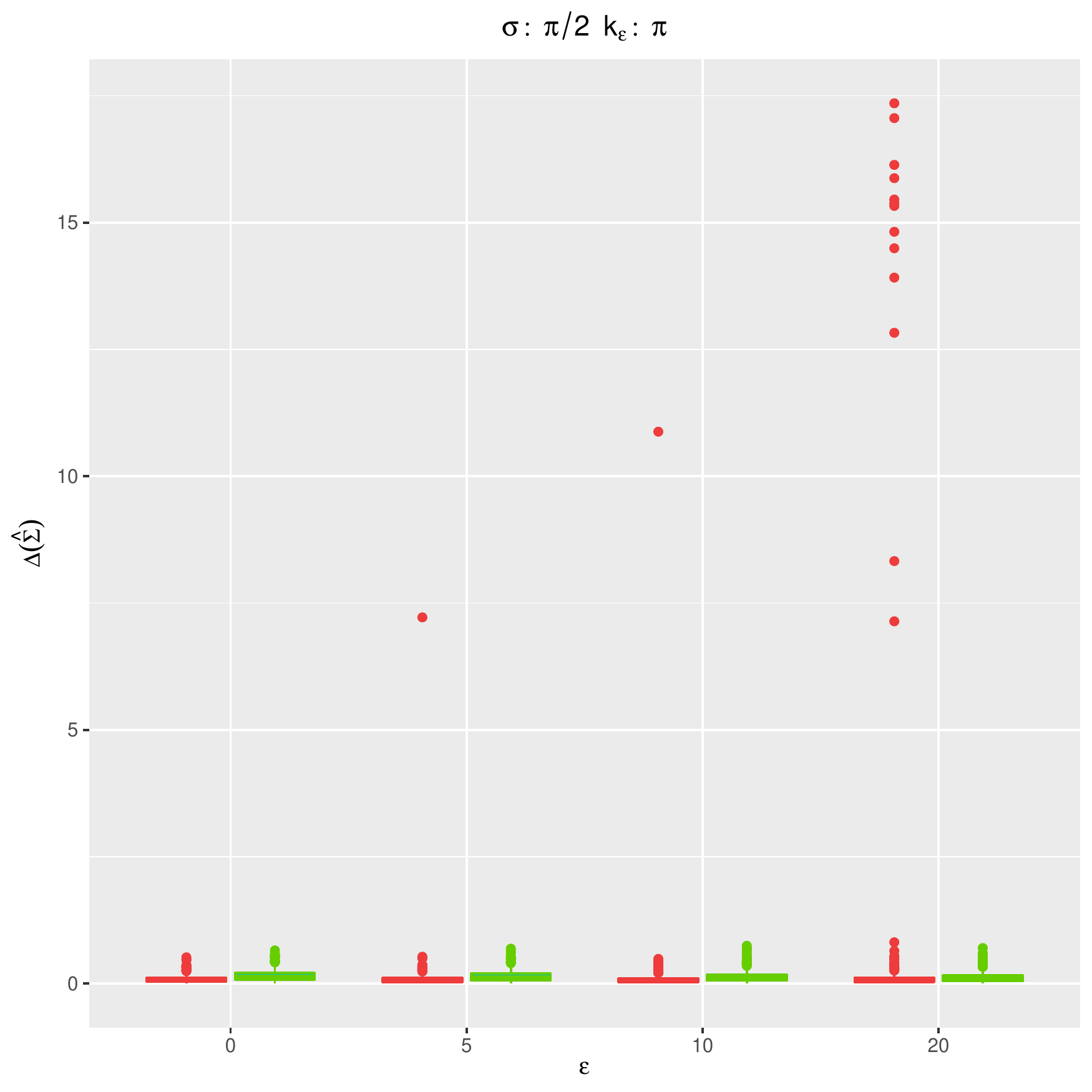} \\
 	\end{center}
 	\caption{Distribution of the divergence measure for $n=50$ and $p=2$ using the weighted CEM (in green) and the CEM (in red). The contamination rate $\epsilon$ is given on the horizontal axis. Increasing contamination size $k_\epsilon$ from left to right, increasing $\sigma$ from top to bottom.}
 	\label{fig:sm:2}
\end{figure}

\begin{figure}
	 \begin{center}
	 \includegraphics[height=0.3\textheight, width=0.3\textwidth]{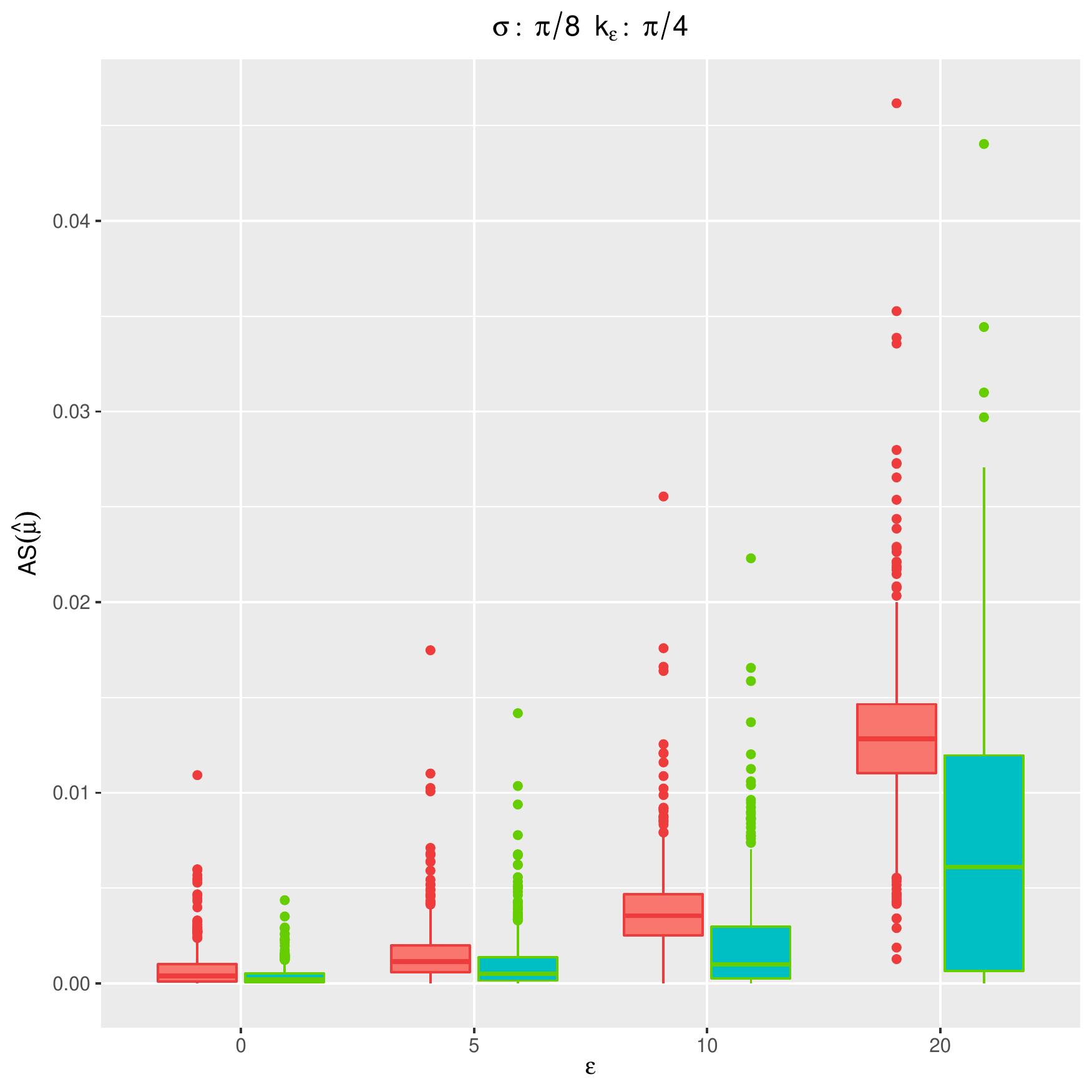} 
	 \includegraphics[height=0.3\textheight, width=0.3\textwidth]{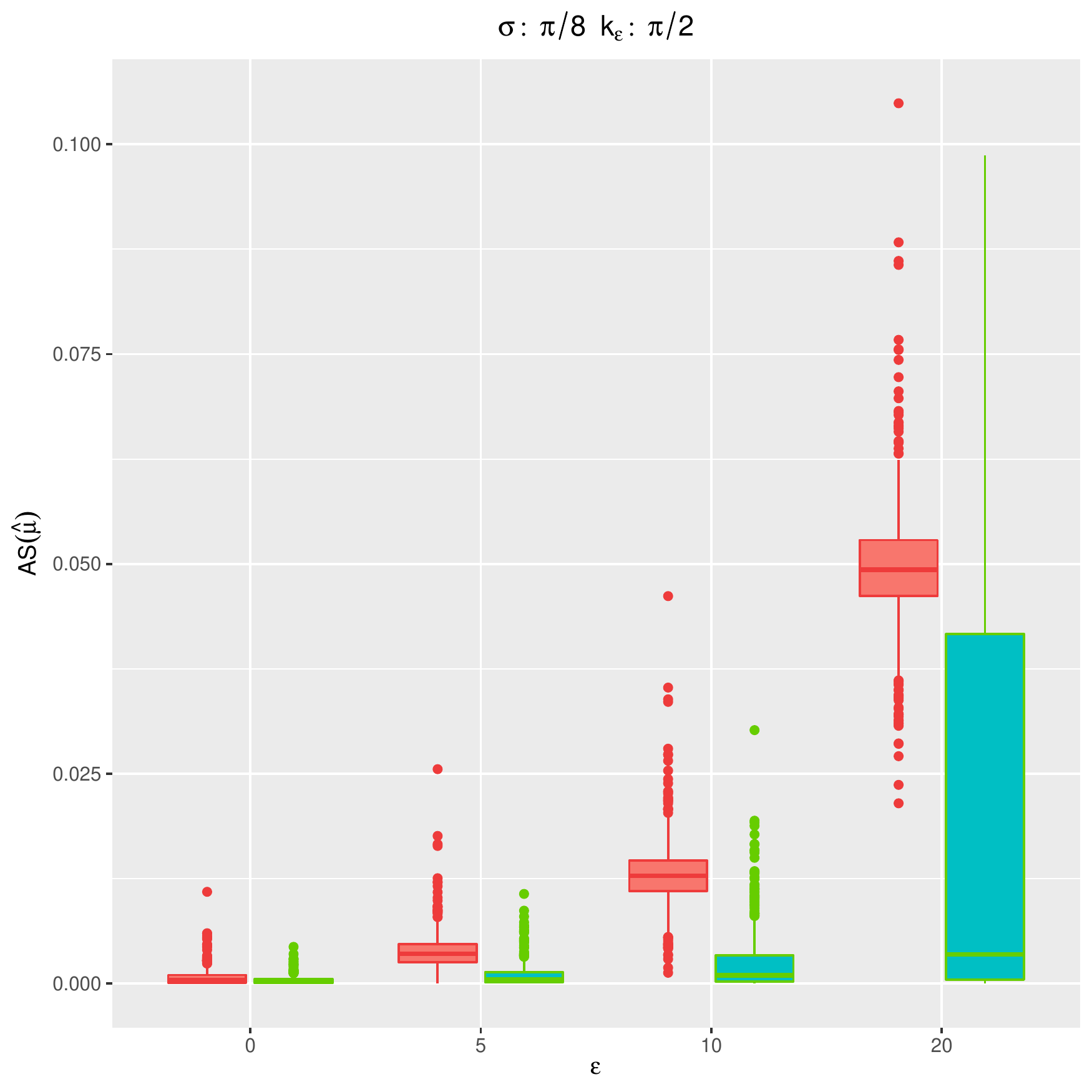} 
	 \includegraphics[height=0.3\textheight, width=0.3\textwidth]{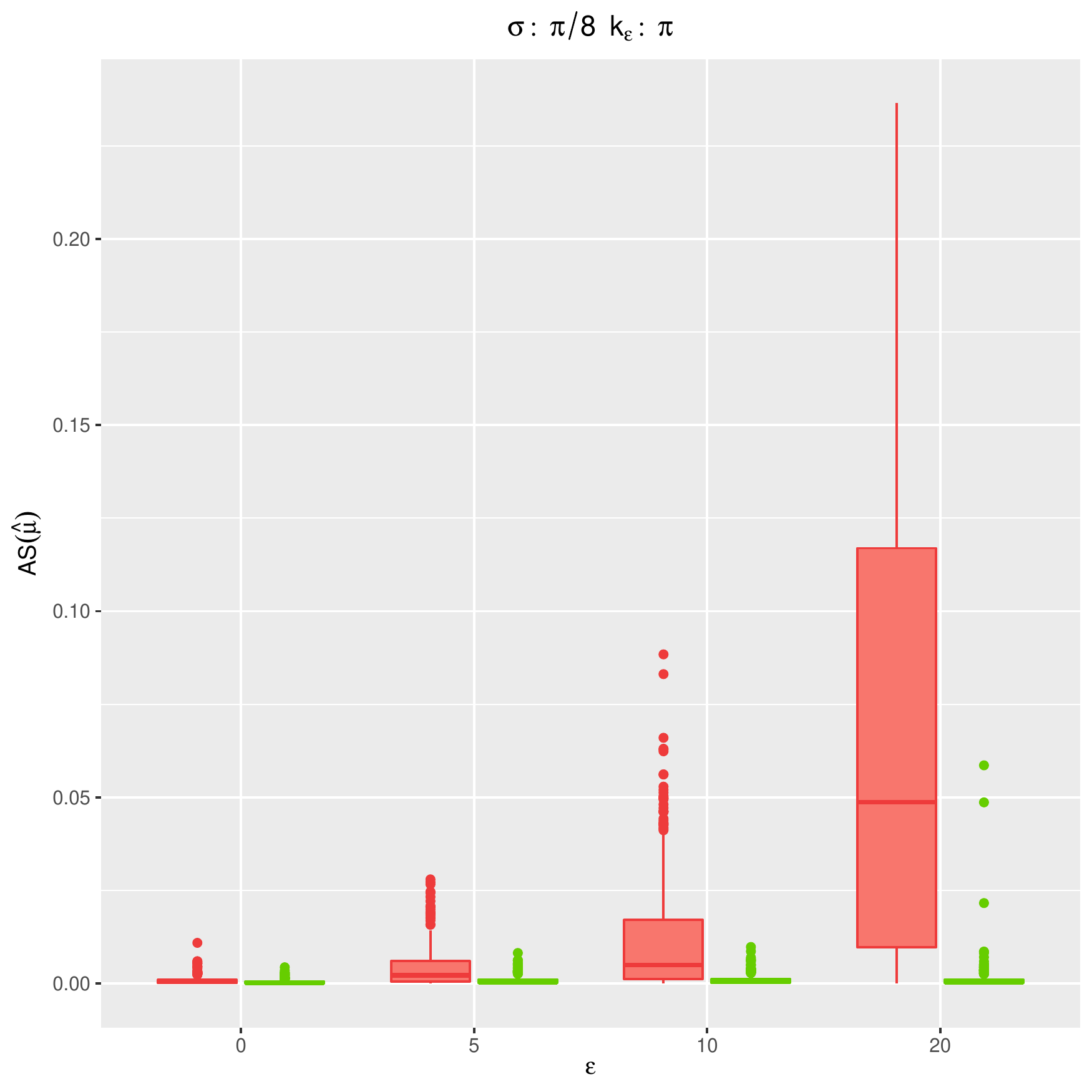} \\
	\includegraphics[height=0.3\textheight, width=0.3\textwidth]{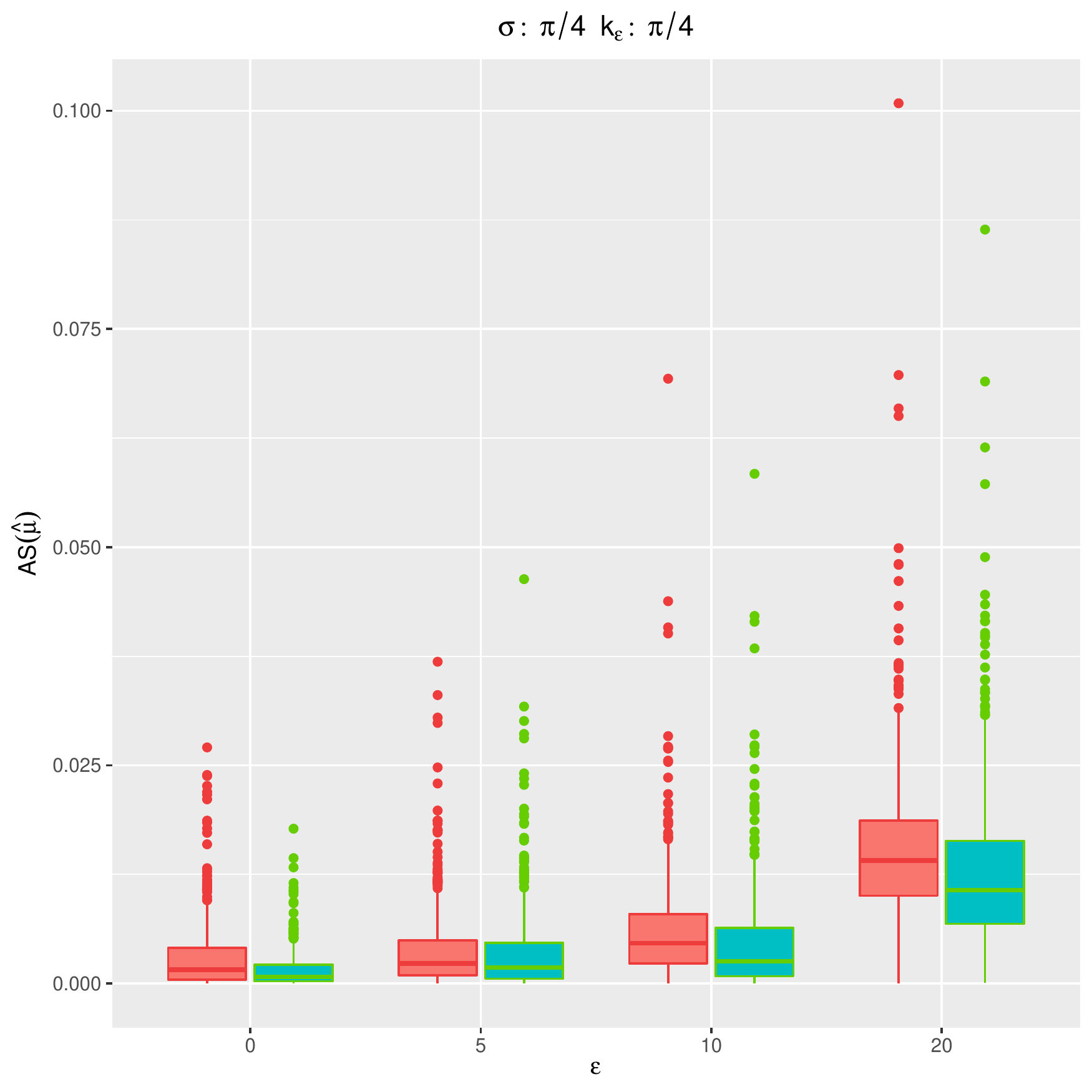} 
	 \includegraphics[height=0.3\textheight, width=0.3\textwidth]{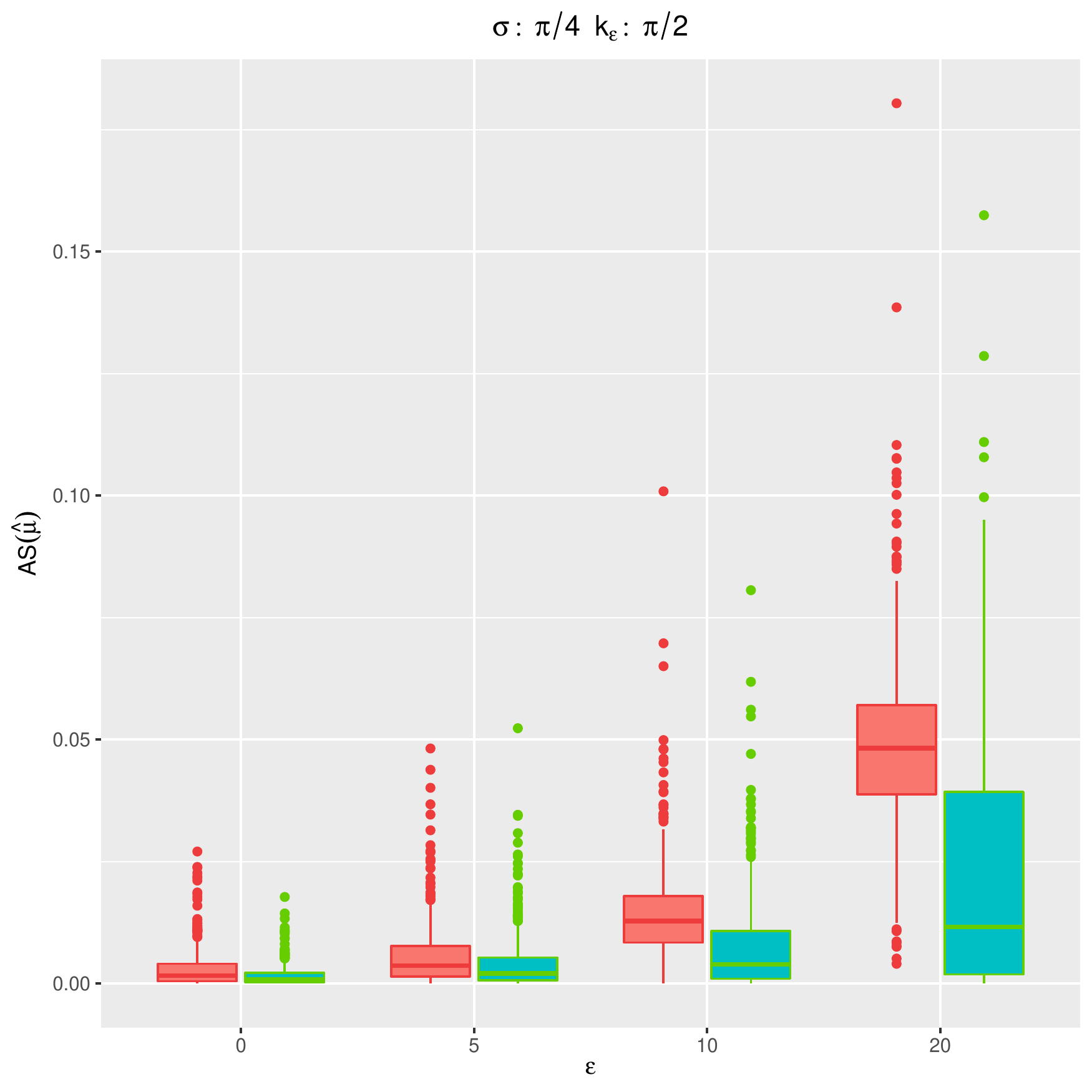} 
	 \includegraphics[height=0.3\textheight, width=0.3\textwidth]{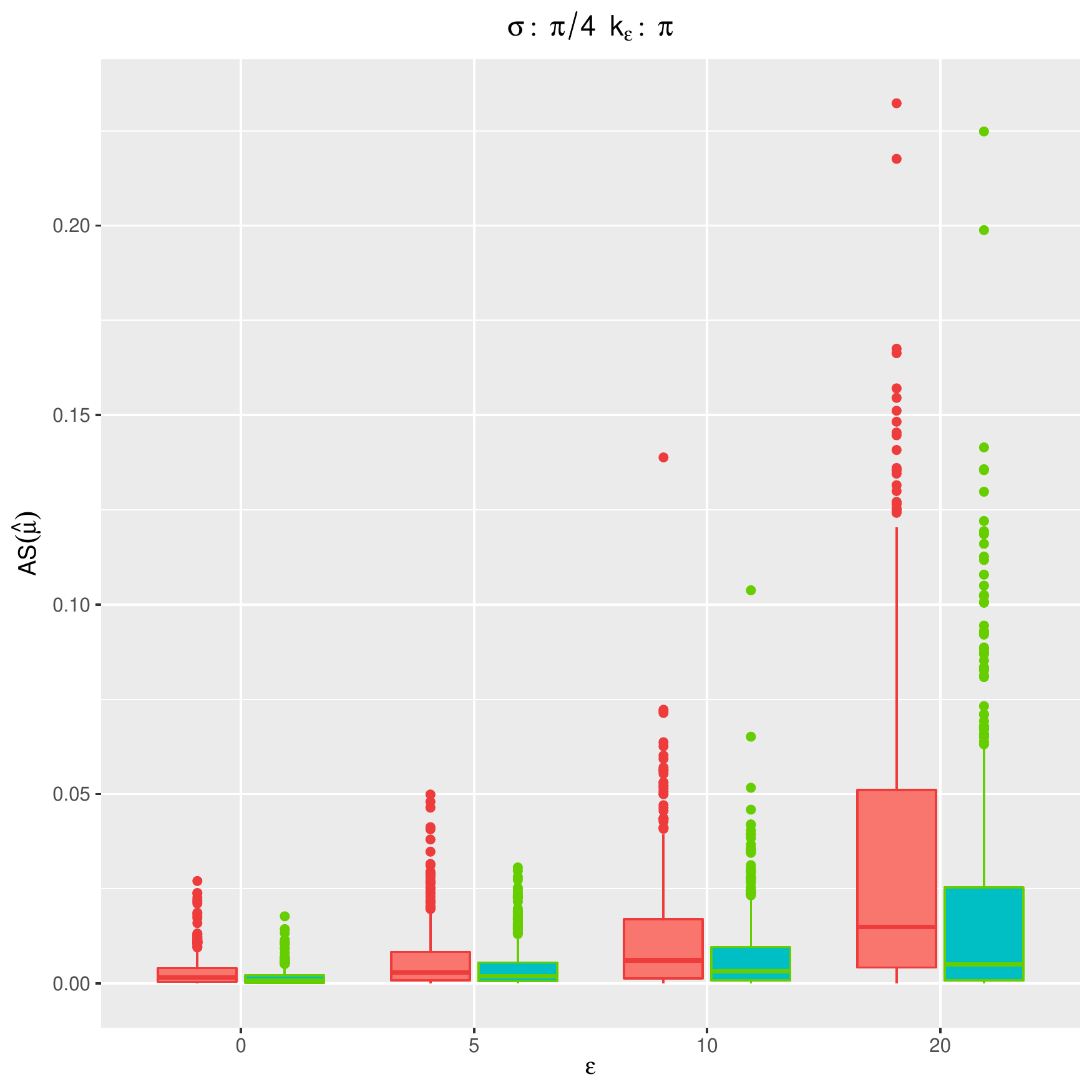} \\
	 \includegraphics[height=0.3\textheight, width=0.3\textwidth]{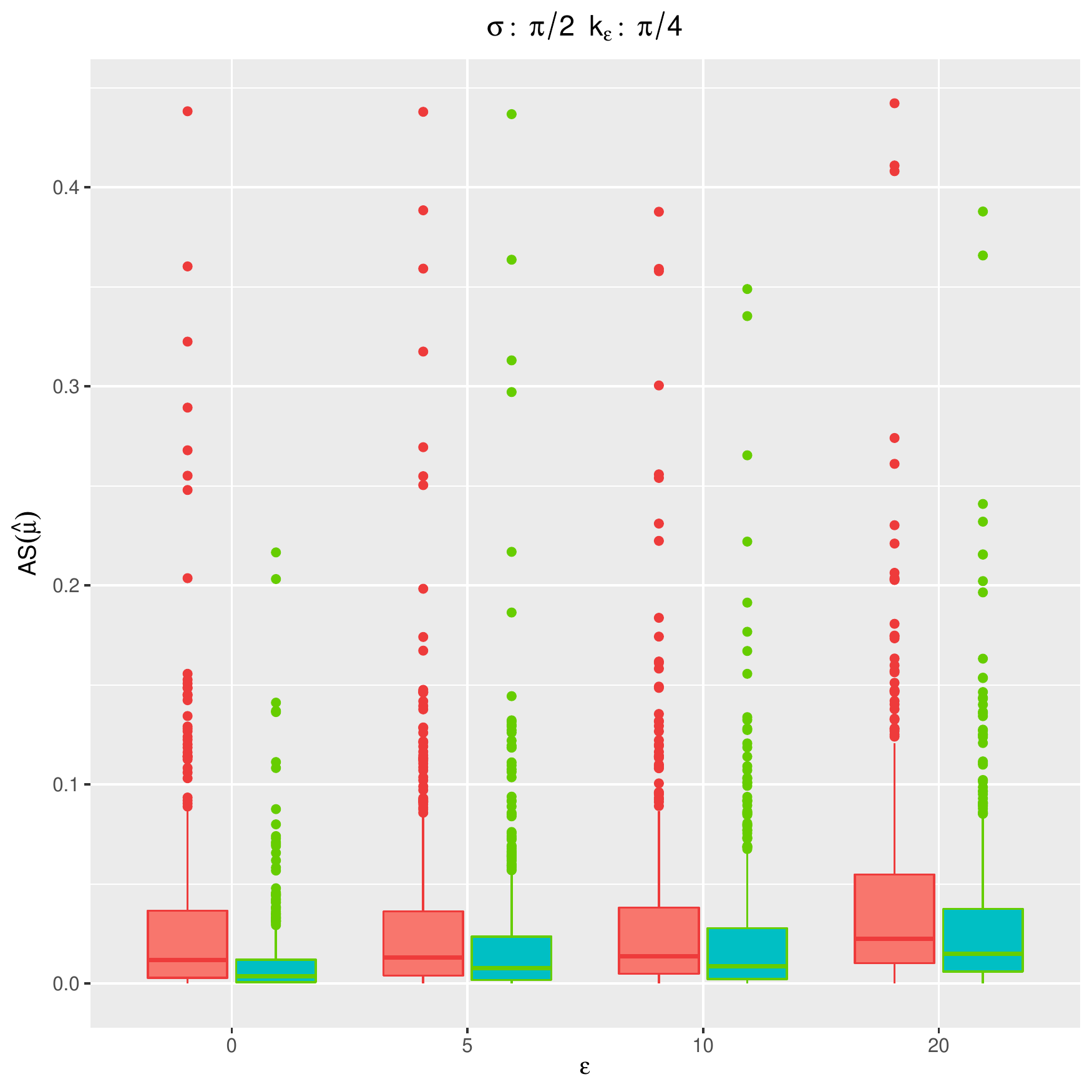} 
	 \includegraphics[height=0.3\textheight, width=0.3\textwidth]{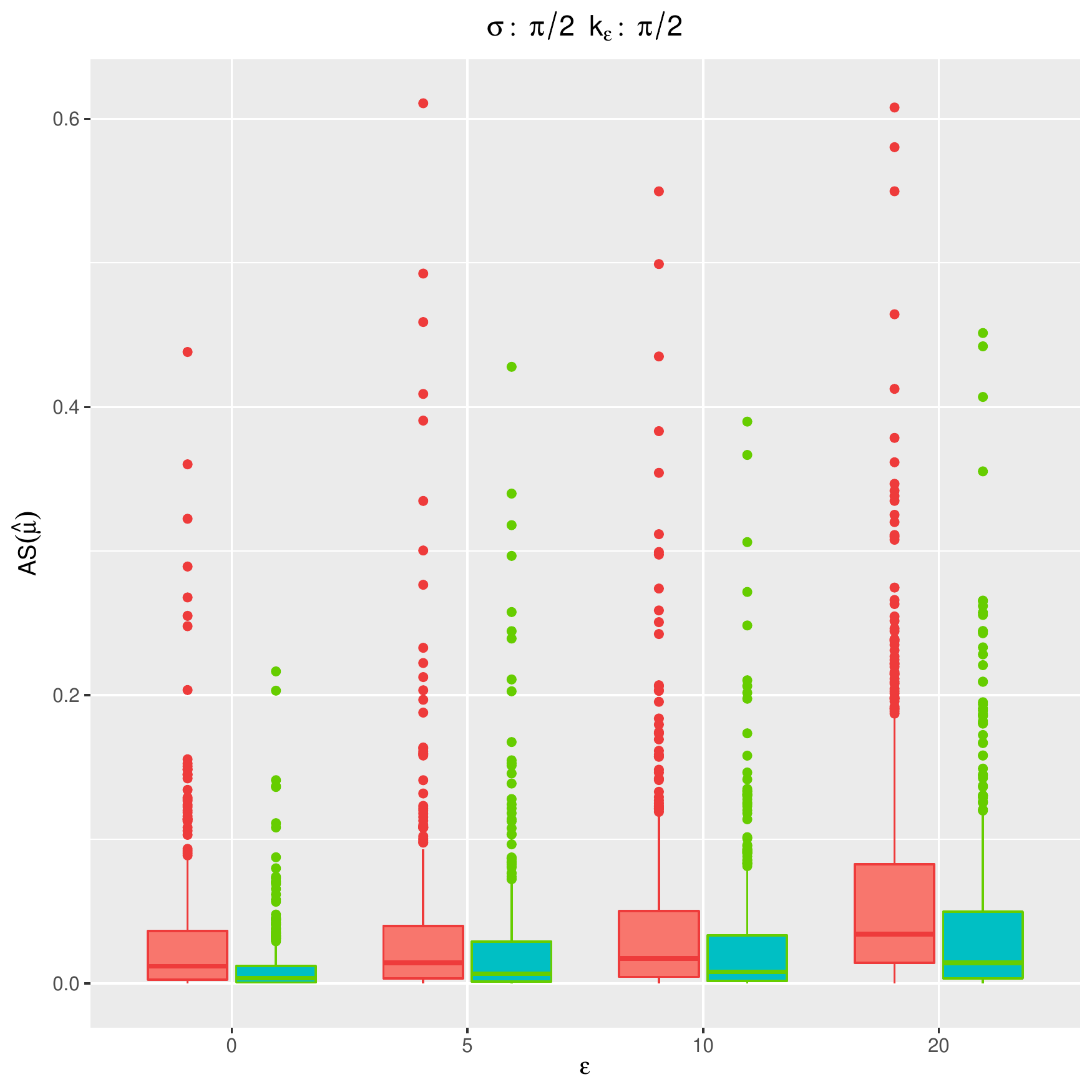} 
	 \includegraphics[height=0.3\textheight, width=0.3\textwidth]{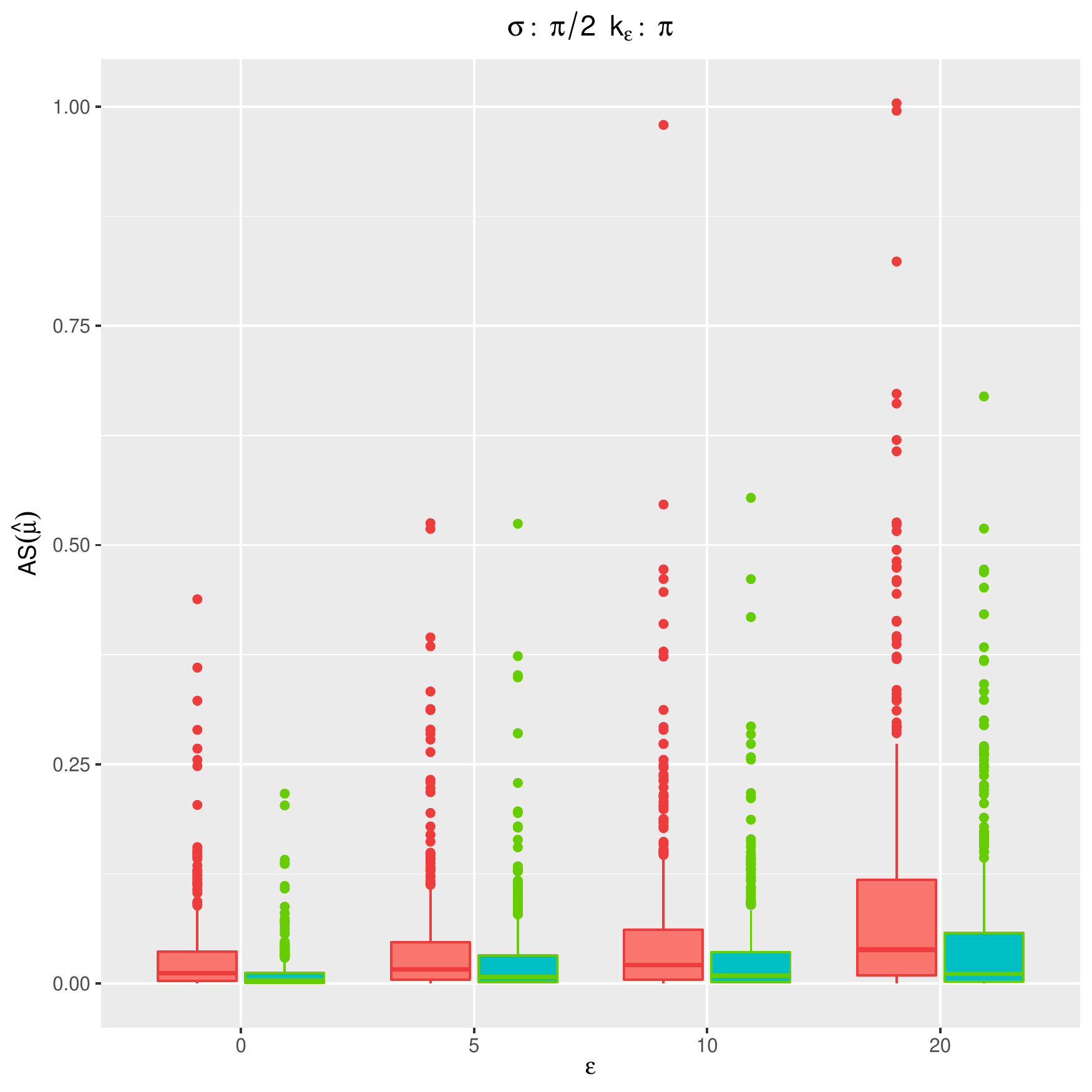} \\
	 \end{center}
	 \caption{Distribution of angle separation for $n=100$ and $p=2$ using weighted CEM (in green) and the CEM (in red). The contamination rate $\epsilon$ is given on the horizontal axis. Increasing contamination size $k_\epsilon$ from left to right, increasing $\sigma$ from top to bottom.}
	 \label{fig:sm:3}
	\end{figure}

\begin{figure}
 	\begin{center}
 		\includegraphics[height=0.3\textheight, width=0.3\textwidth]{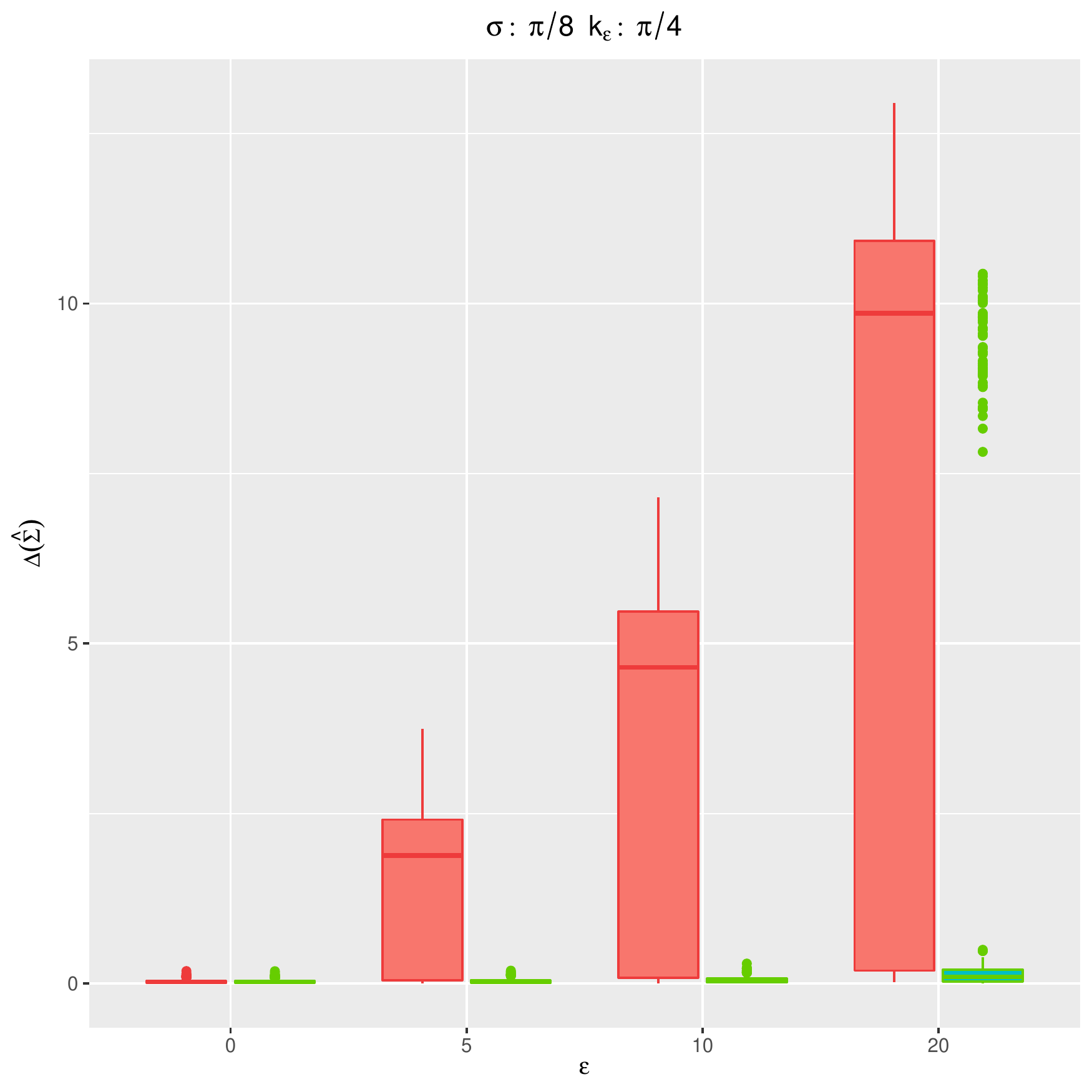} 
 		\includegraphics[height=0.3\textheight, width=0.3\textwidth]{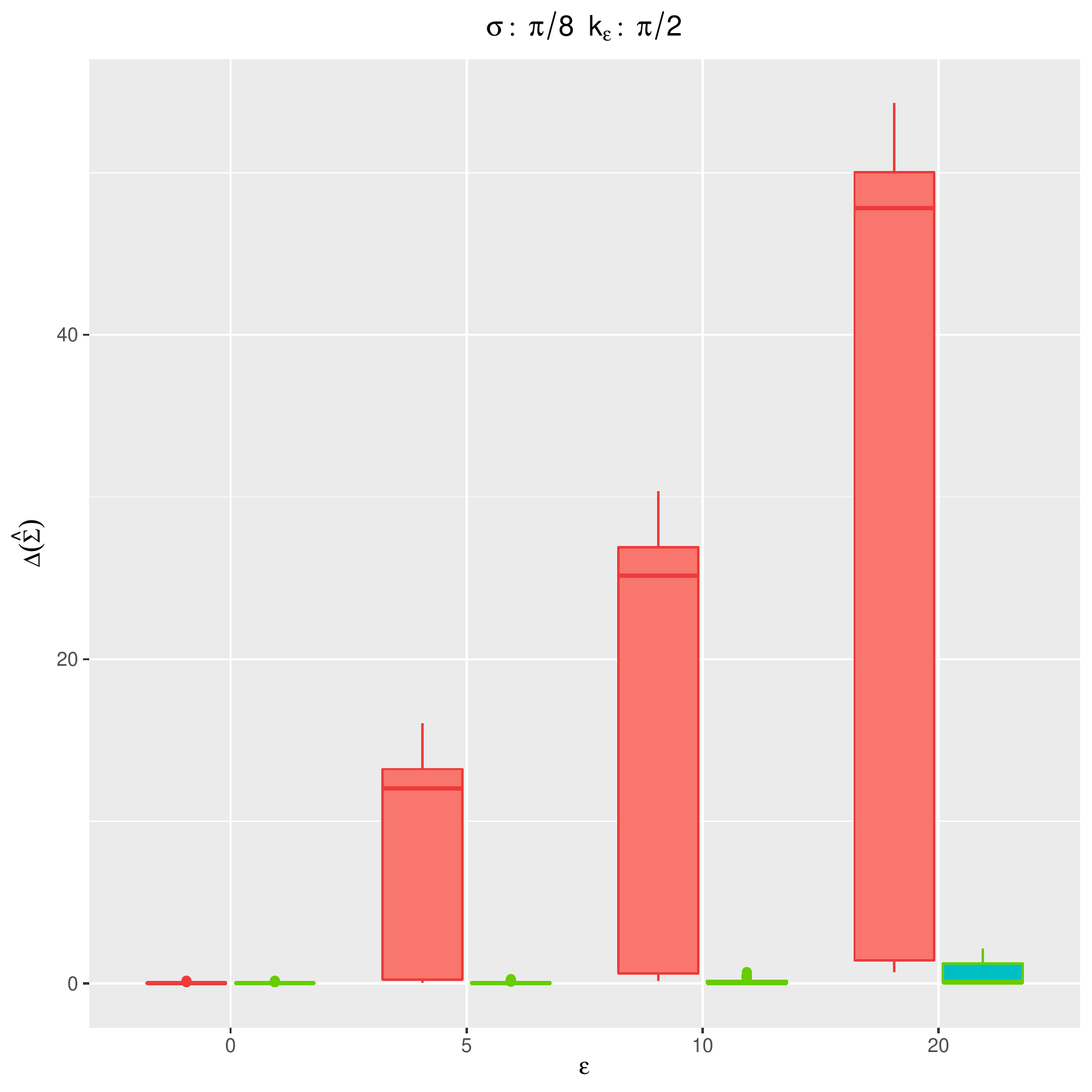} 
 		\includegraphics[height=0.3\textheight, width=0.3\textwidth]{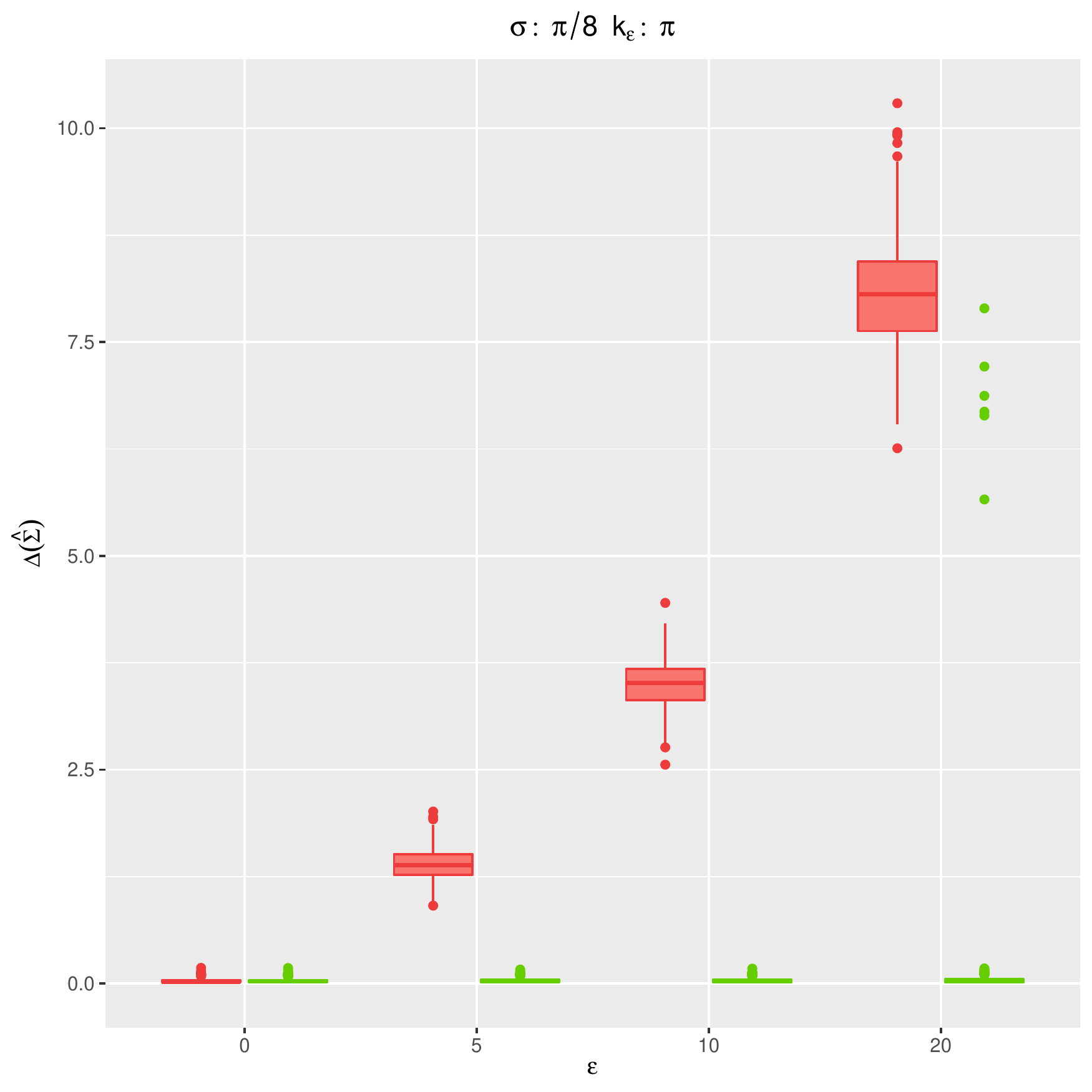} \\
 		\includegraphics[height=0.3\textheight, width=0.3\textwidth]{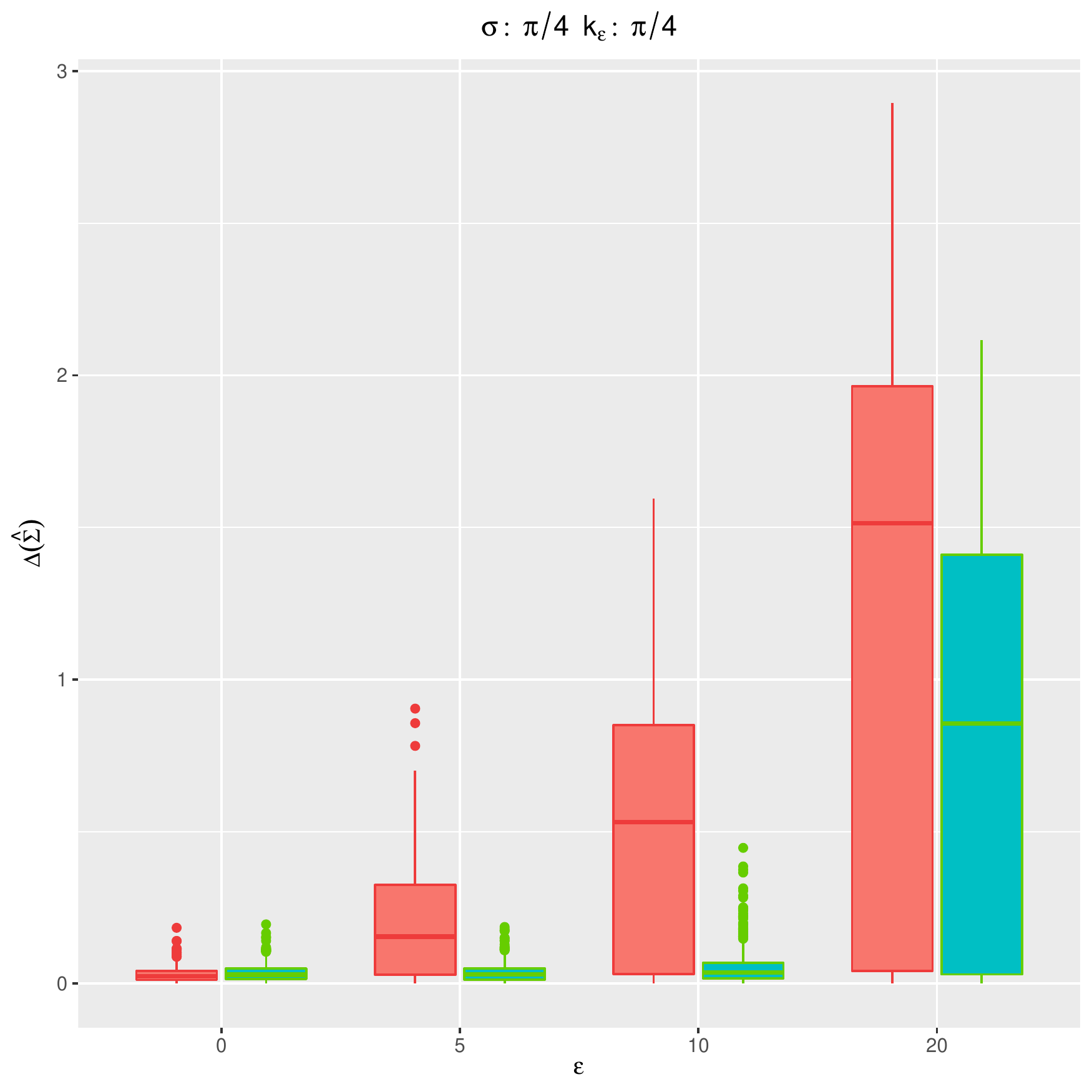} 
 		\includegraphics[height=0.3\textheight, width=0.3\textwidth]{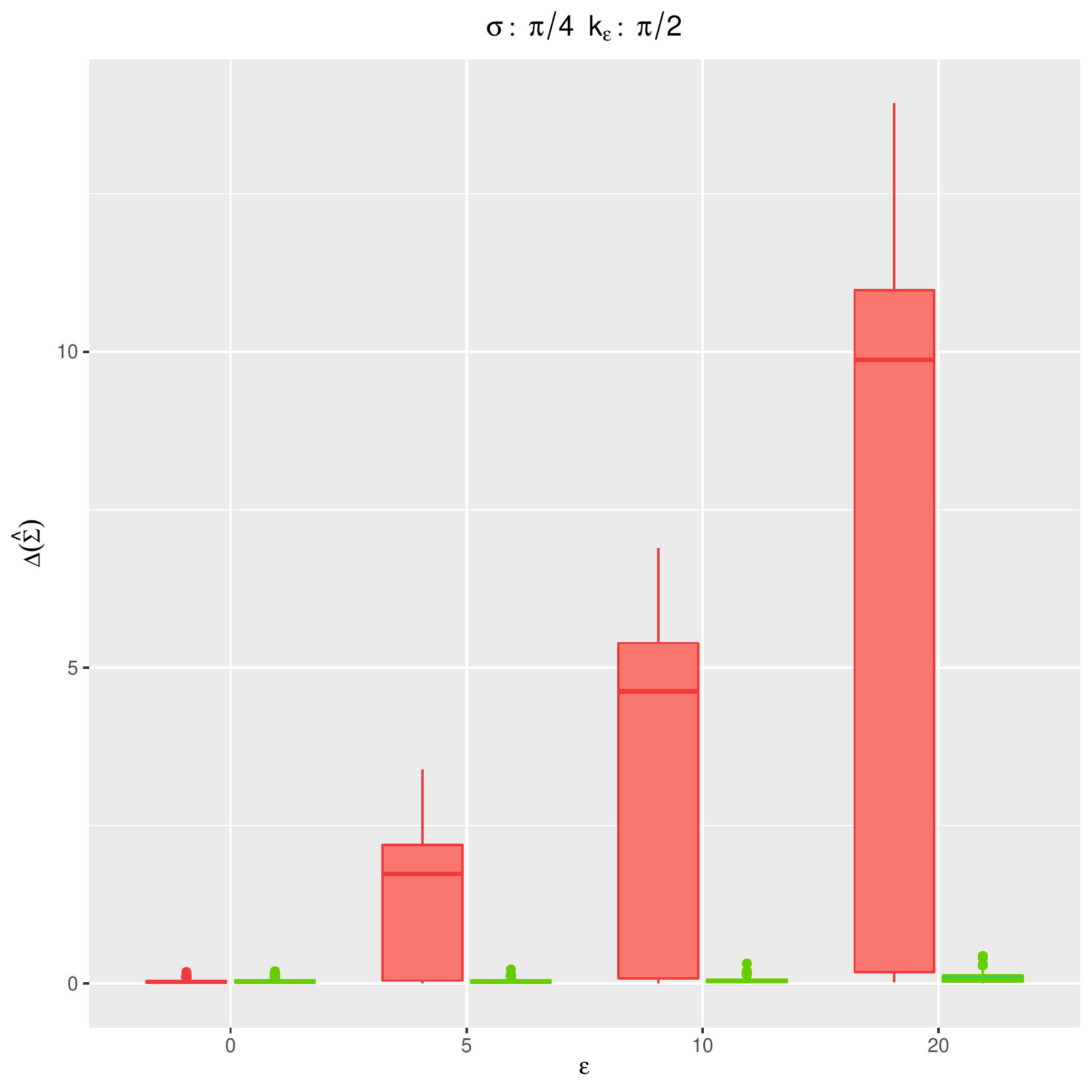} 
 		\includegraphics[height=0.3\textheight, width=0.3\textwidth]{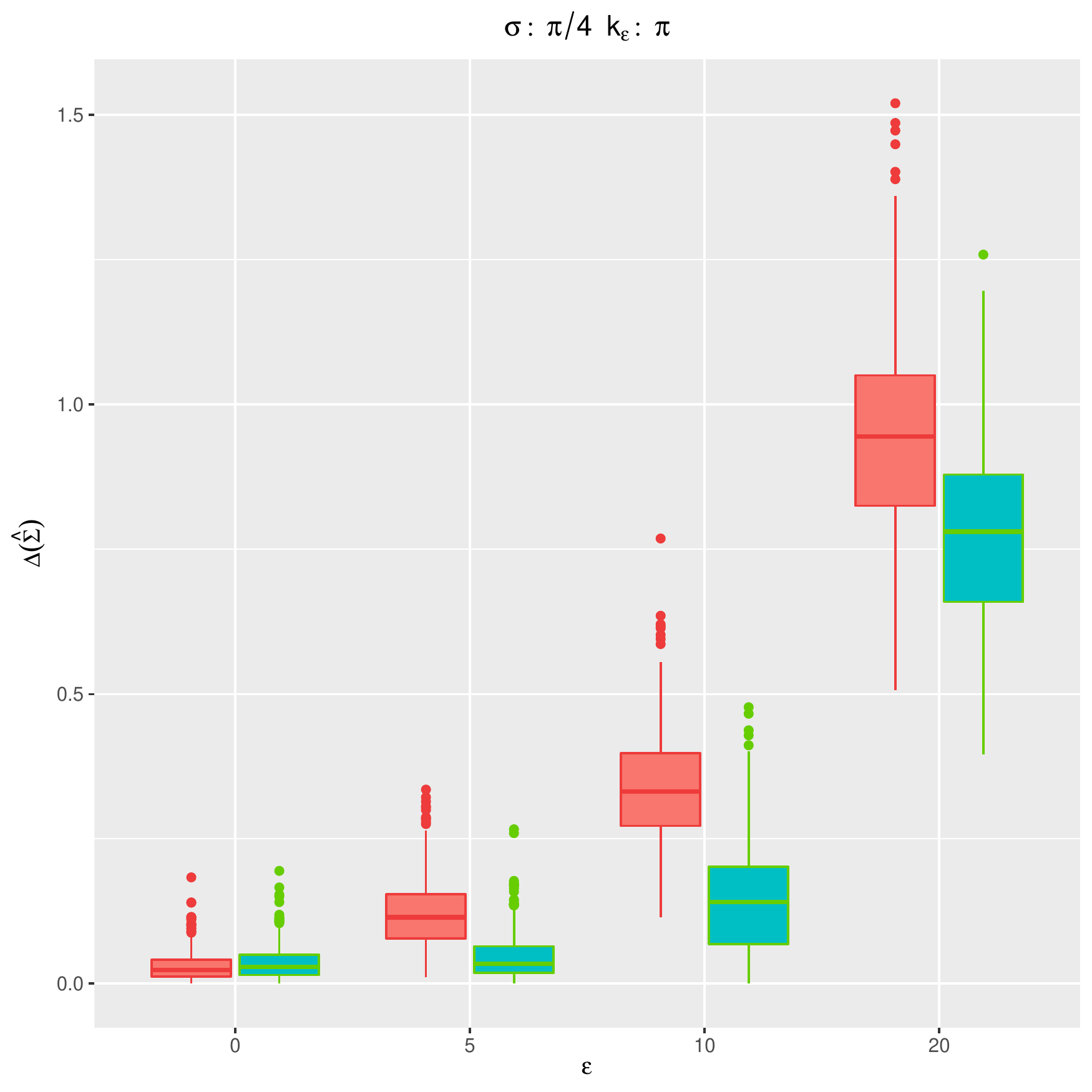} \\
 		\includegraphics[height=0.3\textheight, width=0.3\textwidth]{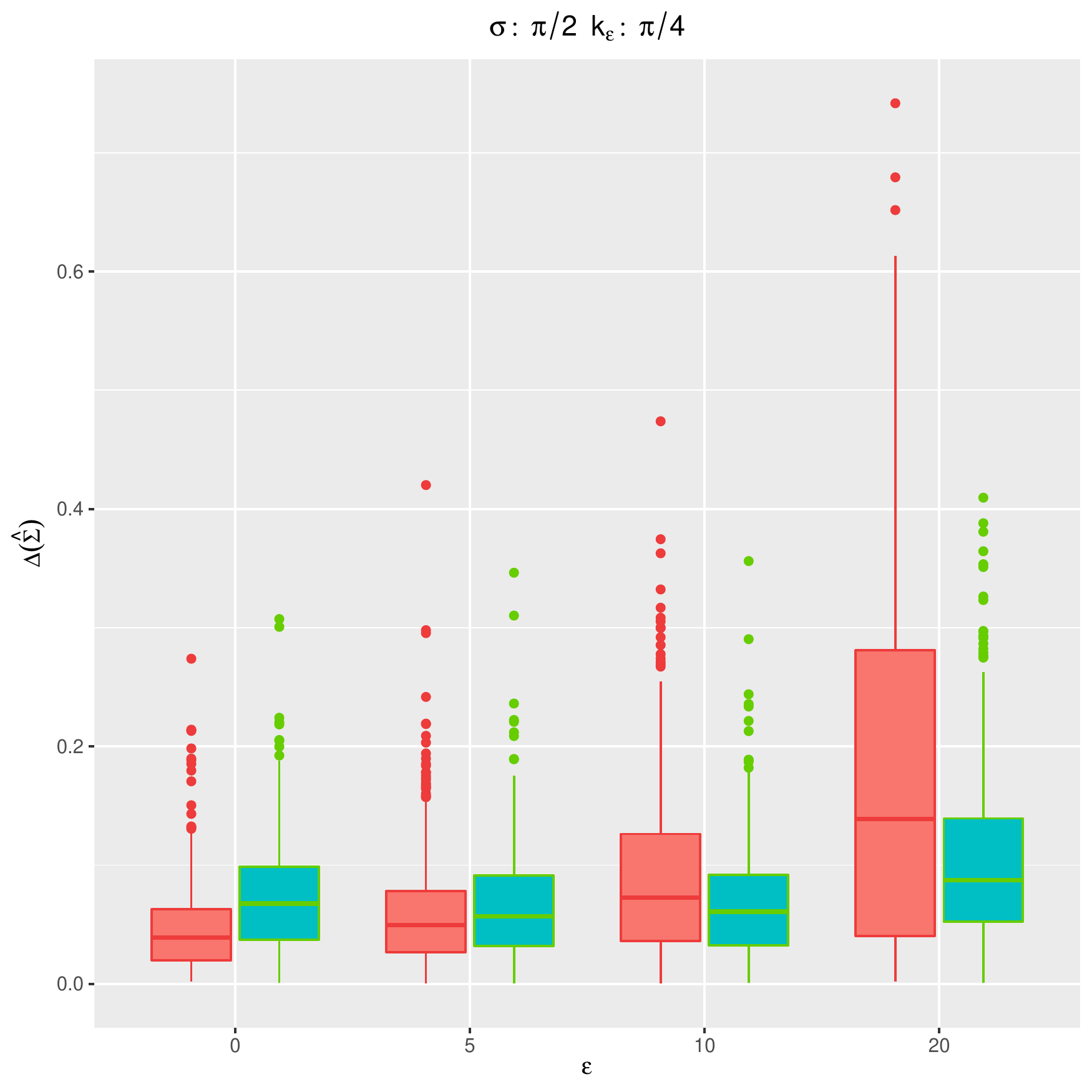} 
 		\includegraphics[height=0.3\textheight, width=0.3\textwidth]{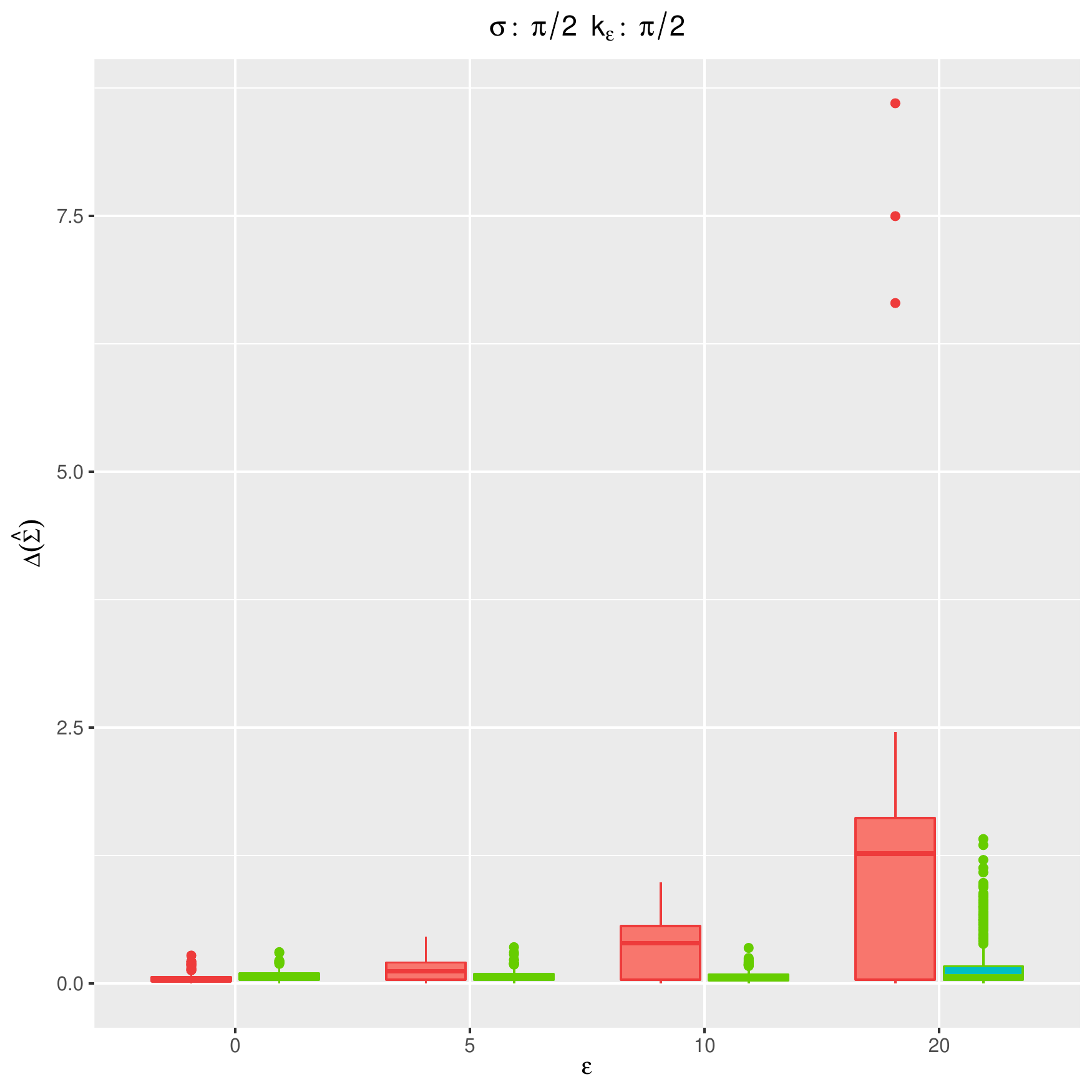} 
 		\includegraphics[height=0.3\textheight, width=0.3\textwidth]{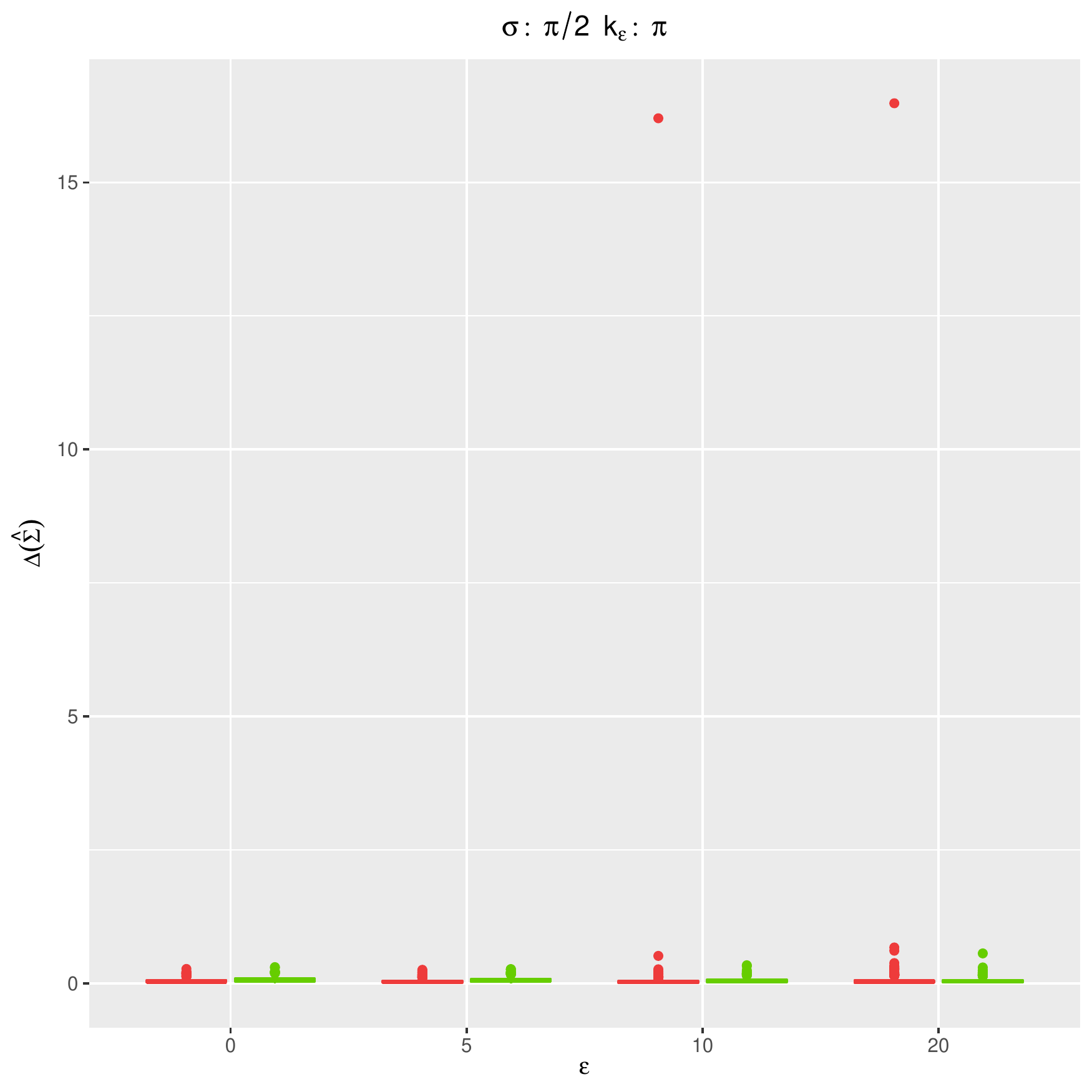} \\
 	\end{center}
 	\caption{Distribution of the divergence measure for $n=100$ and $p=2$ using the weighted CEM (in green) and the CEM (in red). The contamination rate $\epsilon$ is given on the horizontal axis. Increasing contamination size $k_\epsilon$ from left to right, increasing $\sigma$ from top to bottom.}
 	\label{fig:sm:4}
 \end{figure}

\begin{figure}
	 \begin{center}
	 \includegraphics[height=0.3\textheight, width=0.3\textwidth]{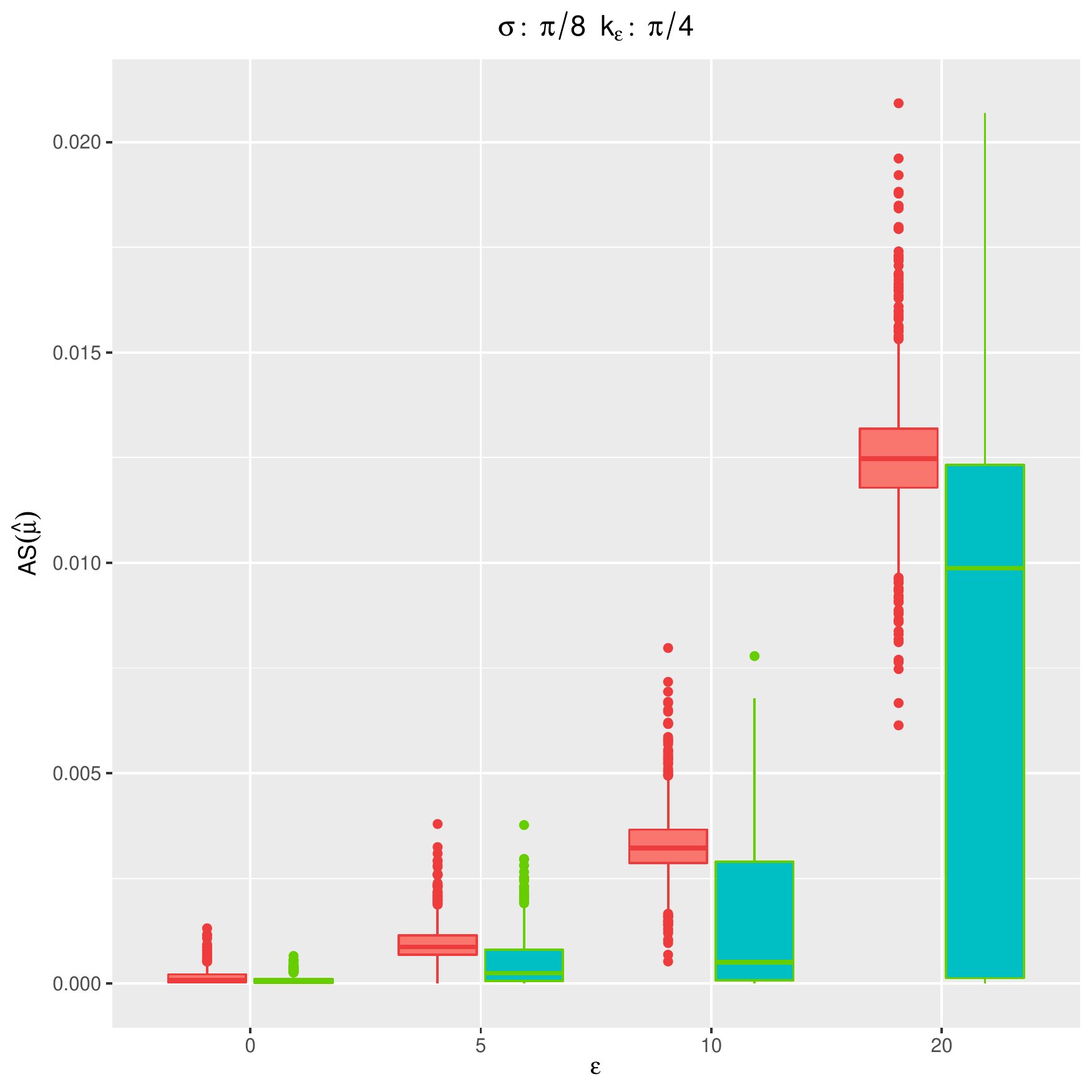} 
	 \includegraphics[height=0.3\textheight, width=0.3\textwidth]{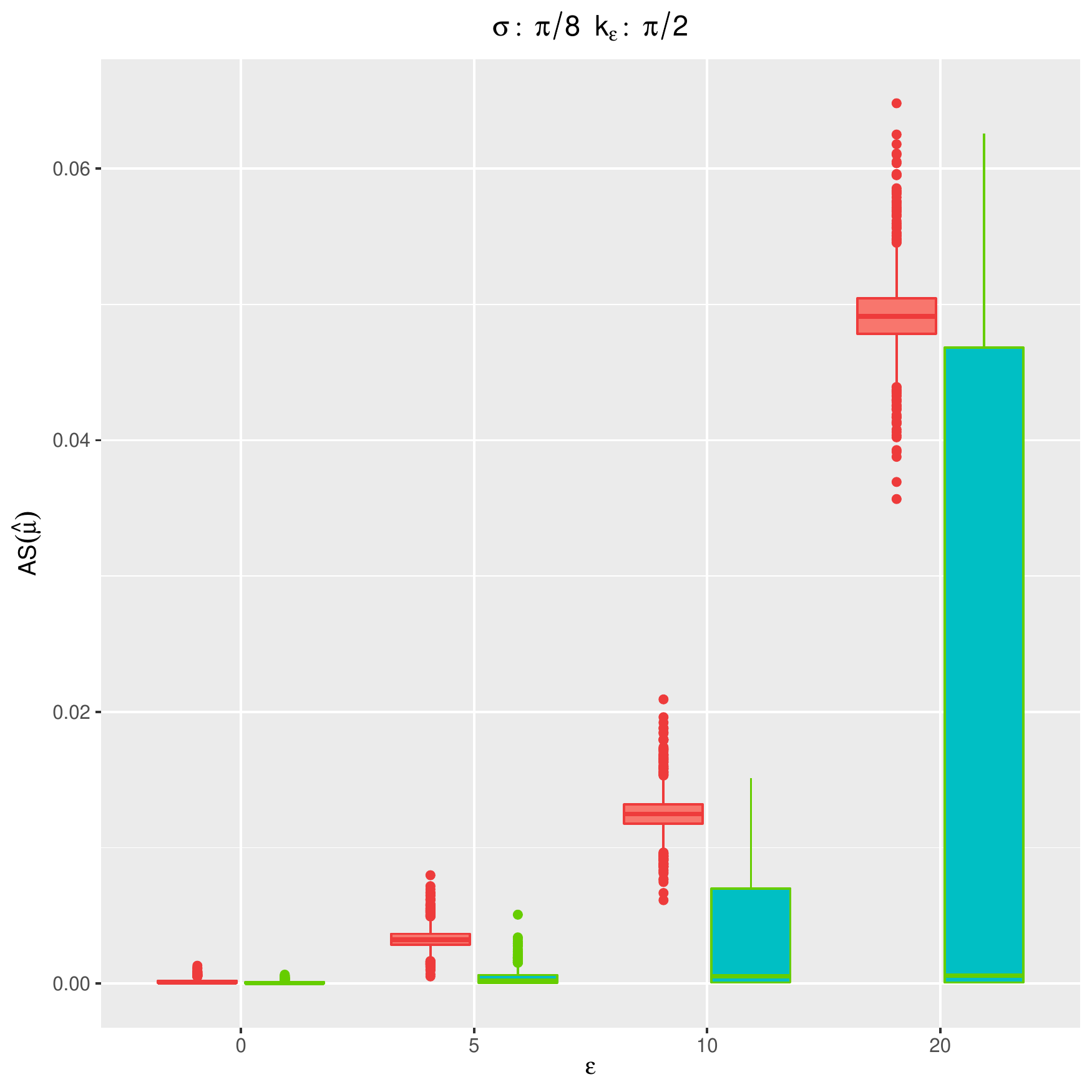} 
	 \includegraphics[height=0.3\textheight, width=0.3\textwidth]{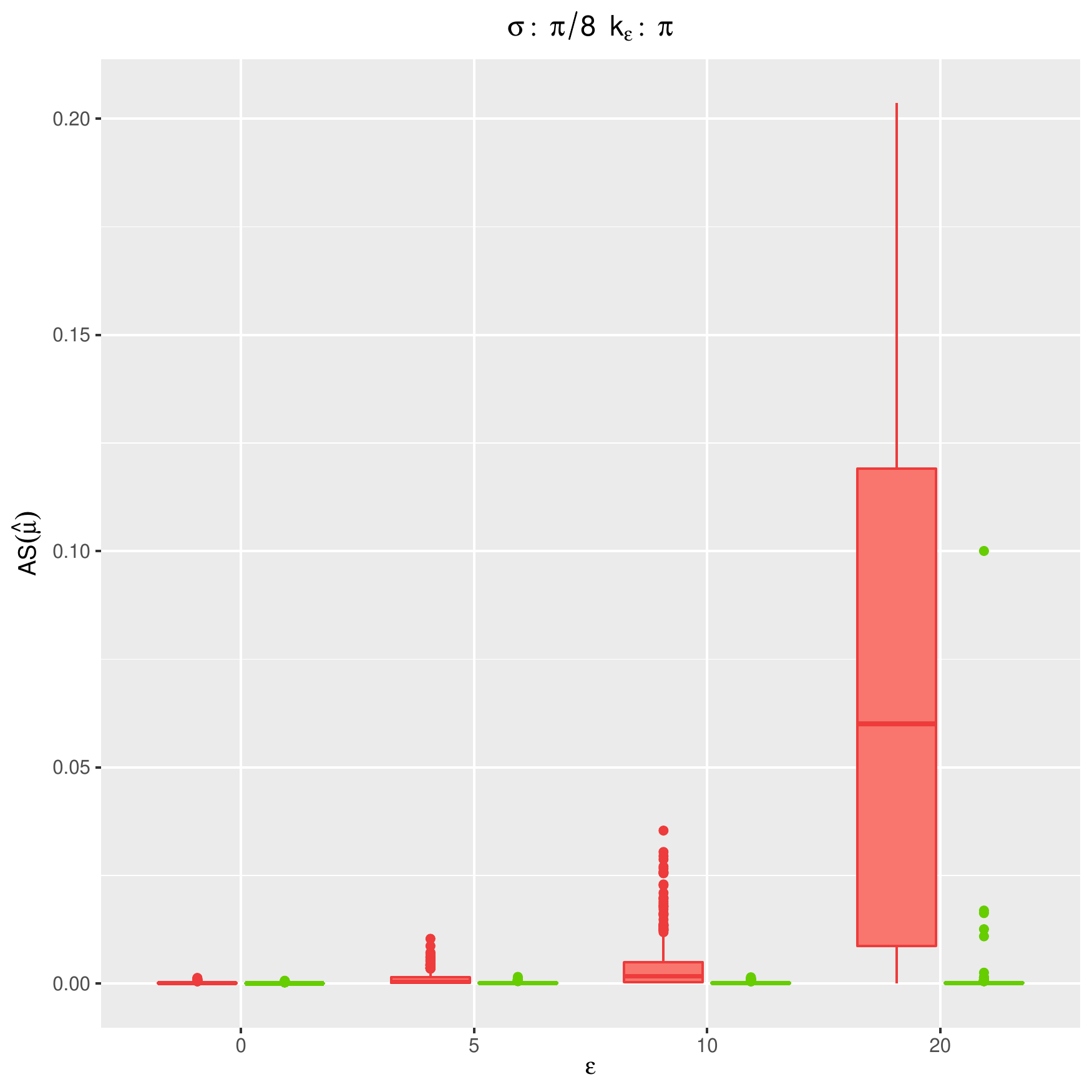} \\
	\includegraphics[height=0.3\textheight, width=0.3\textwidth]{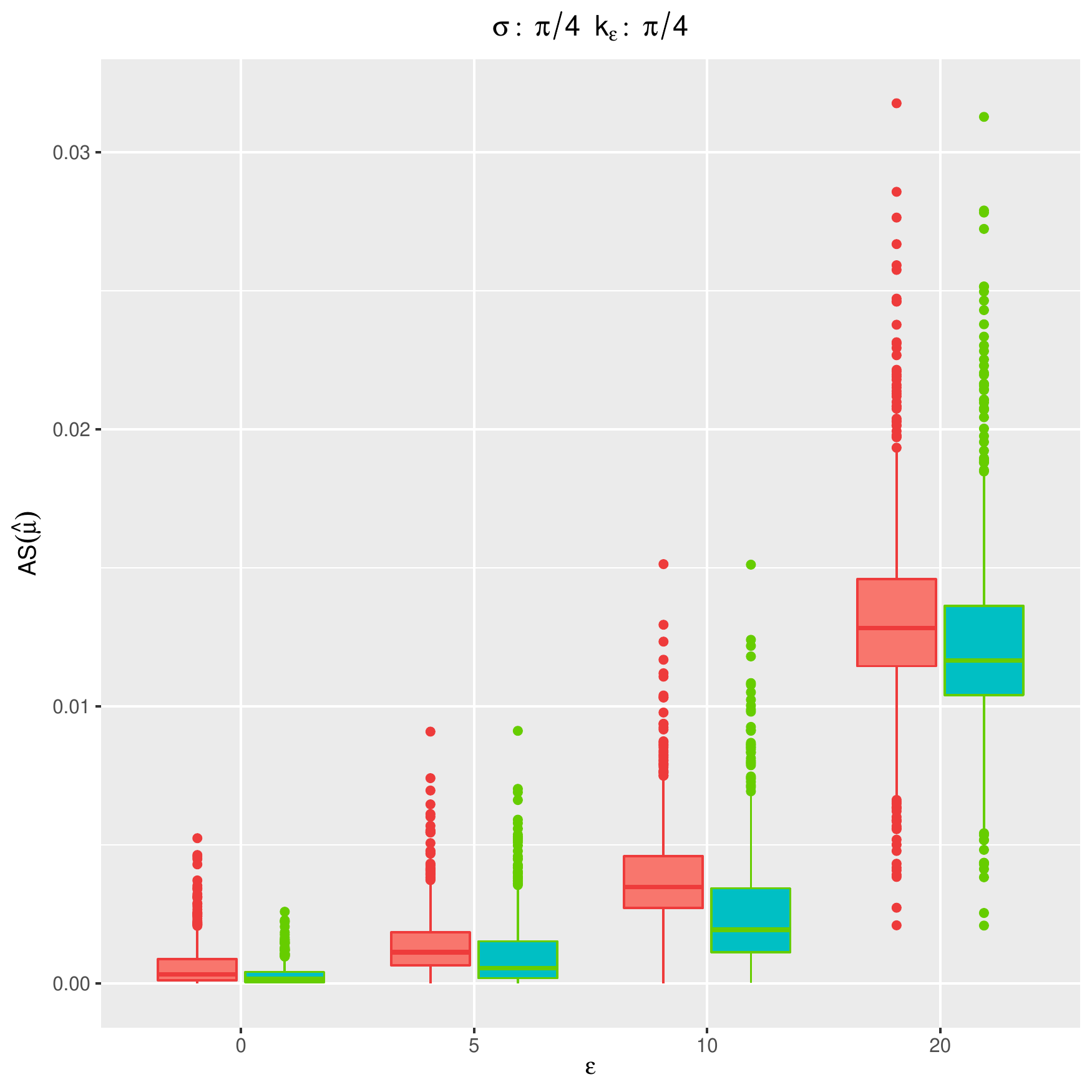} 
	 \includegraphics[height=0.3\textheight, width=0.3\textwidth]{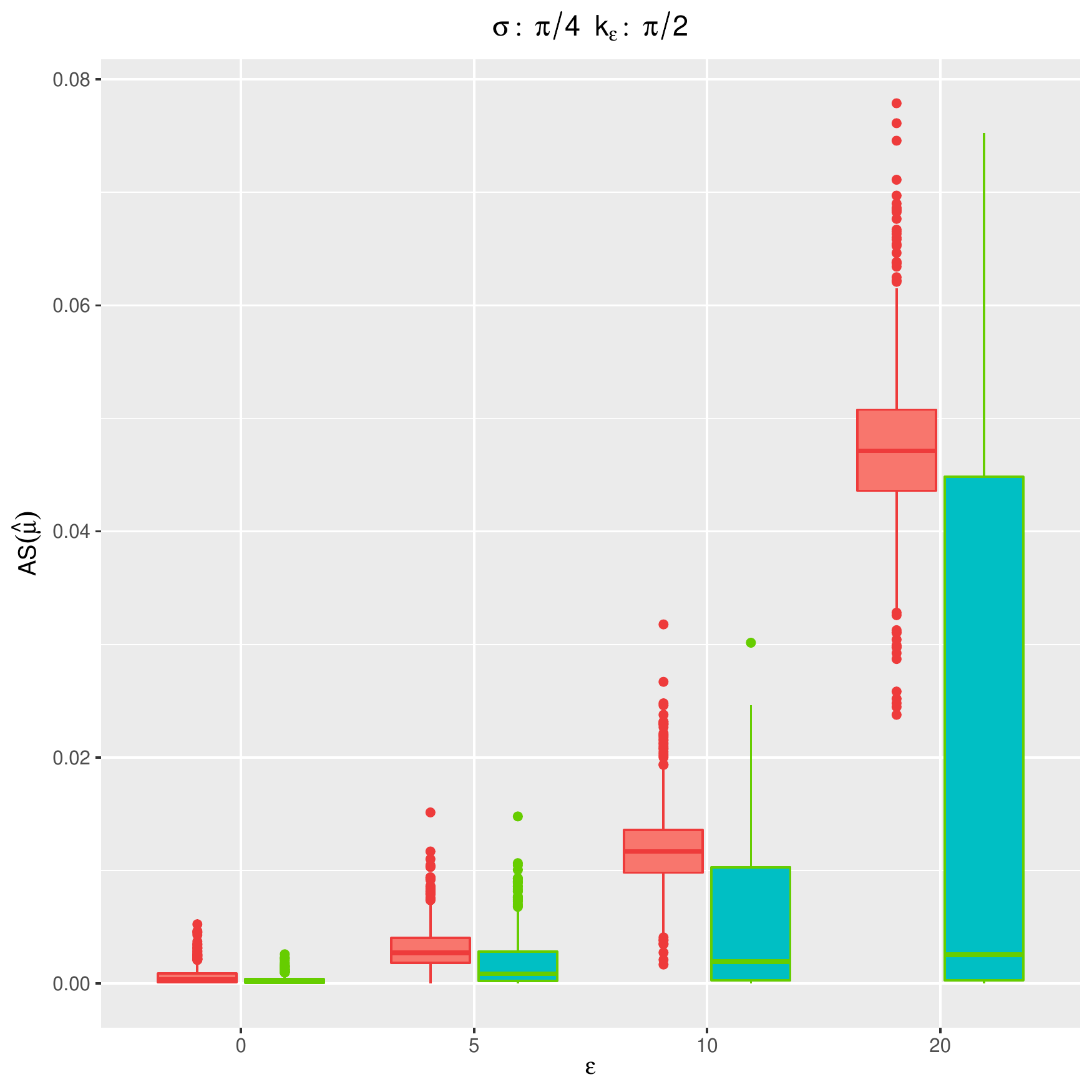} 
	 \includegraphics[height=0.3\textheight, width=0.3\textwidth]{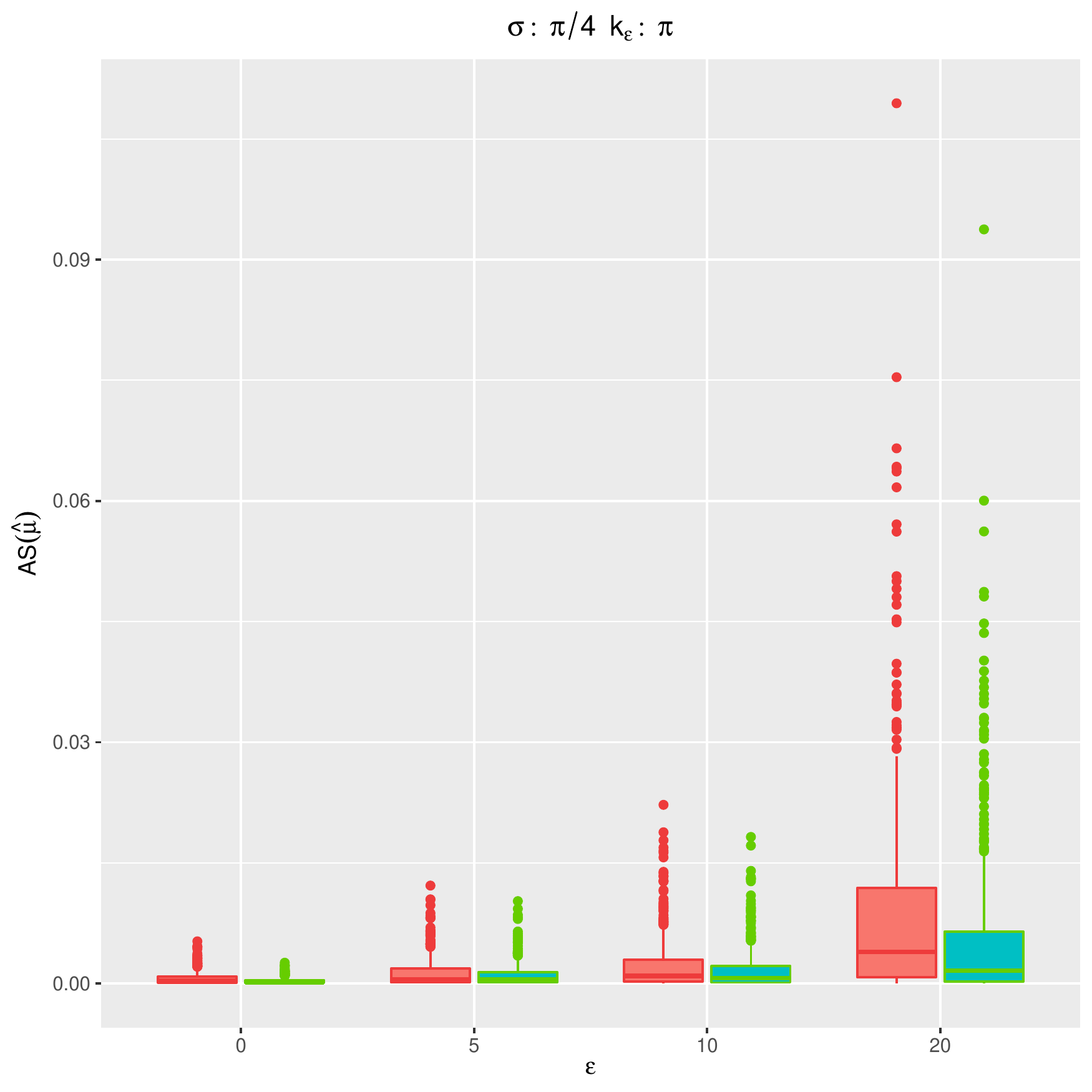} \\
	 \includegraphics[height=0.3\textheight, width=0.3\textwidth]{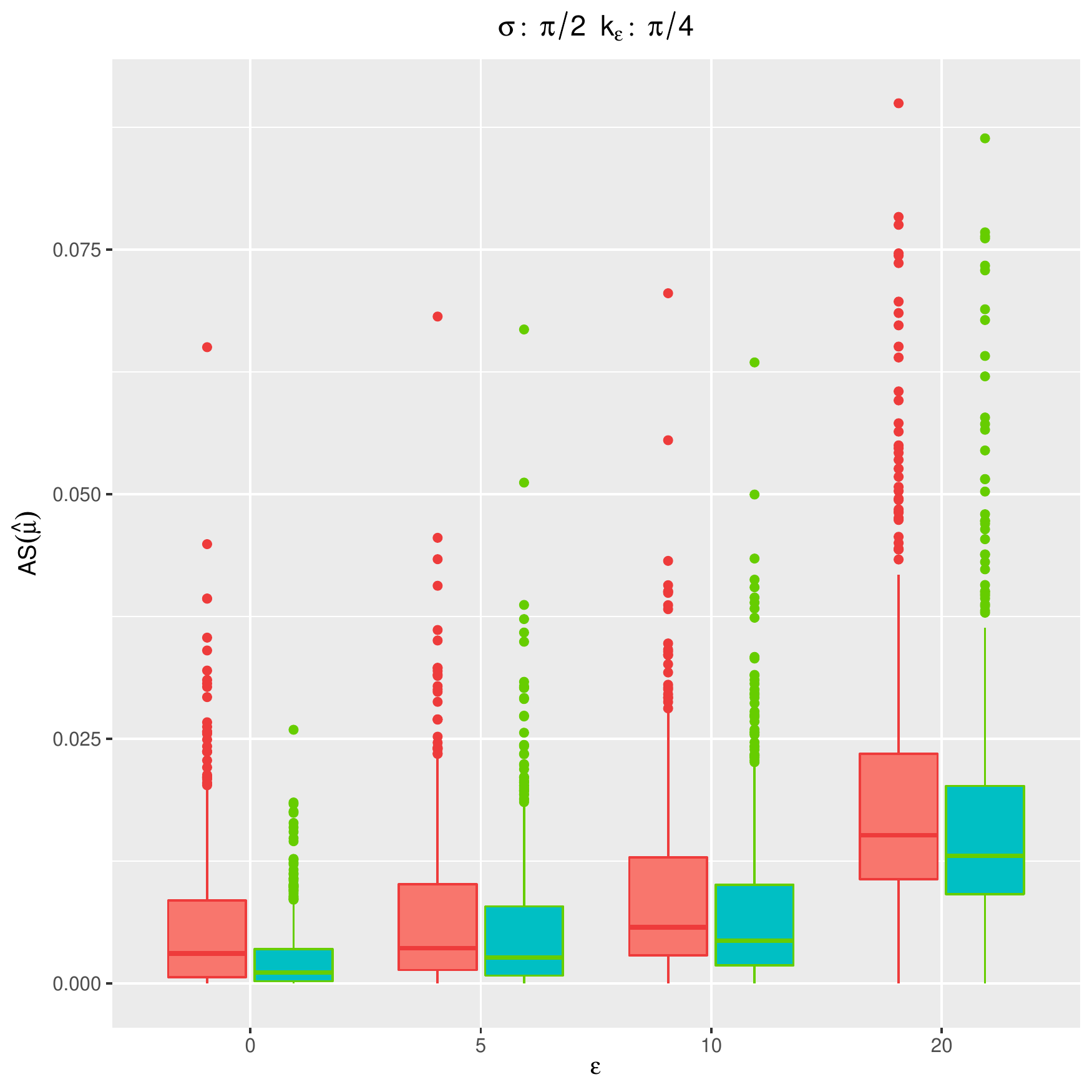} 
	 \includegraphics[height=0.3\textheight, width=0.3\textwidth]{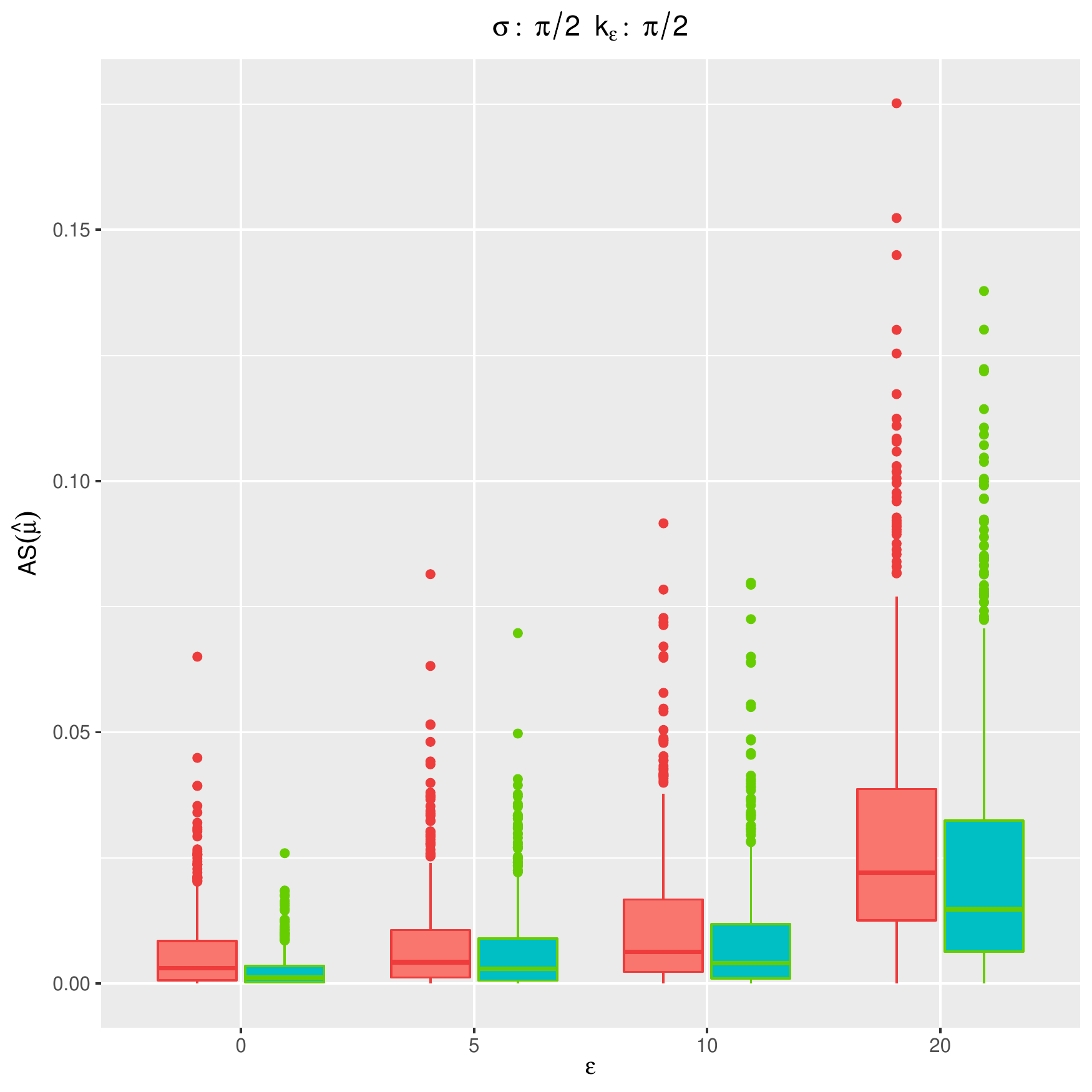} 
	 \includegraphics[height=0.3\textheight, width=0.3\textwidth]{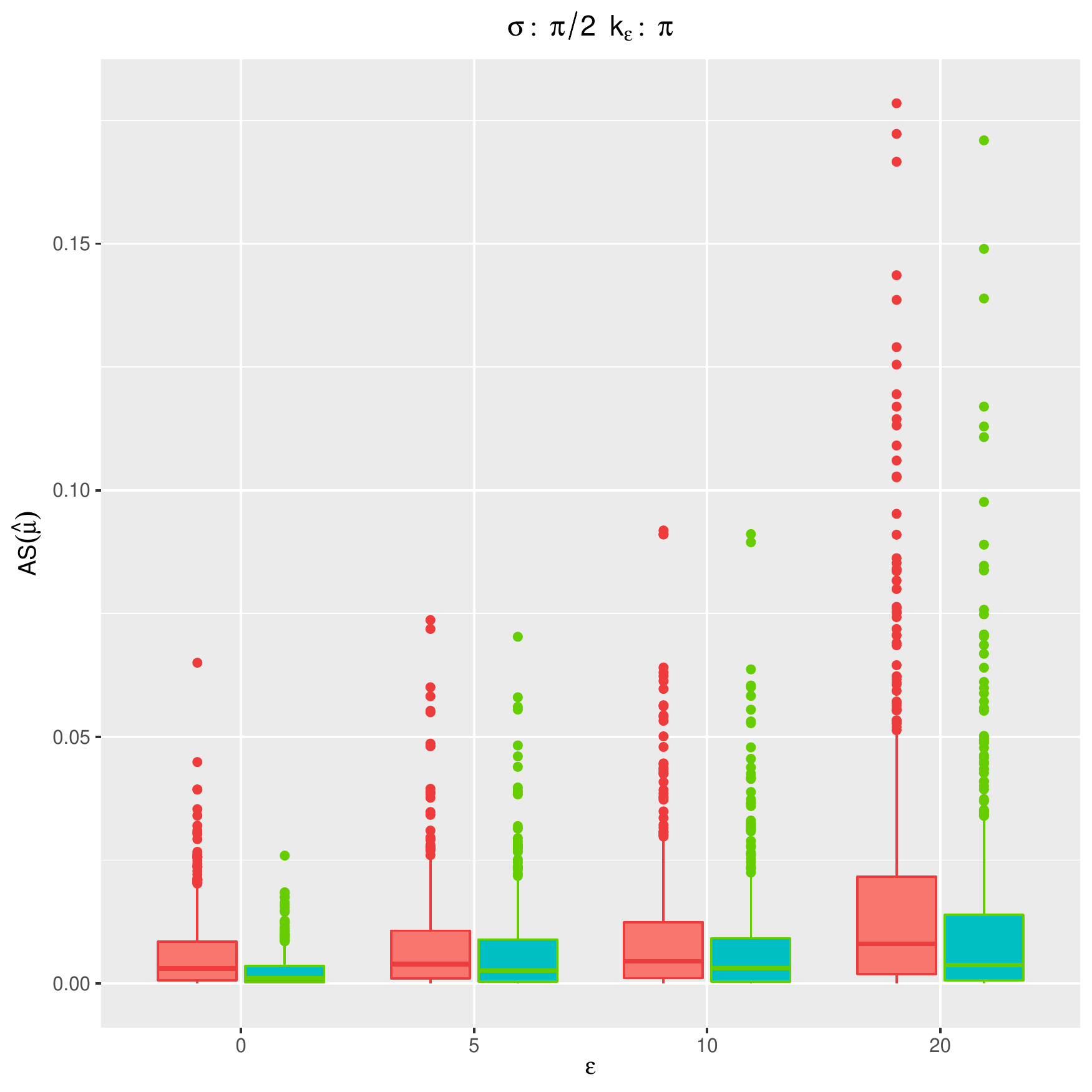} \\
	 \end{center}
	 \caption{Distribution of angle separation for $n=500$ and $p=2$ using weighted CEM (in green) and the CEM (in red). The contamination rate $\epsilon$ is given on the horizontal axis. Increasing contamination size $k_\epsilon$ from left to right, increasing $\sigma$ from top to bottom.}
	 \label{fig:sm:5}
	\end{figure}

\begin{figure}
 	\begin{center}
 		\includegraphics[height=0.3\textheight, width=0.3\textwidth]{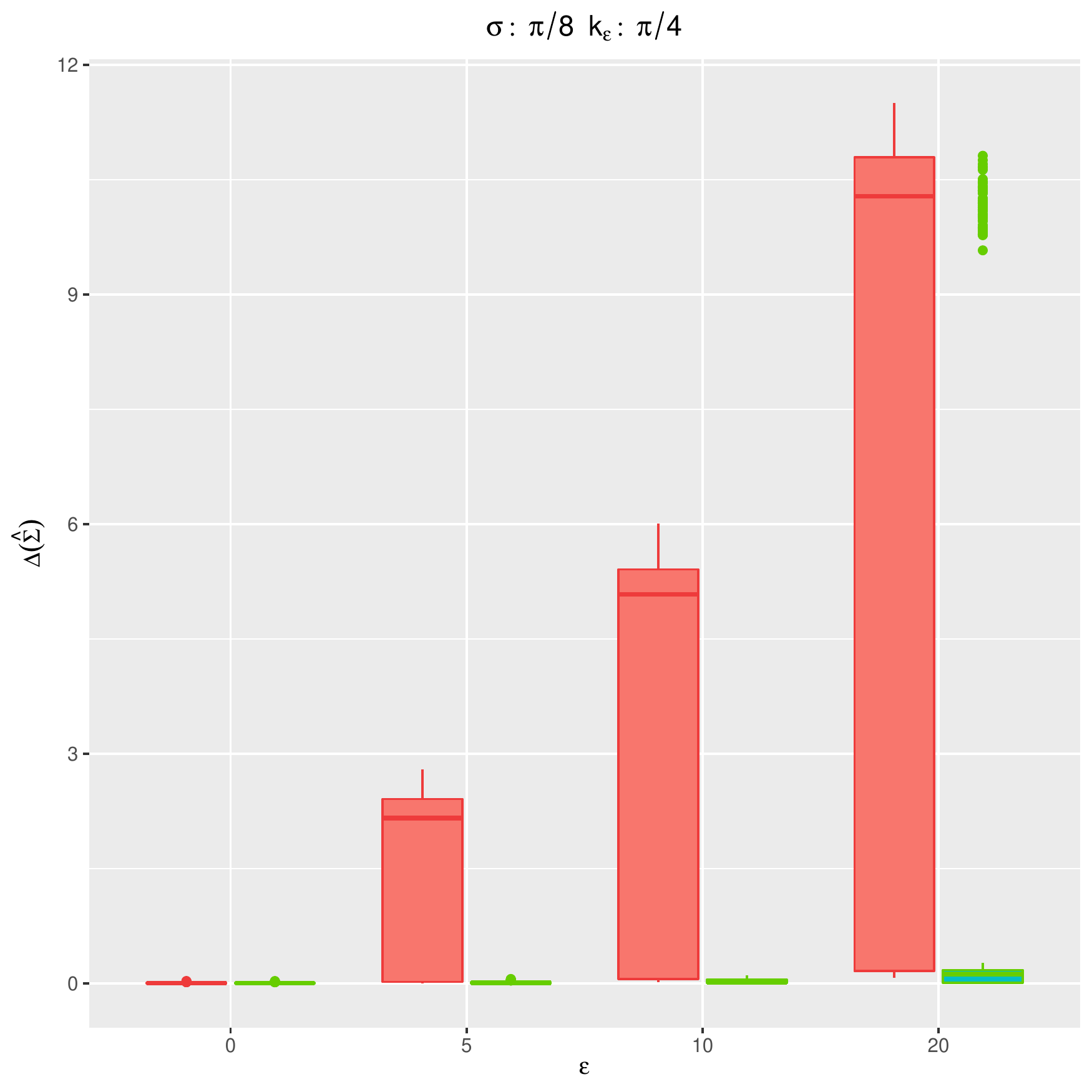} 
 		\includegraphics[height=0.3\textheight, width=0.3\textwidth]{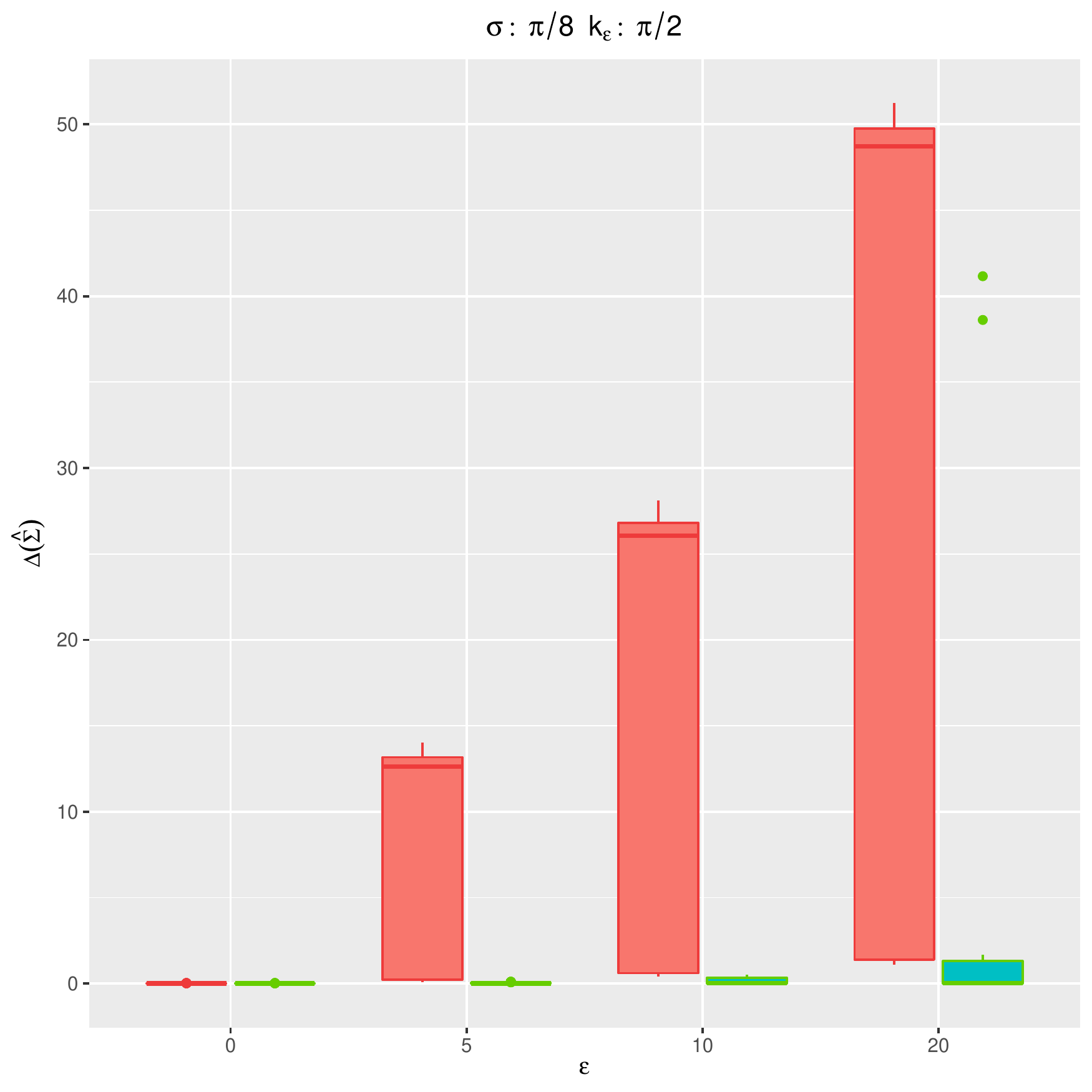} 
 		\includegraphics[height=0.3\textheight, width=0.3\textwidth]{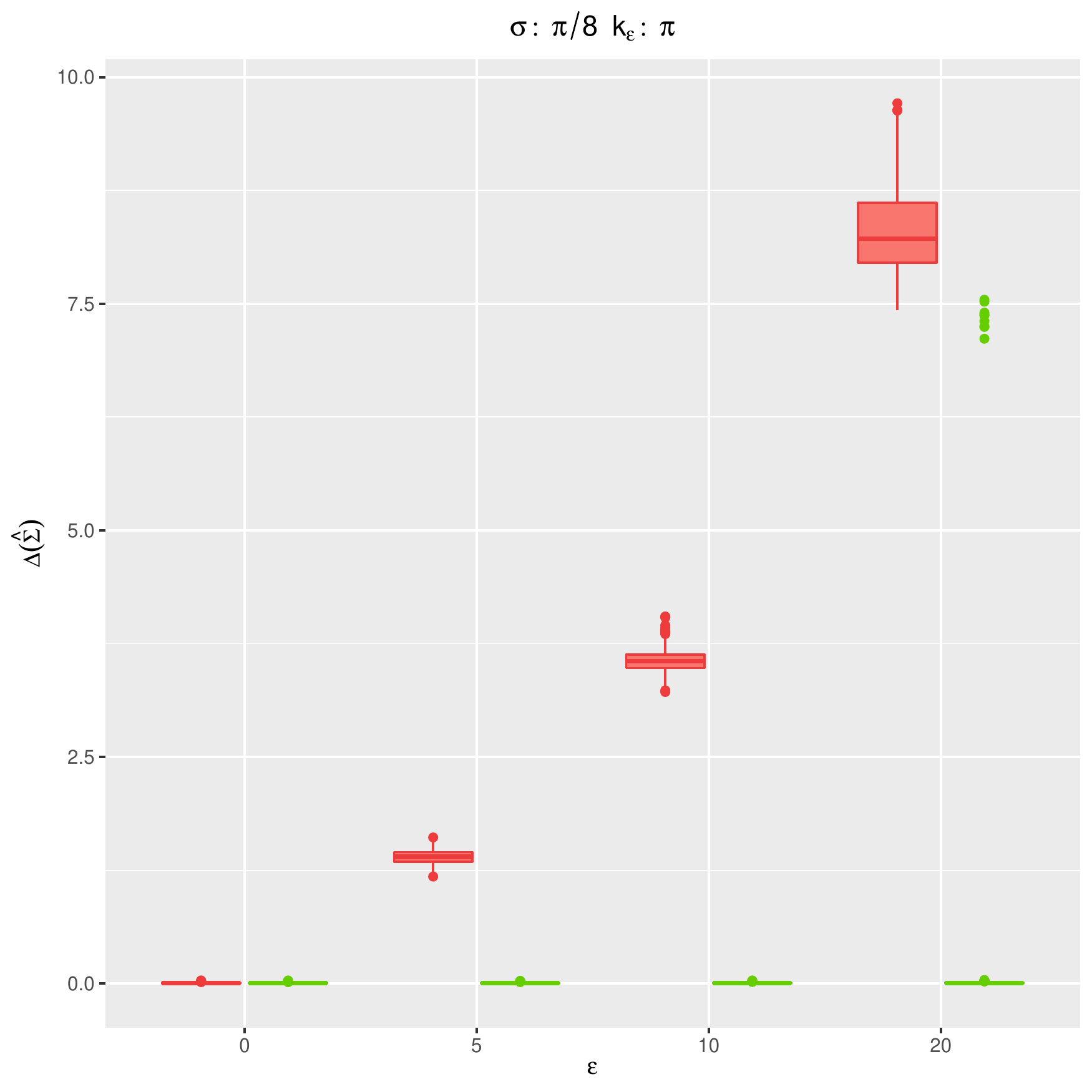} \\
 		\includegraphics[height=0.3\textheight, width=0.3\textwidth]{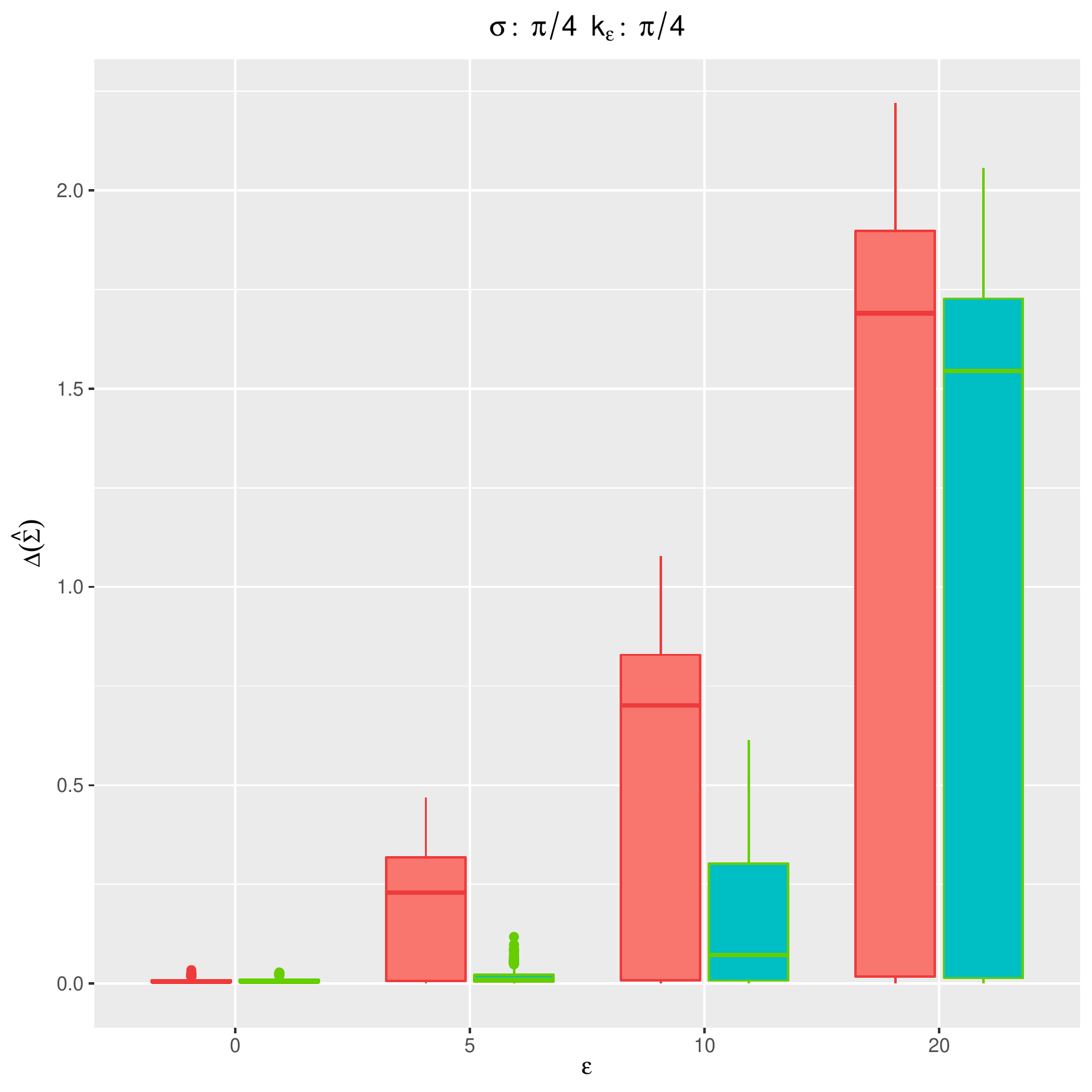} 
 		\includegraphics[height=0.3\textheight, width=0.3\textwidth]{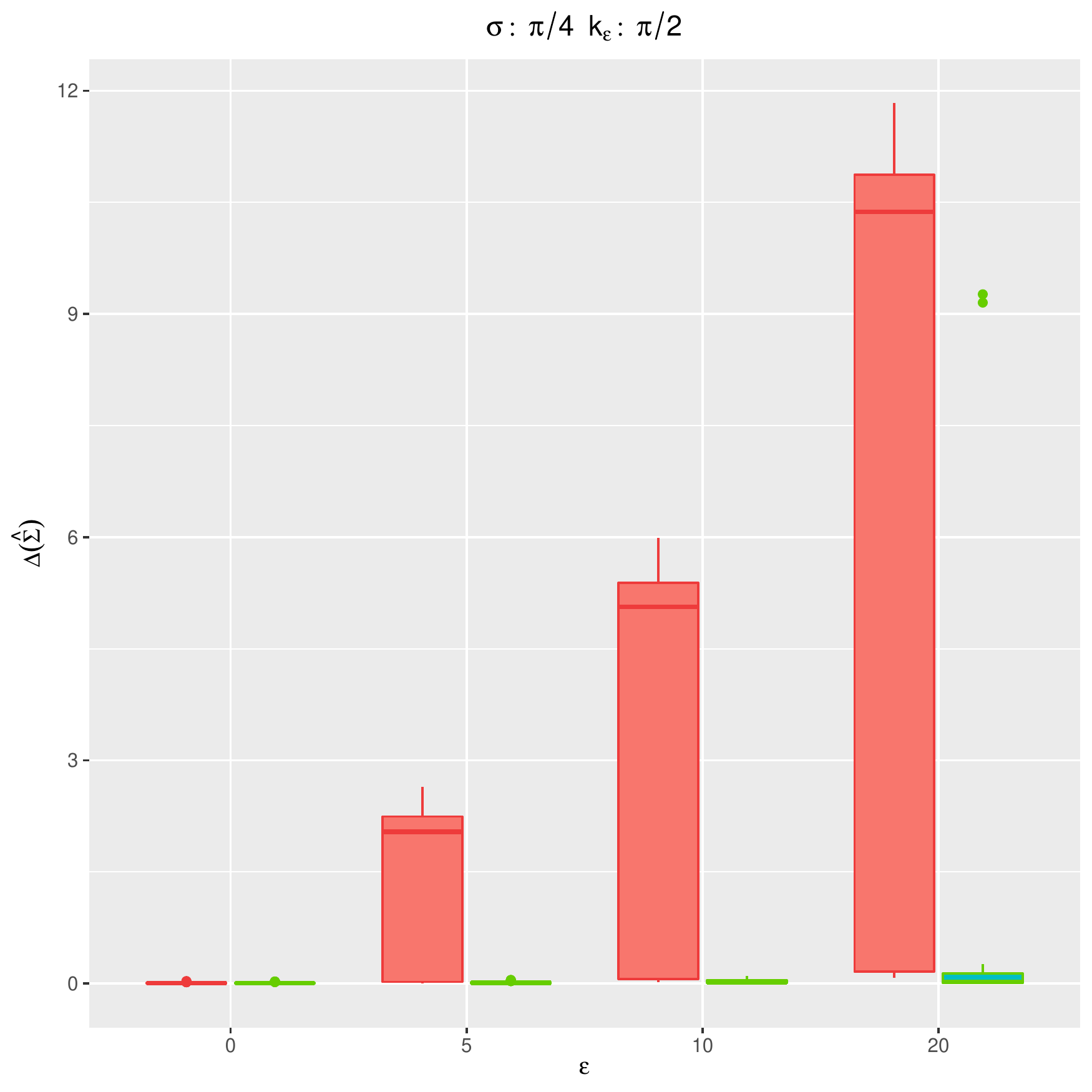} 
 		\includegraphics[height=0.3\textheight, width=0.3\textwidth]{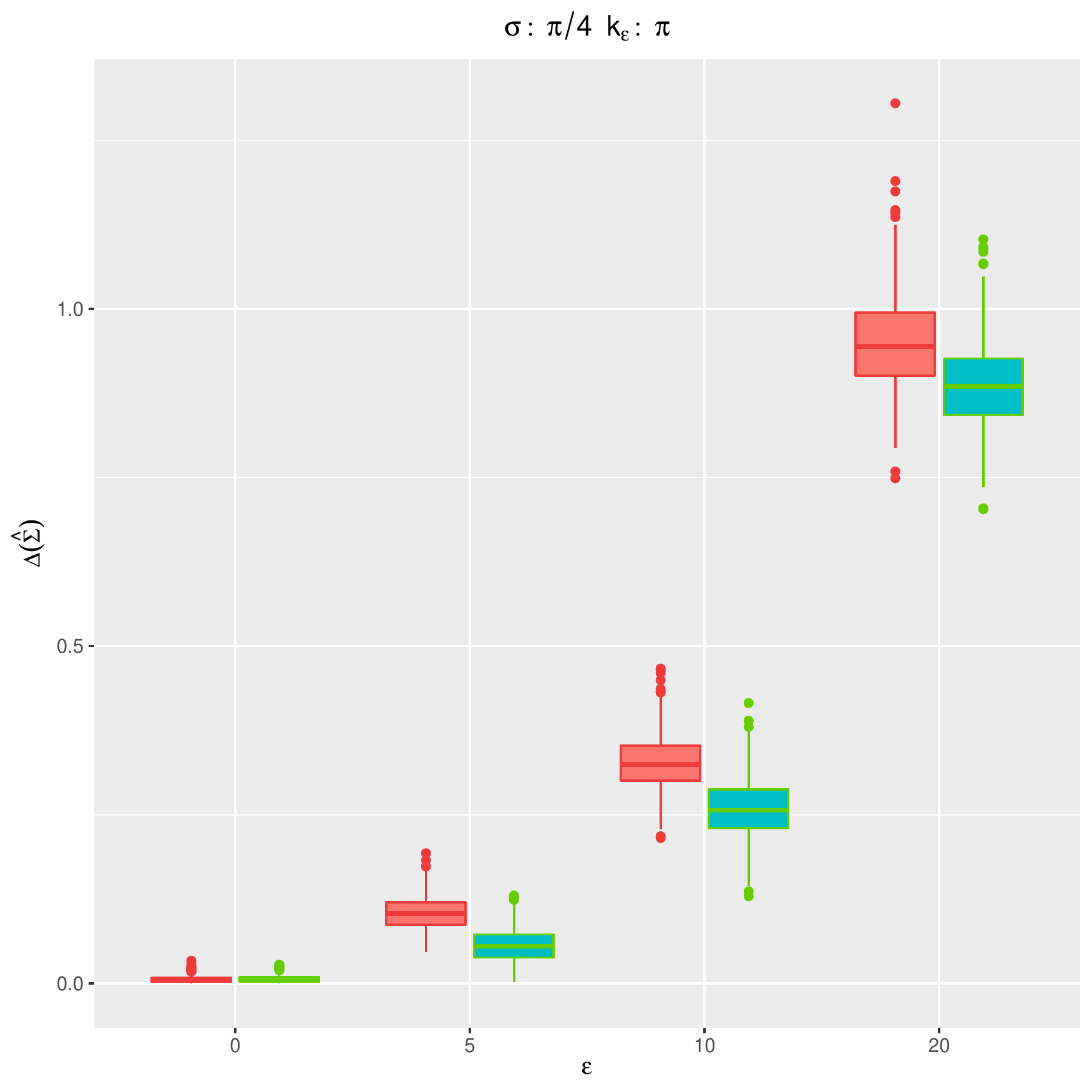} \\
 		\includegraphics[height=0.3\textheight, width=0.3\textwidth]{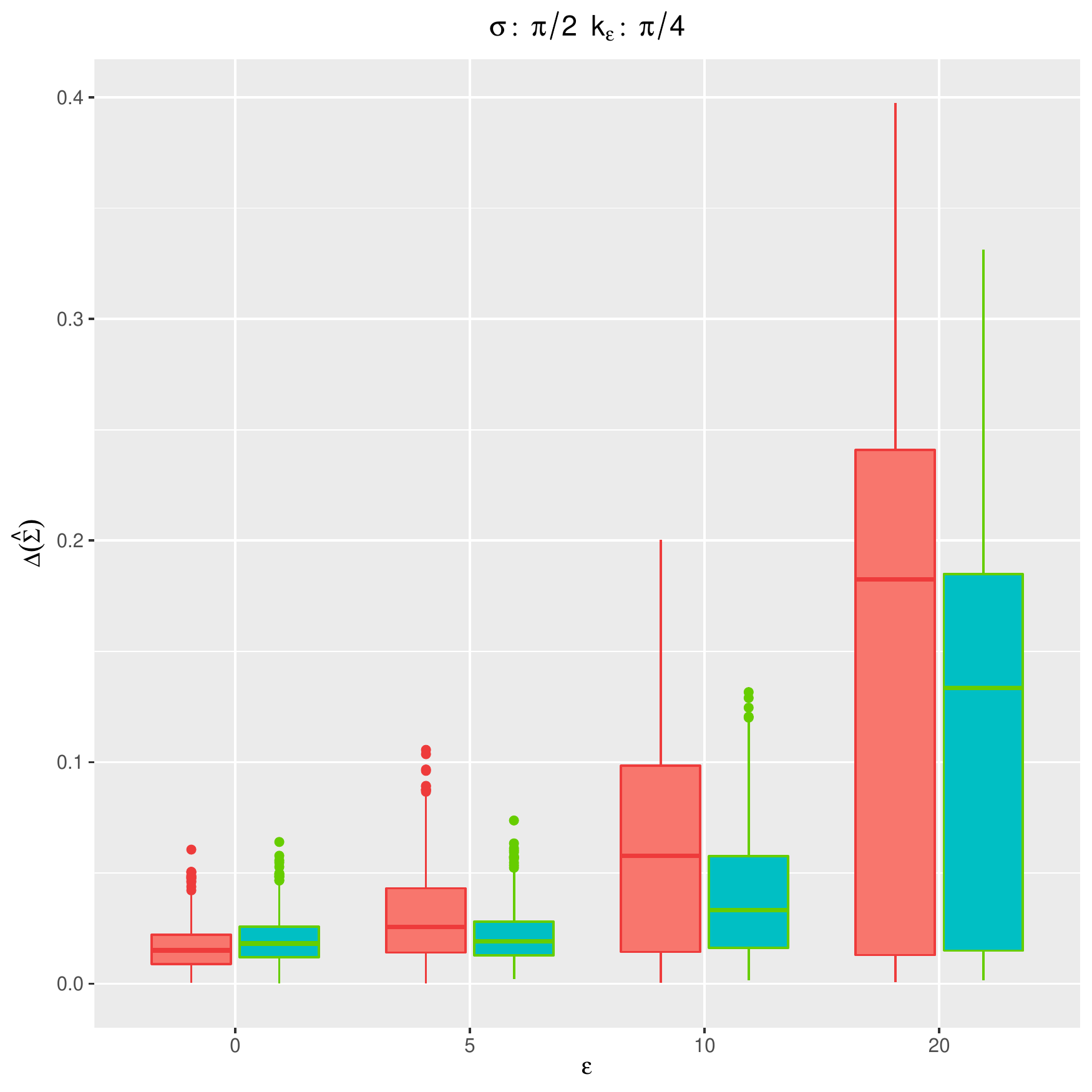} 
 		\includegraphics[height=0.3\textheight, width=0.3\textwidth]{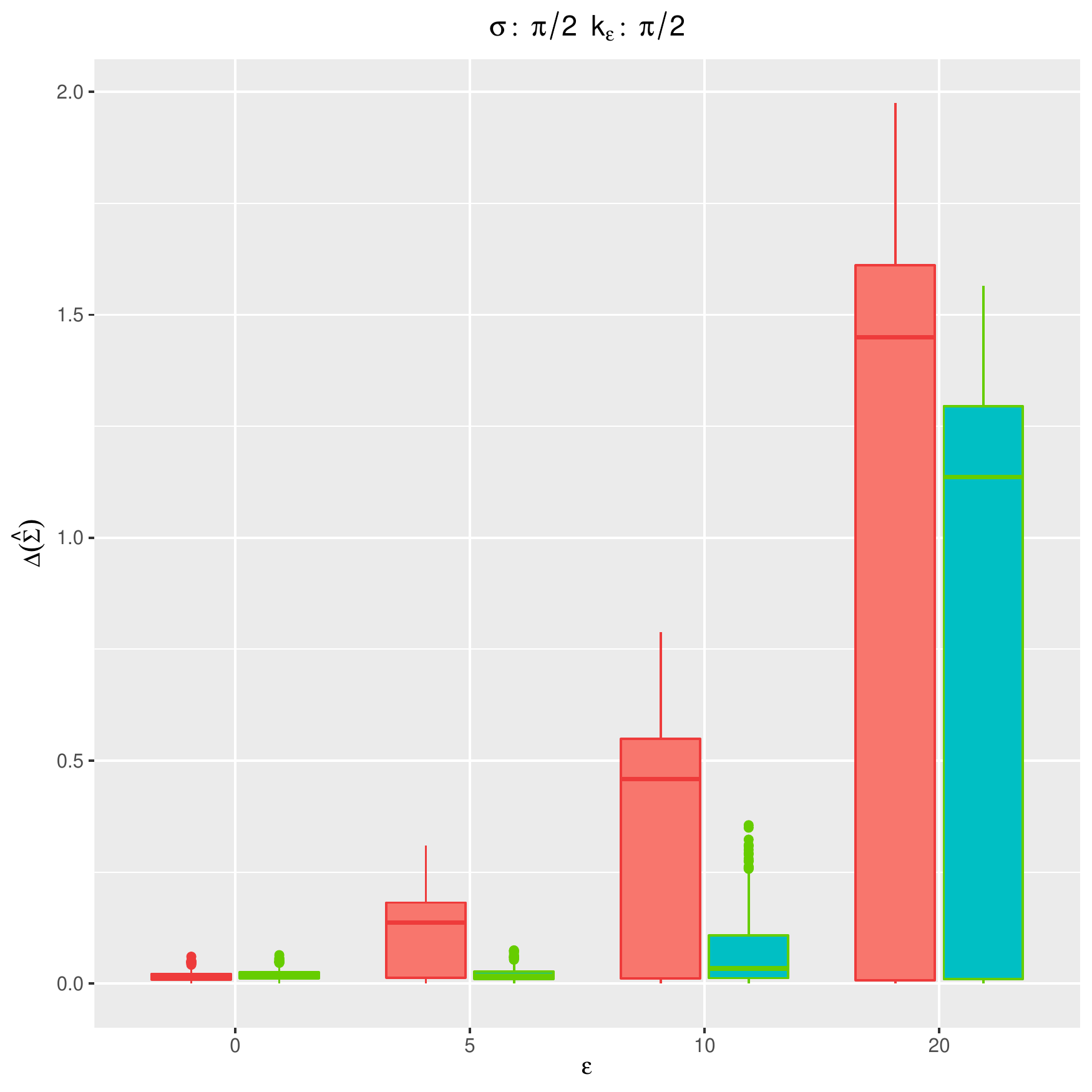} 
 		\includegraphics[height=0.3\textheight, width=0.3\textwidth]{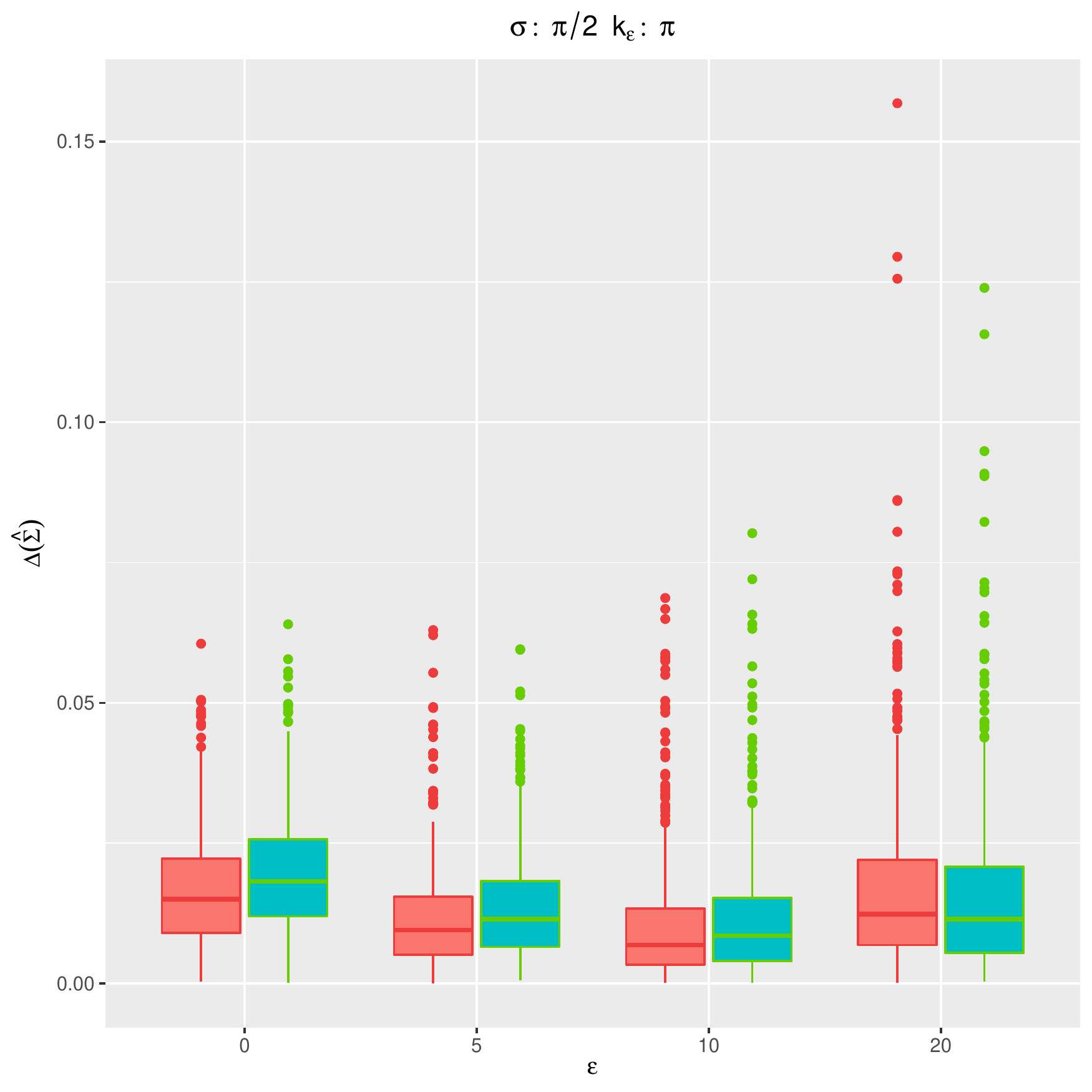} \\
 	\end{center}
 	\caption{Distribution of the divergence measure for $n=500$ and $p=2$ using the weighted CEM (in green) and the CEM (in red). The contamination rate $\epsilon$ is given on the horizontal axis. Increasing contamination size $k_\epsilon$ from left to right, increasing $\sigma$ from top to bottom.}
 	\label{fig:sm:6}
 \end{figure}

        Figure \ref{fig:sm:1}, Figure \ref{fig:sm:3} and Figure \ref{fig:sm:5} show the angle separation whereas Figure \ref{fig:sm:2}, Figure \ref{fig:sm:4} and Figure \ref{fig:sm:6} display the measure of accuracy in estimating the variance-covariance components for $p=2$ and $n = 50, 100, 500$, respectively.
        
\begin{figure}
	 \begin{center}
	 \includegraphics[height=0.3\textheight, width=0.3\textwidth]{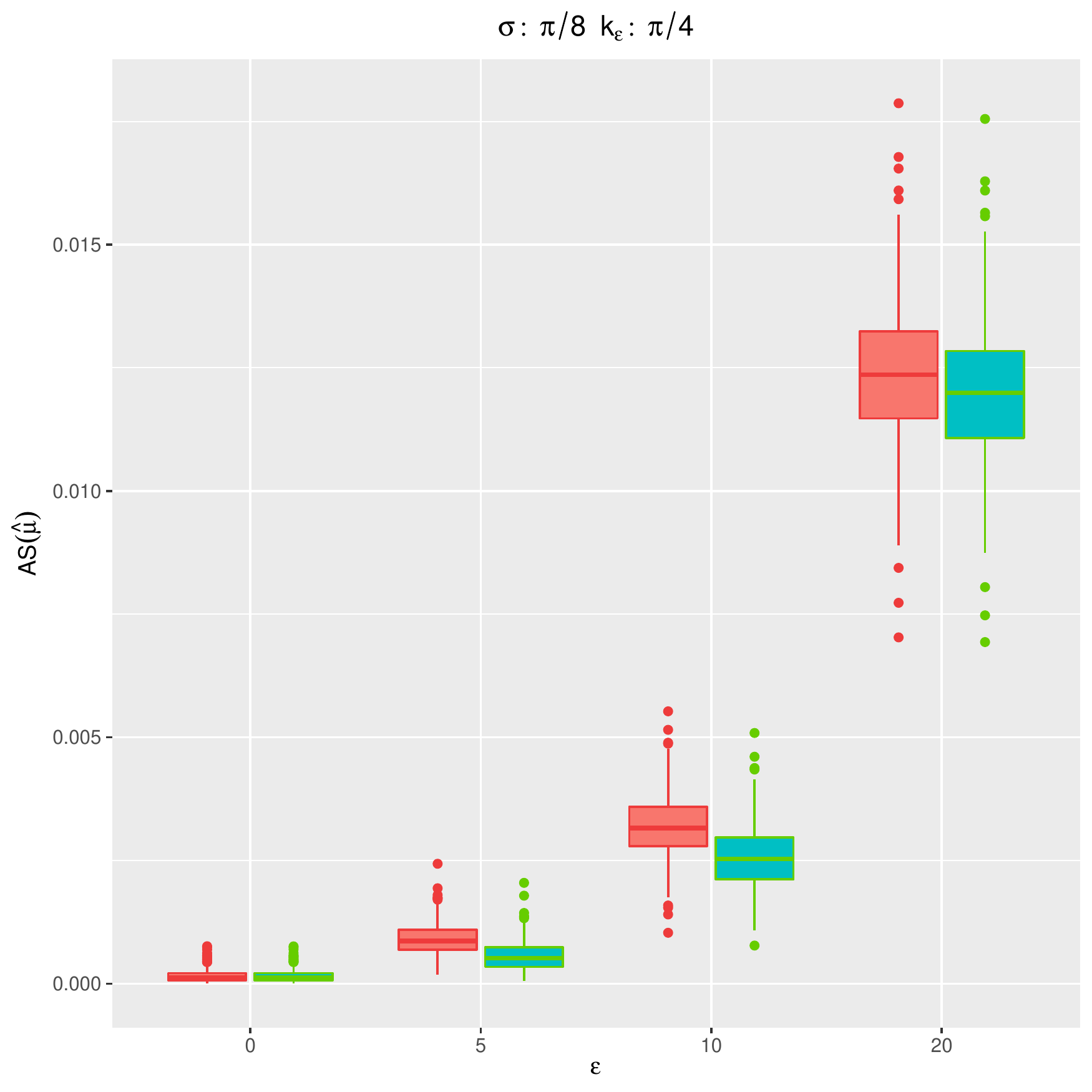} 
	 \includegraphics[height=0.3\textheight, width=0.3\textwidth]{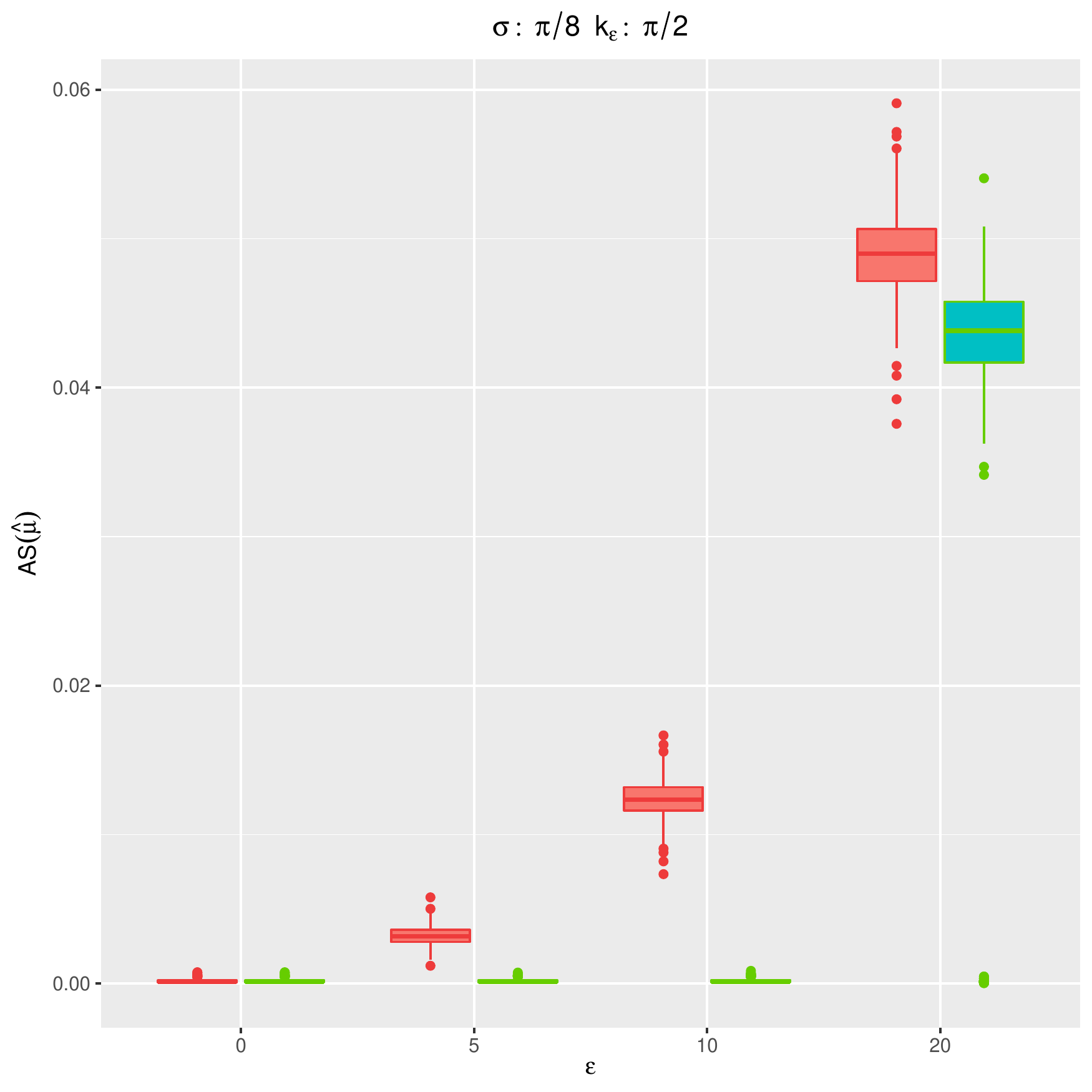} 
	 \includegraphics[height=0.3\textheight, width=0.3\textwidth]{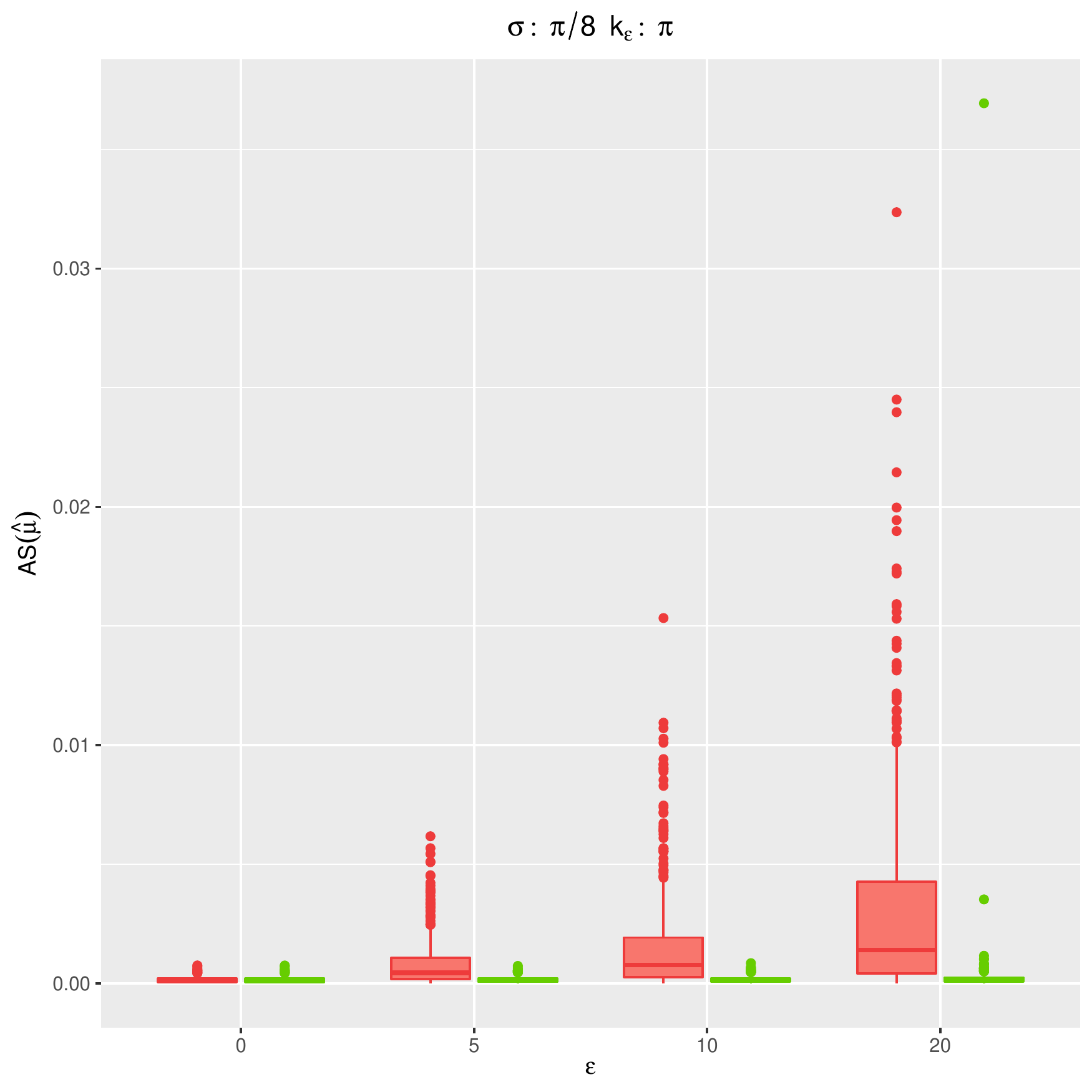} \\
	\includegraphics[height=0.3\textheight, width=0.3\textwidth]{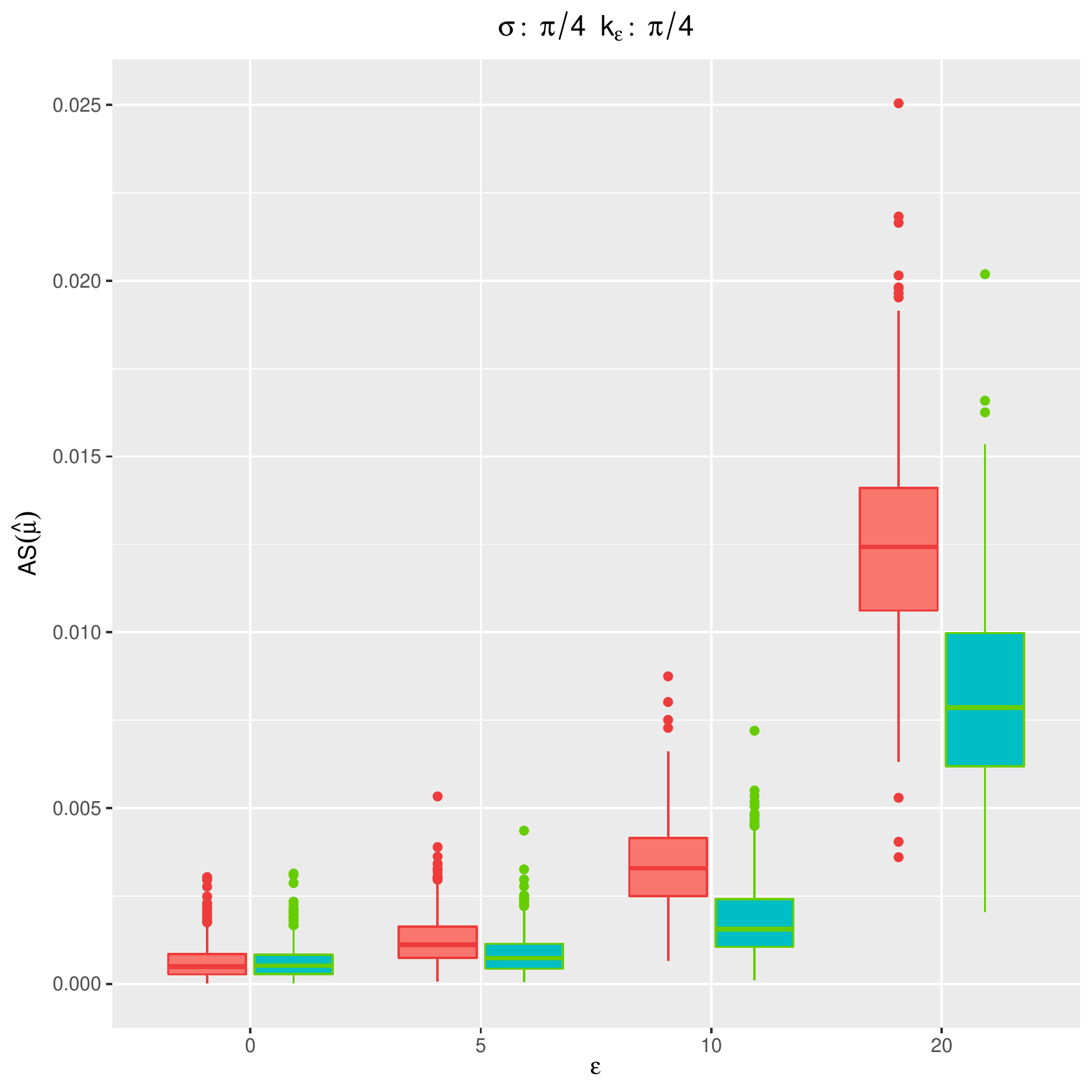} 
	 \includegraphics[height=0.3\textheight, width=0.3\textwidth]{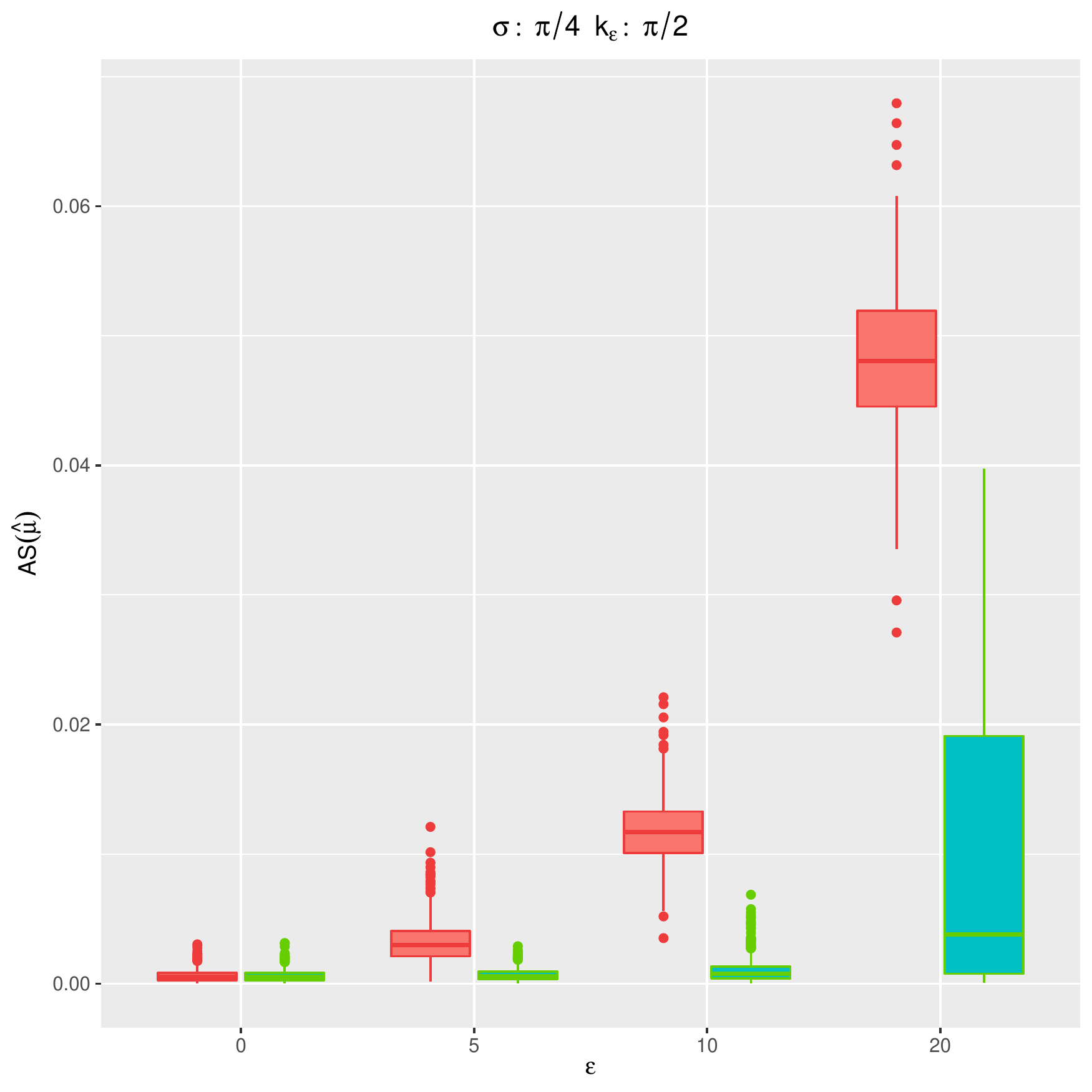} 
	 \includegraphics[height=0.3\textheight, width=0.3\textwidth]{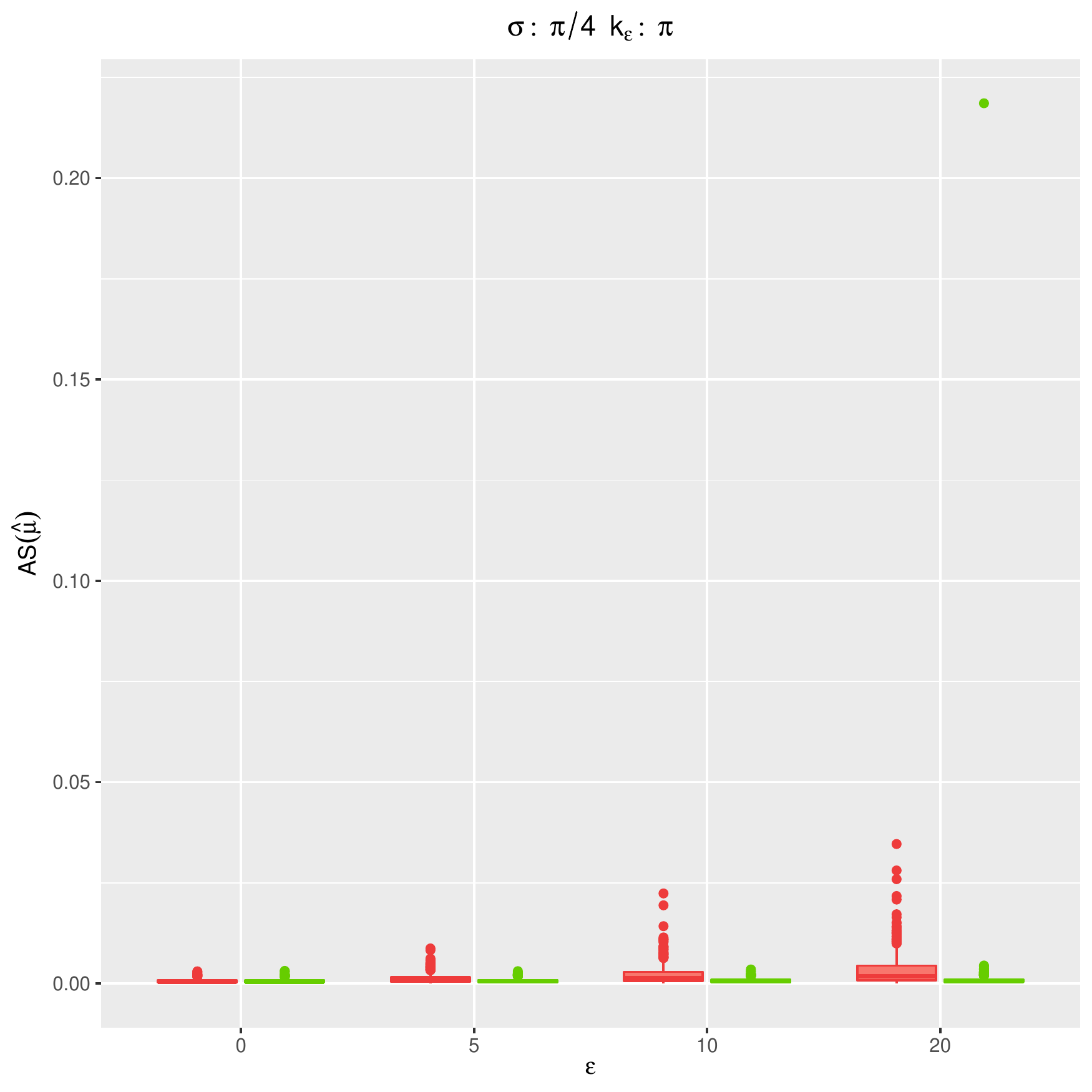} \\
	 \includegraphics[height=0.3\textheight, width=0.3\textwidth]{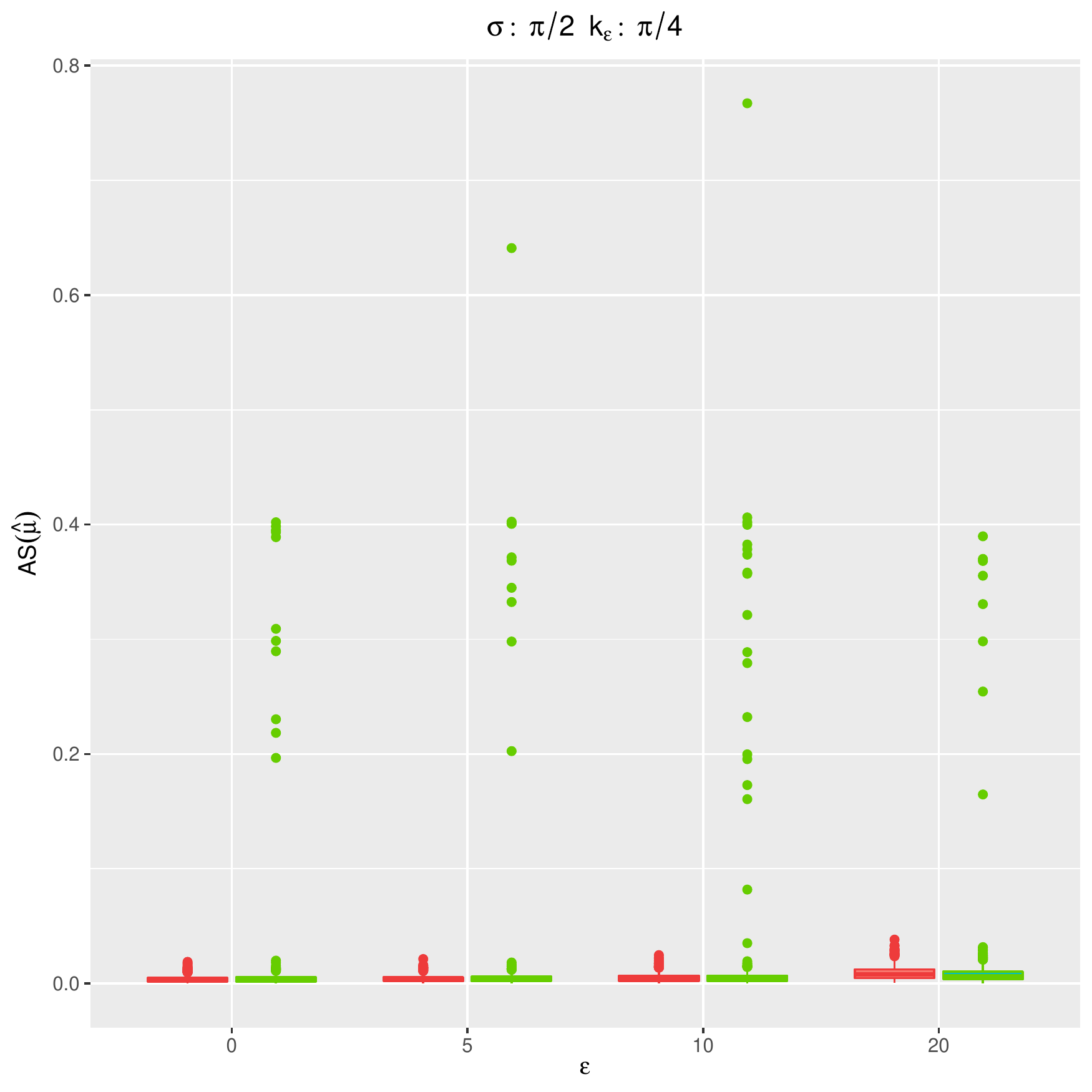} 
	 \includegraphics[height=0.3\textheight, width=0.3\textwidth]{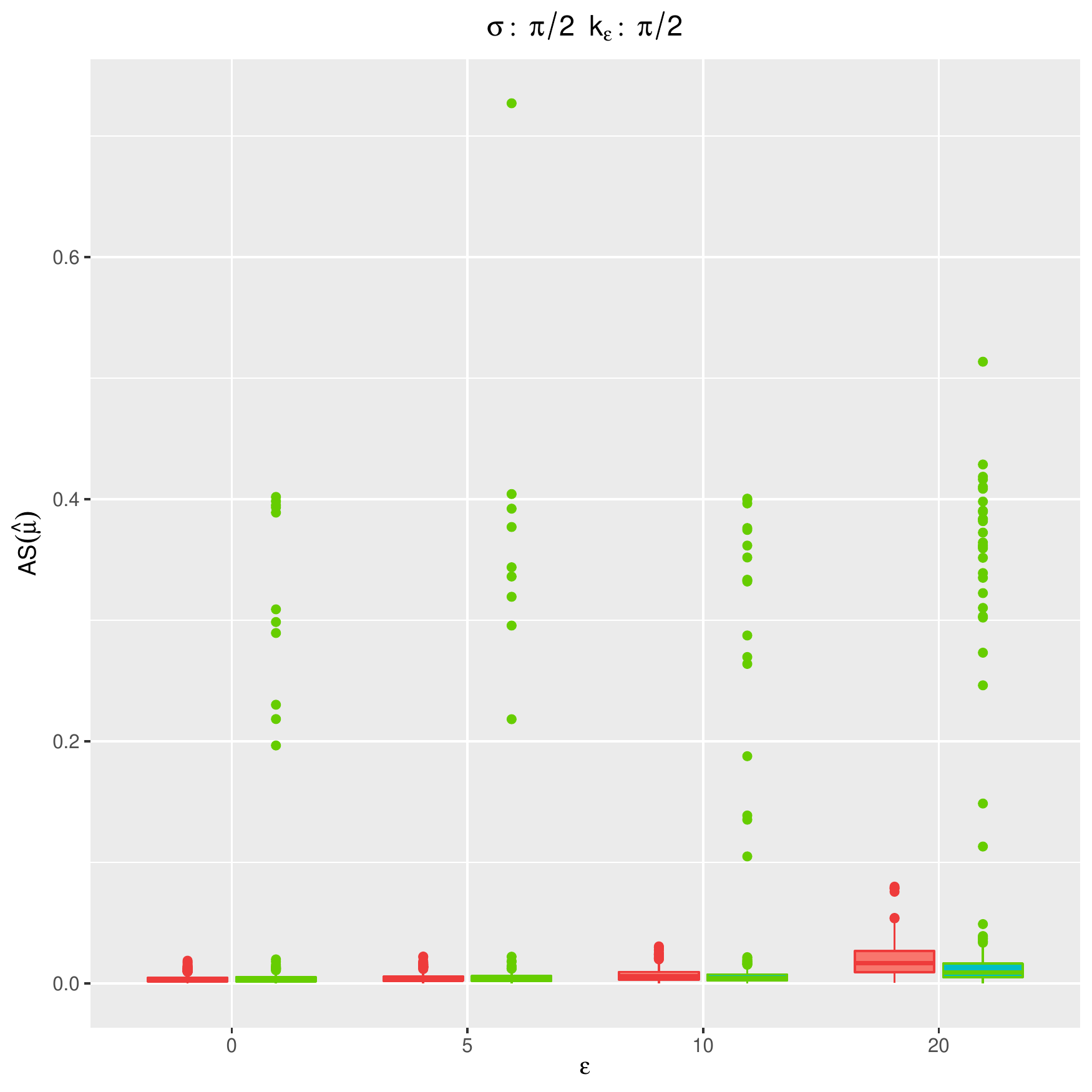} 
	 \includegraphics[height=0.3\textheight, width=0.3\textwidth]{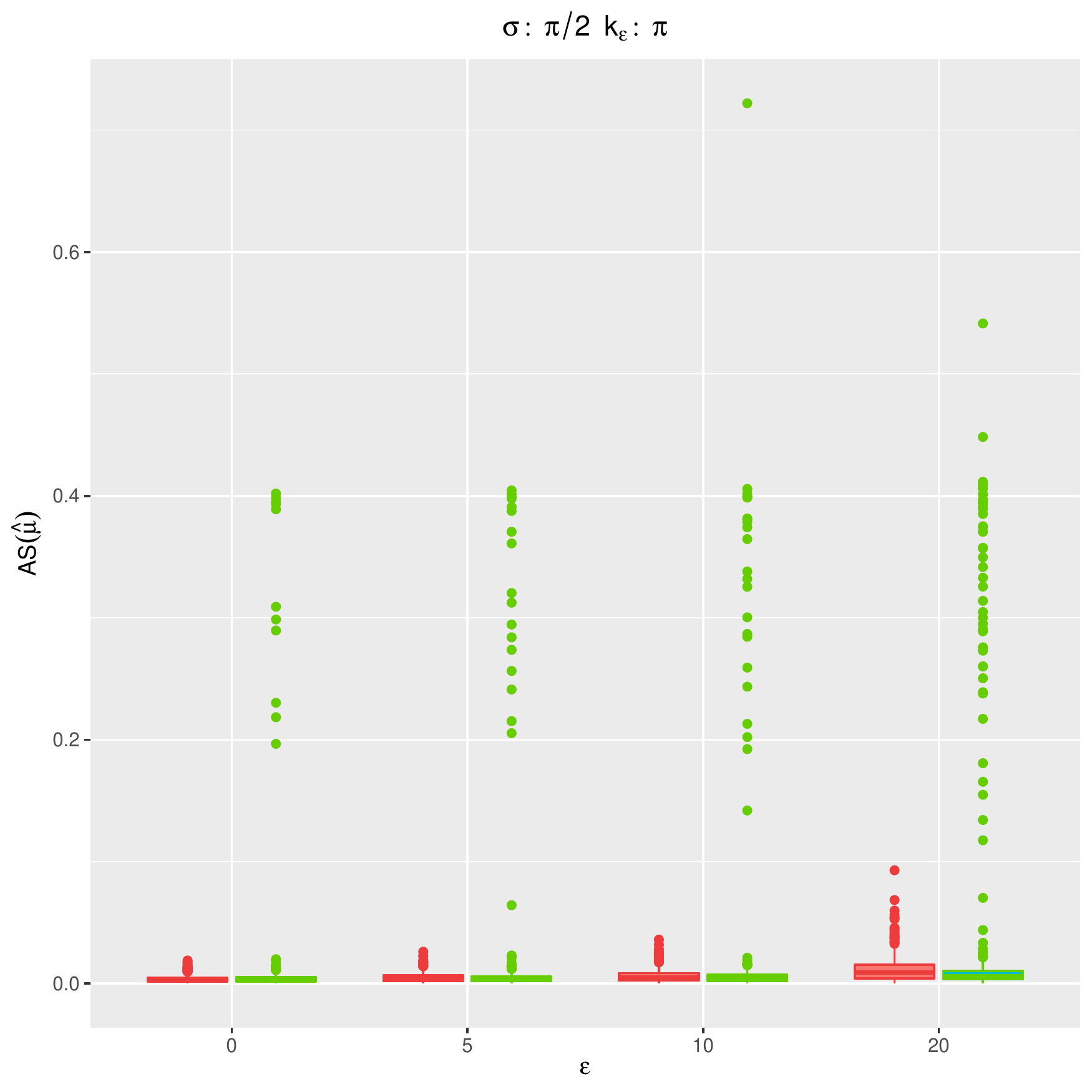} \\
	 \end{center}
	 \caption{Distribution of angle separation for $n=500$ and $p=5$ using weighted CEM (in green) and the CEM (in red). The contamination rate $\epsilon$ is given on the horizontal axis. Increasing contamination size $k_\epsilon$ from left to right, increasing $\sigma$ from top to bottom.}
	 \label{fig:sm:7}
\end{figure}

\begin{figure}
 	\begin{center}
 		\includegraphics[height=0.3\textheight, width=0.3\textwidth]{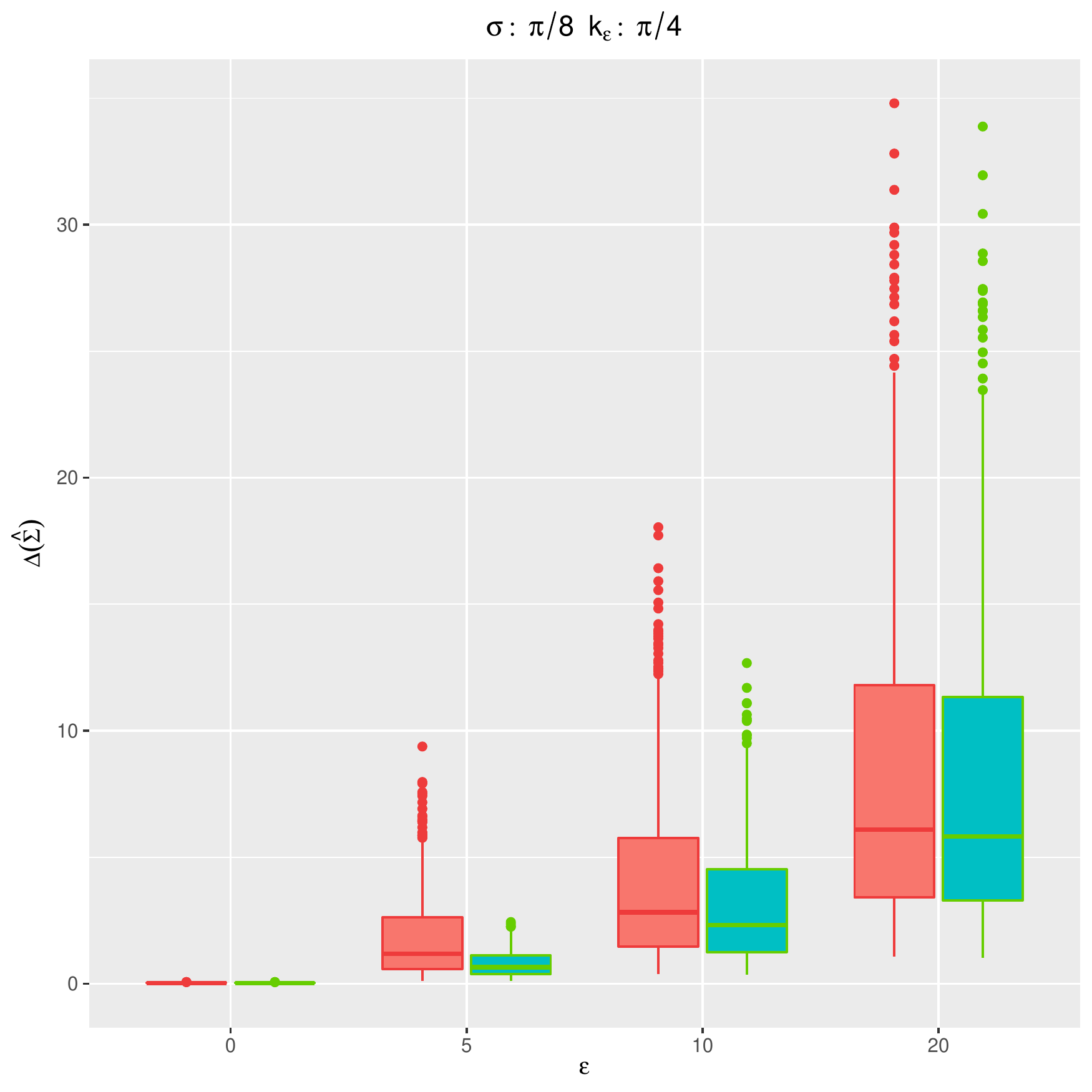} 
 		\includegraphics[height=0.3\textheight, width=0.3\textwidth]{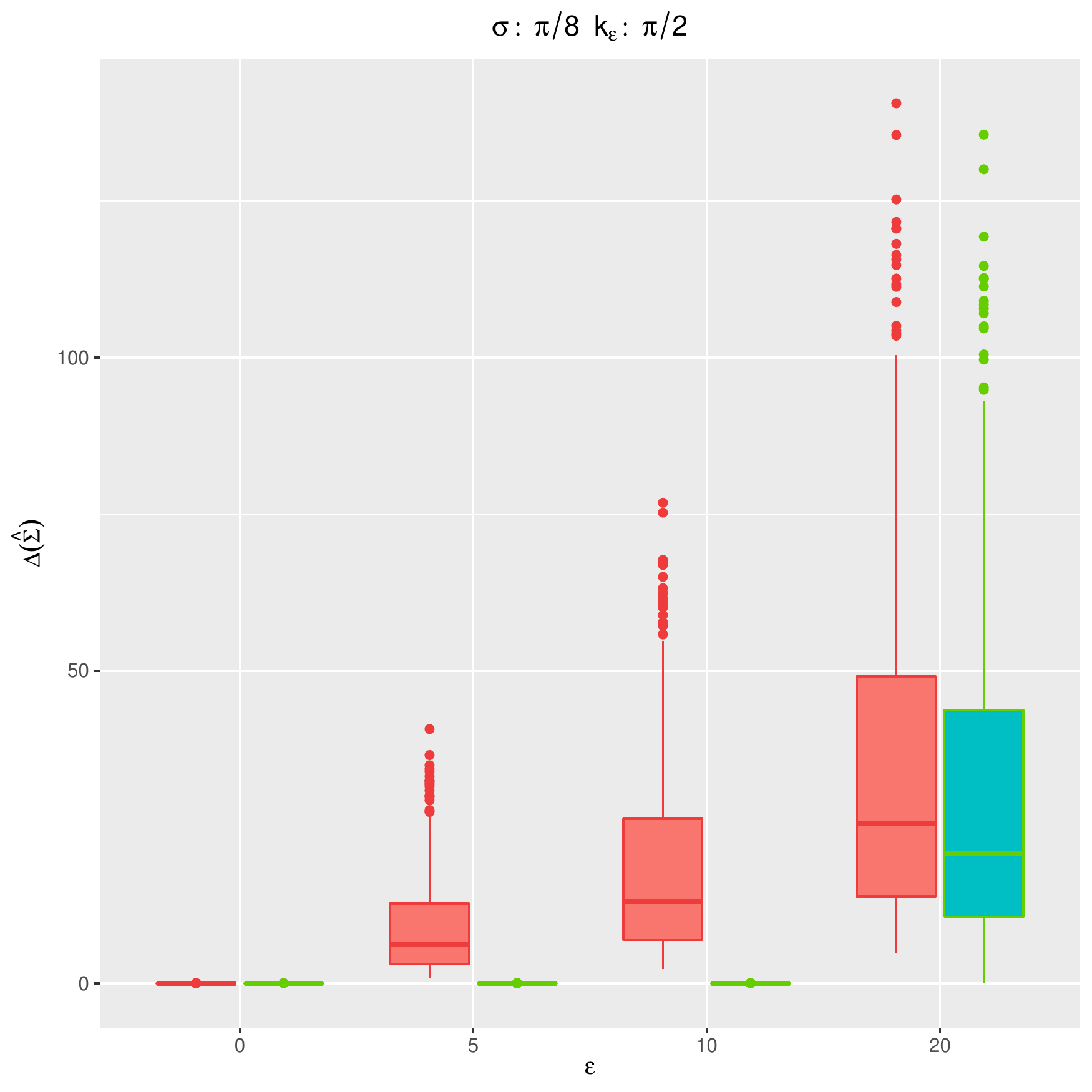} 
 		\includegraphics[height=0.3\textheight, width=0.3\textwidth]{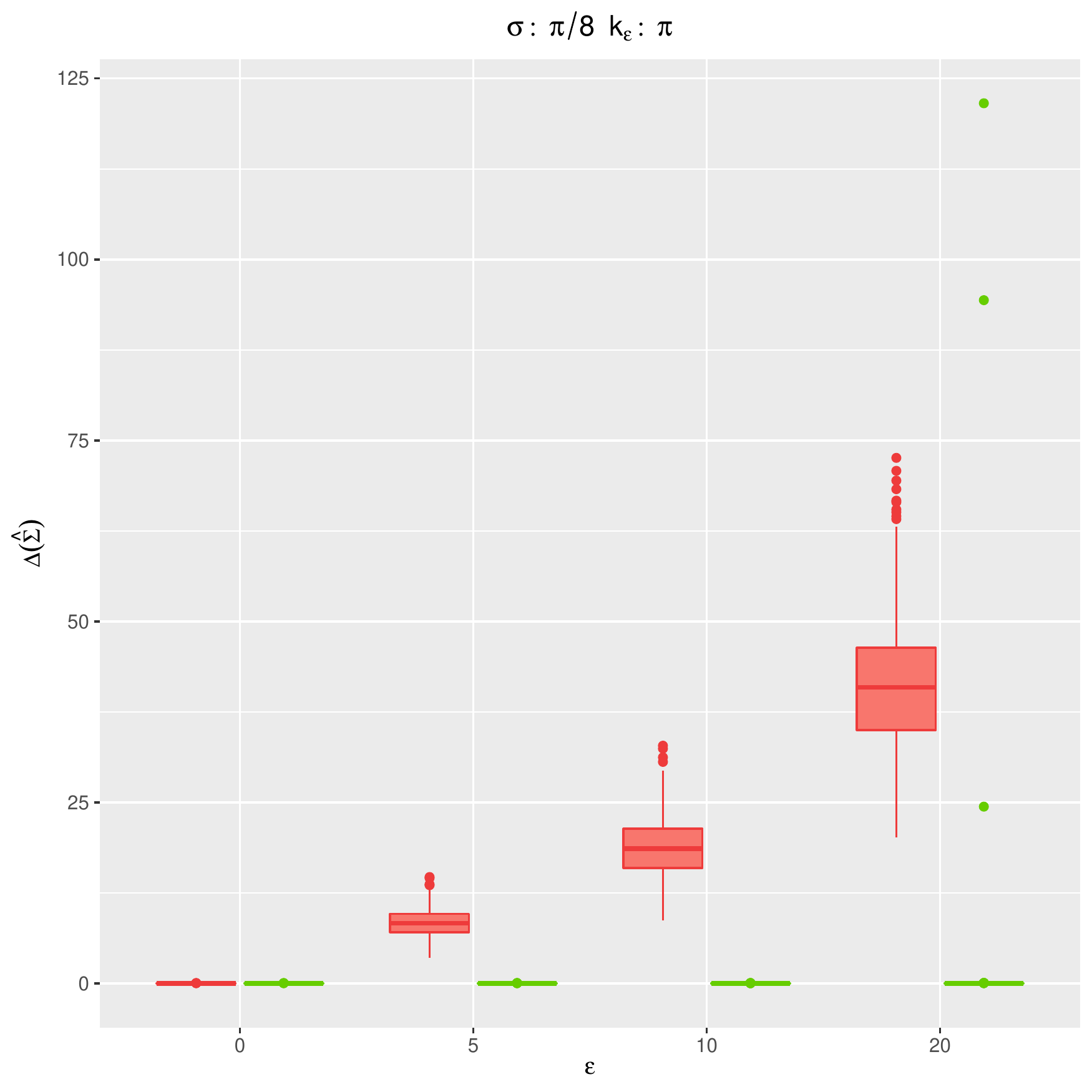} \\
 		\includegraphics[height=0.3\textheight, width=0.3\textwidth]{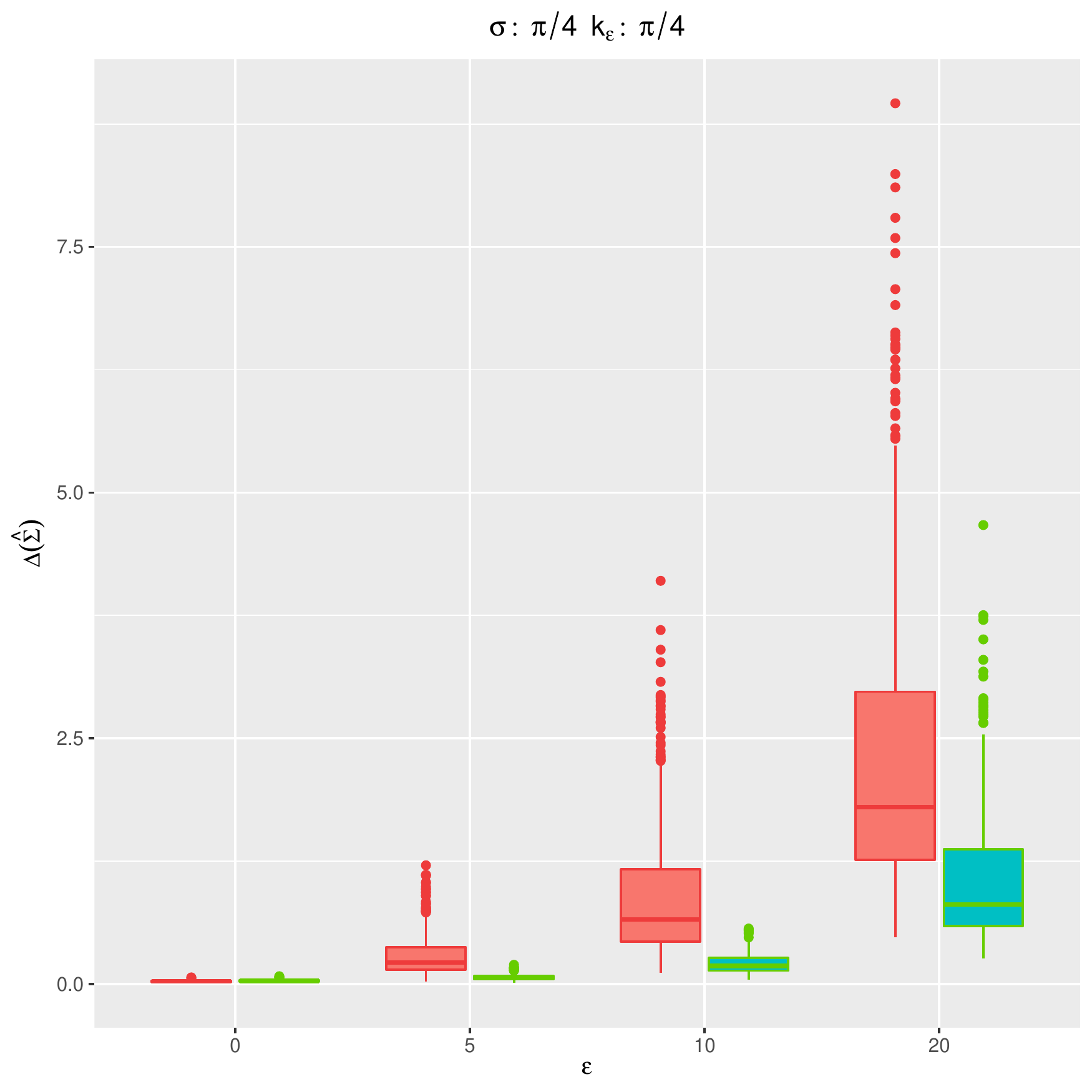} 
 		\includegraphics[height=0.3\textheight, width=0.3\textwidth]{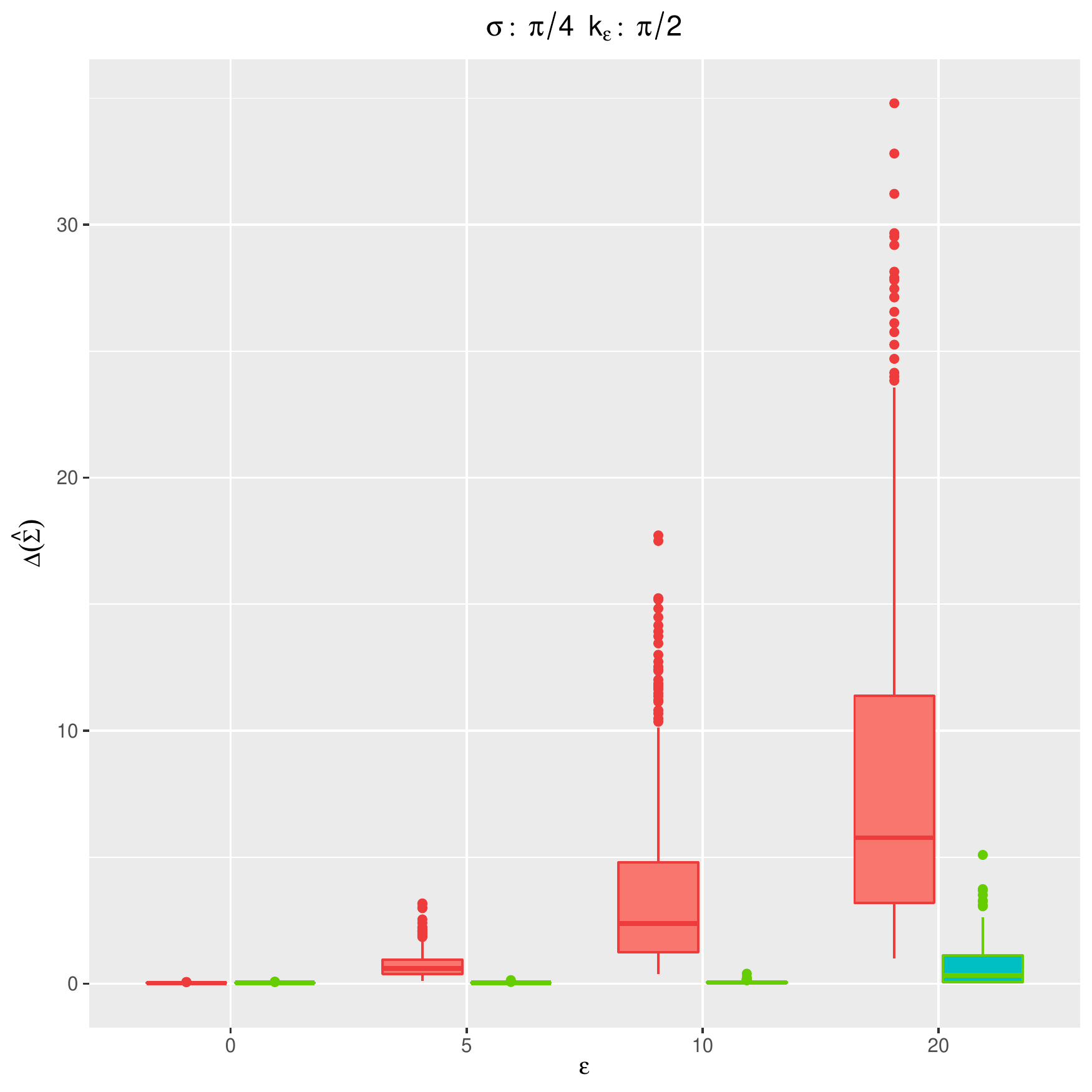} 
 		\includegraphics[height=0.3\textheight, width=0.3\textwidth]{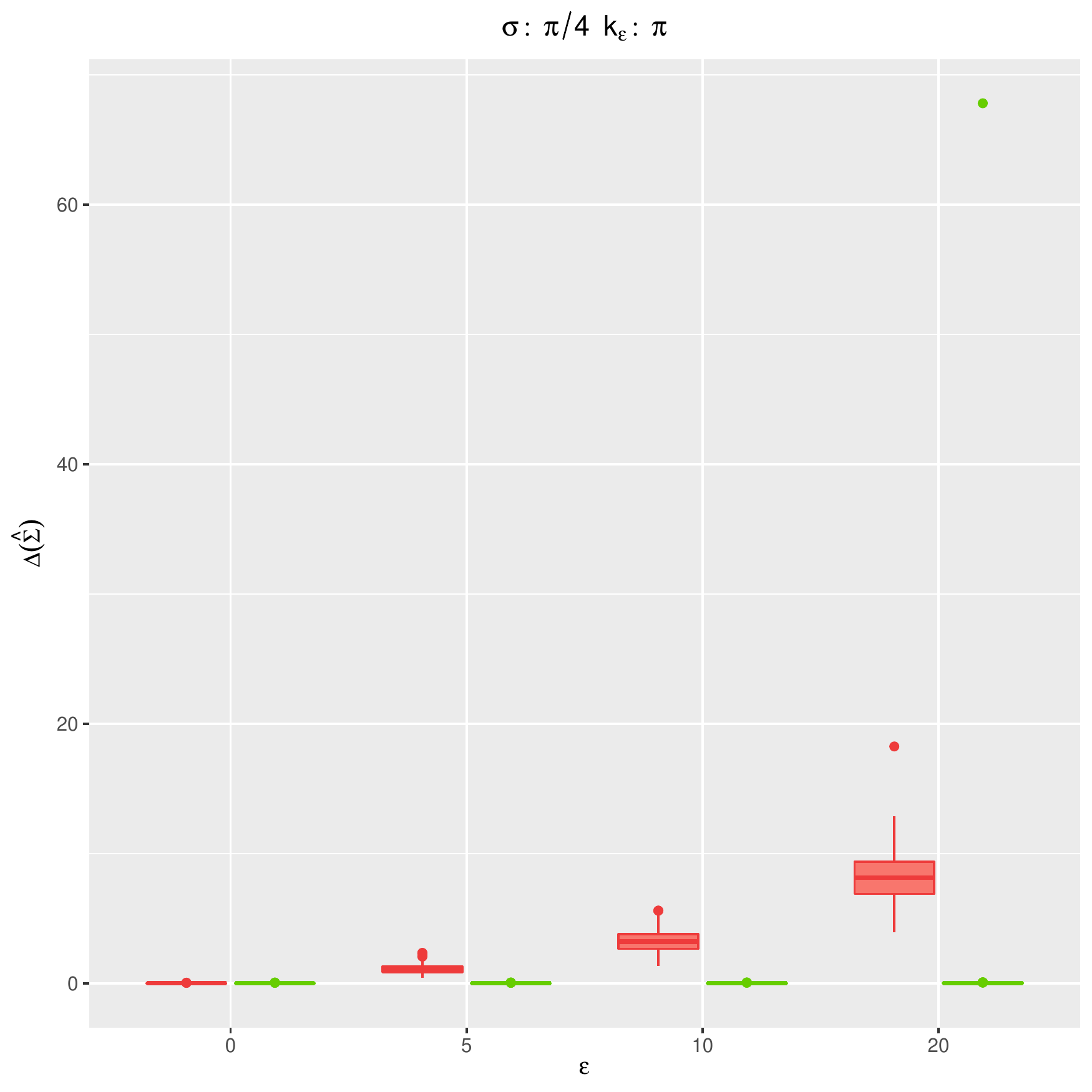} \\
 		\includegraphics[height=0.3\textheight, width=0.3\textwidth]{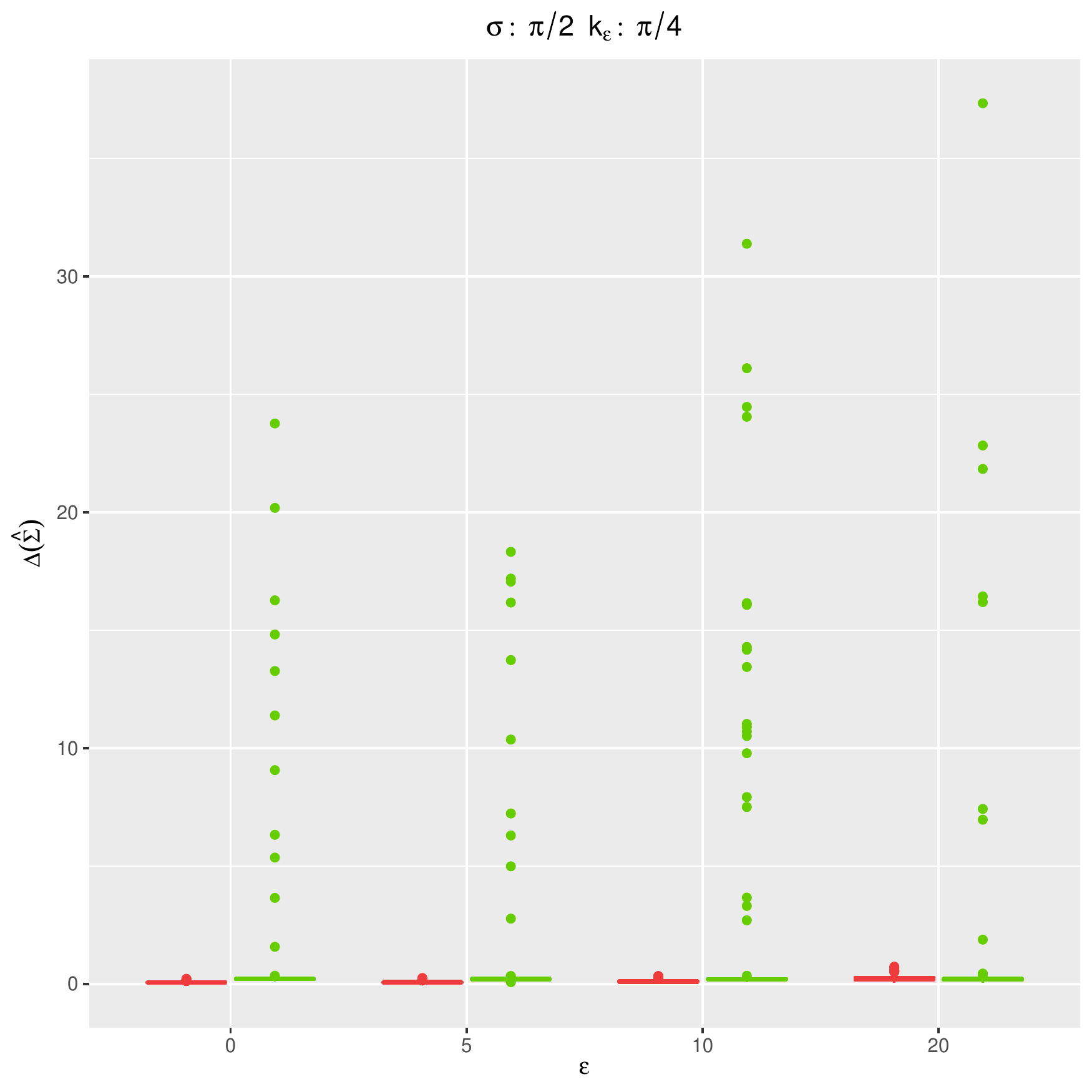} 
 		\includegraphics[height=0.3\textheight, width=0.3\textwidth]{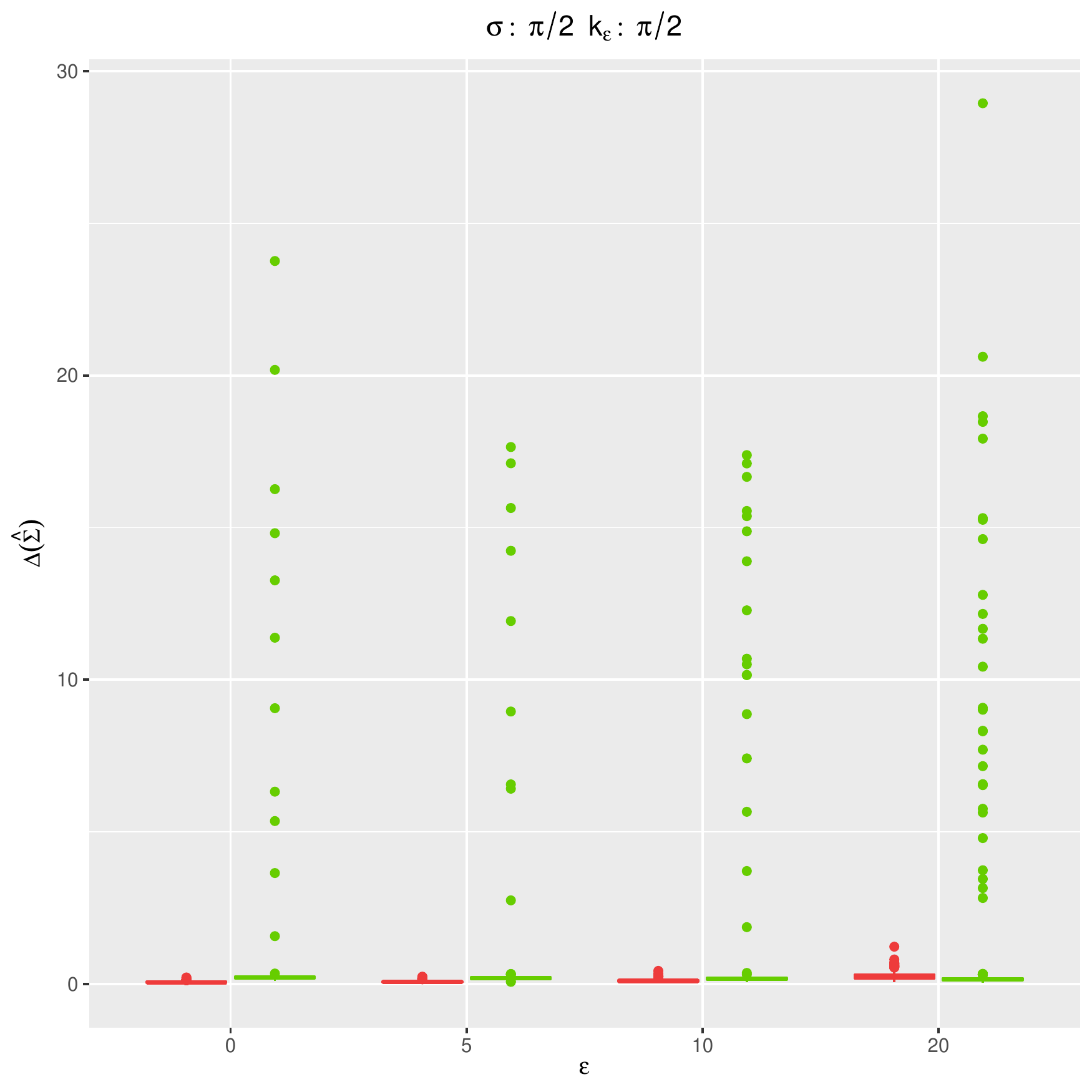} 
 		\includegraphics[height=0.3\textheight, width=0.3\textwidth]{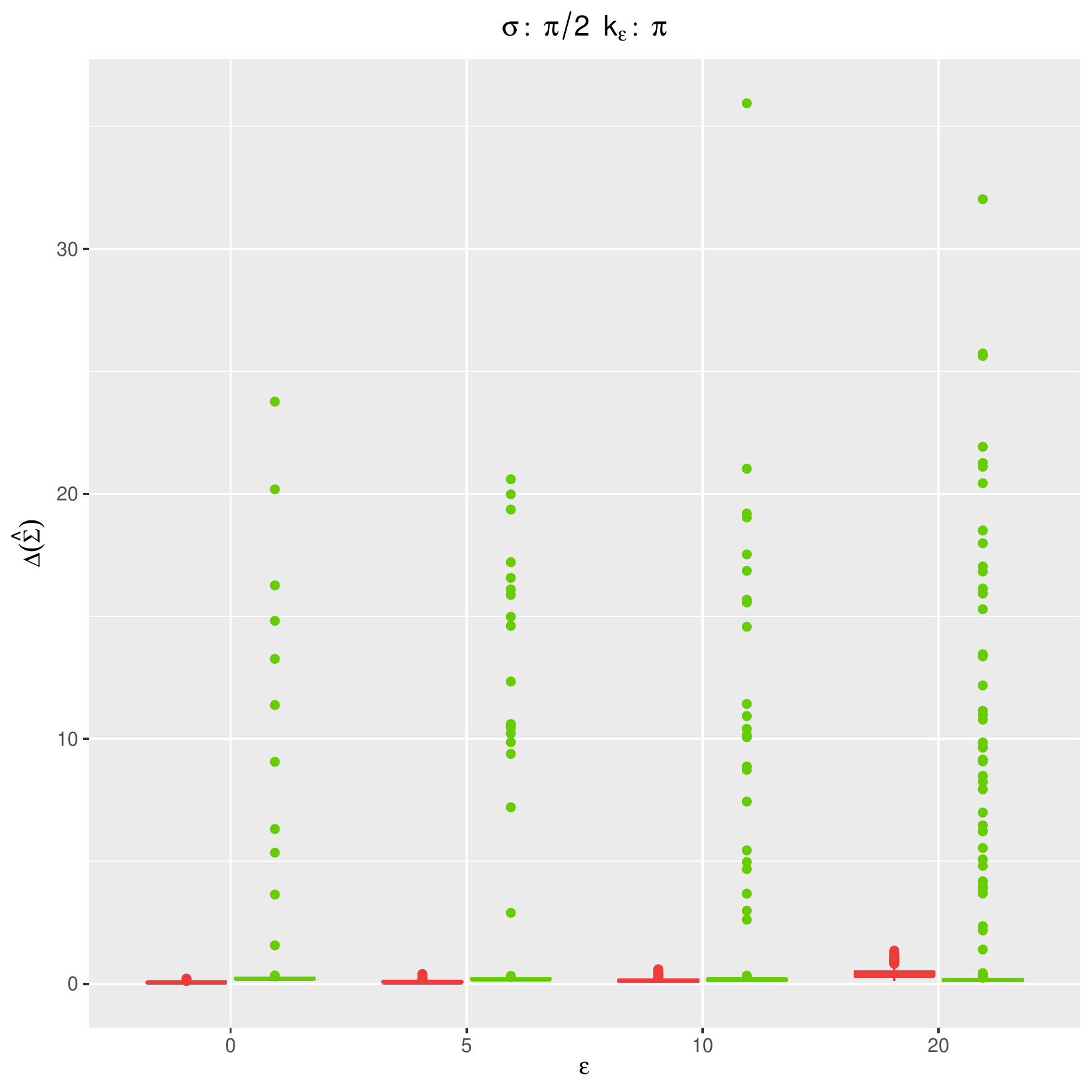} \\
 	\end{center}
 	\caption{Distribution of the divergence measure for $n=500$ and $p=5$ using the weighted CEM (in green) and the CEM (in red). The contamination rate $\epsilon$ is given on the horizontal axis. Increasing contamination size $k_\epsilon$ from left to right, increasing $\sigma$ from top to bottom.}
 	\label{fig:sm:8}
\end{figure}

  Figure \ref{fig:sm:7} show the angle separation whereas Figure \ref{fig:sm:8} display the measure of accuracy in estimating the variance-covariance components for $p=5$ and $n = 500$, respectively.
 
\bibliography{robtorus}